\documentclass[aps,ams,prb,twocolumn]{revtex4-1}
\usepackage{longtable}
\usepackage{subfigure}
\usepackage{amsmath,amssymb,amsfonts,bm}
\usepackage{graphicx}
\pdfoutput=1
\usepackage{dcolumn}
\usepackage{color}

\def\KeyWord#1{$\backslash$\IfColor{$\!\!$\textRed{#1}\textBlack}{#1}$\!\!$}

\newcommand{\be}{\begin{equation} }
\newcommand{\ee}{\end{equation} }
\newcommand{\ba}{\begin{eqnarray} }
\newcommand{\ea}{\end{eqnarray} }
\newcommand{\n}{\nonumber \\ }

\def\shs#1{{\color{blue} #1}}

\def\chainmail{Chain-Mail }
\def\mycup{\cup }
\newcommand{\mac}{\mathcal}
\def\em{\it}

\begin{document}
\include{epsfx}
\title{Space-Time Geometry of Topological phases}
\author{F. J. Burnell$^{1,2}$ and Steven H. Simon$^1$ \\ $^1$Rudolf Peierls Centre for Theoretical Physics,  University of Oxford, Oxford OX1 3NP, UK
\\ $^2$All Souls College, Oxford, UK}
\date{\today}

\begin{abstract}
The 2+1 dimensional lattice models of Levin and Wen [PRB 71, 045110 (2005)] provide the most general known microscopic construction of topological phases of matter.  Based heavily on the mathematical structure of category theory, many of the special properties of these models are not obvious.  In the current paper, we present a geometrical space-time picture of the partition function of the Levin-Wen models which can be described as doubles (two copies with opposite chiralities) of underlying Anyon theories.  Our space-time picture describes the partition function as a knot invariant of a complicated link, where both the lattice variables of the microscopic Levin-Wen model and the terms of the hamiltonian are represented as labeled strings of this link.   This complicated link, previously studied in the mathematical literature, and known as \chainmail, can be related directly to known topological invariants of 3-manifolds such as the so called Turaev-Viro invariant and the Witten-Reshitikhin-Turaev invariant.   We further consider quasi-particle excitations of the Levin-Wen models and we see how they can be understood by adding additional strings to the \chainmail link representing quasi-particle world-lines.  Our construction gives particularly important new insight into how a doubled theory arises from these microscopic models. 
\end{abstract}

\maketitle

\section{Introduction}

The intersection between topology and the physics of strongly correlated systems harbors great theoretical and practical interest.  Motivated by the dream of building topologically protected qubits, numerous approaches to finding topological phases of matter in physical materials have been explored.  In this work, we focus on perhaps the most prominent attempt to build theoretical lattice models that exhibit topological properties -- the string net models of Levin and Wen\cite{LW}.  We situate these models in the context of other mathematics and physics-based approaches to topological theories.  In doing so, we seek not only a new understanding of the models themselves, but also a methodology which allows us to generalize these constructions in potentially fruitful ways.

The origins of this field of study extend to the 1980s, when the notion of a
``topological quantum field theory"  (TQFT) was introduced.  These TQFTs were first studied in the context of
gravitation\cite{Witten}
with the goal of
constructing a metric-independent action.  Though
ultimately their success as a theory of gravity was limited to $2+1$
spatial dimensions, the exploration forged several profound connections between topology and
physics.
Witten's groundbreaking work\cite{WittenJones} showed that there is a very deep and powerful link between Chern-Simons gauge theory, conformal field theory, and mathematical invariants of knots, links, and 3-manifolds.  Growing from this discovery, as well as from other developments in the the study of conformal field theory\cite{MooreSeiberg}, was the realization that in $2+1$ dimensions, point particles can have exotic (``anyonic") quantum statistics\cite{MooreRead,Frohlich} described by a non-abelian topological field theory.  Such statistics are far richer than those of bosons and fermions, and even beyond the fractional statistics realized by the Jain series of fractional quantum Hall states.

More recently, anyon theories in $2+1$ dimensions have received renewed attention in the context of topological phases of matter\cite{NayakReview} --- phases of strongly interacting matter whose low energy, long wavelength description is a topological quantum field theory, and whose excitations consequently have anyonic statistics.   Proposals made by Freedman\cite{FreedmanPNAS} and Kitaev\cite{KitaevToric}
that such phases of matter could be used to build powerful quantum computers provide a practical motive for attempting to realize these in the lab.

While there has been some evidence\cite{NayakReview} that certain quantum Hall states\cite{MooreRead} may be realizations of extremely nontrivial topological phases of matter, suitable for quantum computation, experiments in this field are difficult and progress has remained challenging.  As a result researchers have begun to consider what other system might be found which could display similar exotic statistics.
Earlier work on lattice spin models displaying fractional statistics\cite{SondhiRVB, KitaevToric} inspired attempts\cite{KitaevVeryLongPaper,LW,FendleyTopological} to construct  lattice model Hamiltonians exhibiting nontrivial topological phases.  Among these constructions, the Levin-Wen models\cite{LW} are of particular importance
as they can exhibit the most general range of possible quasi-particle
statistics of any known Hamiltonian lattice theory.
Indeed, Hamiltonians of this kind are thought to exist
for a large class of achiral anyon theories known as quantum doubles.  This class encompasses many of the previously studied anyon lattice models such as the Toric code\cite{KitaevToric}, many of its generalizations\cite{CastelChamon, FradkinShenker, ArdonneFF} and doubled Chern-Simons theories\cite{FreedmanShtengleetc}.  Further, as the Hamiltonians are exactly solvable, a complete
understanding of the spectra and behavior is possible.  As such,
they serve as a useful prototype for understanding topological
behavior in more complex systems.  Because
these models realize such a wide range of topological phases, they
may serve as a guide in designing new materials which support the
anyonic statistics essential for topological quantum computation\cite{NayakReview}.

In this work, we focus on a large class of Levin-Wen models that are equivalent to doubled anyon theories (those in which each type of anyon appears in two opposite-chirality copies).
We will expose the connections between these models  and certain topological invariants developed in the context of knot theory.  Specifically, we show how a subset of the Levin-Wen models are related to the knot-theoretic topological invariants of 3-manifolds discovered by Witten\cite{WittenJones} and Reshitikhin and Turaev\cite{RT}.  In particular we show how the vacuum partition function of these models can itself be viewed as a knot-invariant, first introduced by Justin Roberts \cite{RobertsThesis}, known as the \chainmail link.  Further, we show how quasi-particle excitations can be incorporated into this topological description.
On the one hand, the inclusion of quasi-particles, or defects, into 3-manifold invariants such as the \chainmail link, is new to the mathematics community (see, however, Refs. \onlinecite{Martins1}, \onlinecite{Martins2});  whereas on the other hand, the use of these particular mathematical invariants to describe space-time processes of physical systems is new to the physics community.

Our approach gives a direct visualization of space-time processes and how they can be interpreted as knots that represent configurations of quantum numbers on a lattice.  In so doing we obtain a number of new ways of looking at these systems.
First, this re-interpretation of the Levin-Wen partition function renders explicit the connection between these
models and the (continuum) topological field theories, such as Chern-Simons theory, familiar in
other areas of physics. Second, we provides a simple proof that
the ground state partition function of these models give exactly the
Turaev-Viro invariant\cite{TuraevViro}, a well-known mathematical quantity (see also Ref.~\onlinecite{Raseti}). Third, our approach is not tied to any particular lattice structure, and makes obvious how Levin-Wen models may be generalized to arbitrary non-trivalent graphs and lattices.
Finally,  our work gives a new perspective on the nature of ``doubling" --- that is, the production of two copies with opposite chirality-- in these models.  Another recent work\cite{Gilsetal} also gives an interesting description of the connection between Levin-Wen models and doubled Chern-Simons theory (which can be generalized to describe other quantum double theories\cite{PictTQFT}),  by expressing the lattice models as models of closed surfaces on two sheets of opposite chirality.   Though this description is superficially quite different from ours -- and indeed, Ref.~\onlinecite{Gilsetal} focuses principally on describing phase transitions {\em out of} from the purely topological limit by studying fluctuations in the surface topology -- the mathematical underpinnings of the two pictures are closely related.

In situating the Levin-Wen models in the context of the
mathematical discourse on topological field theory, we also hope
to provide a useful framework from which to approach topological
theories on the lattice.  This framework is geometrical in
character -- the model can be phrased in a `pictorial' manner in
which closed loops of string in $2+1$ dimensional
spacetime represent both the quantum degrees of freedom of the model and the operators, such as the Hamiltonian, that act on these degrees of freedom.  While this mapping may sound unusual, it has some distinct advantages.  Notably, once this mapping is made, a large amount of tensor calculus is reduced to trivial geometric manipulations.
We find that certain properties of the
Hamiltonian -- namely, topological invariance and the fact that the Hamiltonian is composed of commuting projectors, and hence is
exactly solvable -- are  reflected in special allowed
geometrical re-arrangements of these strings which leave the
partition function invariant.  
In the mathematical context, these
re-arrangements are in both cases related to deformations of the
space-time which preserve its topology.  (Similar re-arrangements can also
be used to coarse-grain the system, giving a natural interpretation of the exactness of 
the tensor renormalization group of \onlinecite{LevinNave, WenGu} for these states.)
In this sense, our
geometrical construction makes apparent certain connections
between the Hamiltonian-driven approach of Levin and Wen\cite{LW},
and the mathematical approach to topological invariants of 3-manifolds via the
study of knots\cite{WittenJones}.

This paper is structured as follows: broadly speaking, section \ref{OverSect} gives a conceptual overview of the general ideas in this paper, sections \ref{LWSect}, \ref{CHSect}, and \ref{QPSect} flesh out the technical details and explain the correspondence to the mathematical literature, while section \ref{NTriSect} briefly discusses applications of these results to generalizing the model.

Specifically, we begin in section \ref{OverSect} by
focusing on Chern-Simons theory as an easy entry point into our work.  We review the salient features of this theory in section \ref{sub:CStheory}. Section \ref{sub:latticeCStheory} explains the connection between the lattice model of Levin and Wen and the continuum Chern-Simons theory, using loops of Wilson lines to represent quantum numbers on the edges of the lattice as well as to represent the operators that act upon these quantum numbers.  We describe roughly how this construction results in an effective theory for the lattice model which is the double of the original Chern-Simons theory. 
Section \ref{LWSect} gives the technical details of our construction of the ground state partition function of the Levin-Wen lattice models.  We begin in section \ref{sub:categories} with a very brief introduction to the structure of anyon theories (or ``Modular Tensor Categories"), and discuss Levin-Wen lattice  models built on this structure in section \ref{sub:LWmodels}.   In section \ref{LWPict} we construct the partition function of the Levin-Wen models and show how to view it as a space-time picture of successive projectors, similar to a Trotter decomposition.
Section \ref{CHSect} shows in detail how this  construction of the Levin-Wen partition function gives precisely the \chainmail invariant introduced by Roberts\cite{Roberts}.  Since the \chainmail invariant is easily shown to be equivalent to another mathematical invariant, known as the Turaev-Viro\cite{TuraevViro} invariant, this makes apparent the explicit connection between the lattice partition function and the Turaev-Viro state-sum.

In section \ref{QPSect} we turn to study quasi-particles in the Levin-Wen models. Section \ref{sub:QPreview} reviews the Levin-Wen construction of quasi-particles in their model; section \ref{sub:QPMTC} shows the analogous construction in our approach, which adds quasi-particle world-lines to the \chainmail link.   This construction gives an interesting perspective on the difference between the right- and left- handed sectors of the doubled theory, which we elaborate on in sect. \ref{subsub:handle} .
In Sect.~\ref{NTriSect}, we outline how the pictorial construction can be carried out on more general lattice geometries, allowing for Hamiltonians with slightly simpler interactions.  Section \ref{Conclusions} summarizes our results and discusses interesting open questions.

This paper also includes appendices detailing certain diagrammatic calculations, as well as discussions of surgery, and categories more general than those discussed in the rest of the paper.  We have separated these sections from the main text for simplicity of presentation.

\section{Conceptual overview: Chern-Simons theory and topological lattice models} \label{OverSect}

We begin with a qualitative description of our most interesting results, using as an example the class of topological quantum field theories most likely to be familiar to physicists -- Chern-Simons theories.

In this section, we outline our answer to a simple but important question: how can a pure Chern-Simons theory arise from a lattice Hamiltonian?  
Though several examples in the literature find an effective long-wavelength Chern-Simons term after integrating out fermions on the lattice \cite{LaughlinZou, WWZ, HaldaneHoney}, our interest is to construct a lattice model in which the {\em local} variables can be expressed in terms of the Chern-Simons gauge field.  As observed by Ref.~\onlinecite{FreedmanShtengleetc}, the commutation relations of the Chern-Simons gauge field obstruct such a description unless the theory is doubled; consistent with this, our construction yields only doubled theories.  

The usual prescription for putting a Maxwellian gauge theory on the lattice places a gauge field on every edge and recovers the continuum limit by taking the lattice constant to zero. 
Though similar prescriptions can be carried out for Abelian Chern-Simons theories\cite{Semenoff}, conventional lattice formulations valid for general non-Abelian gauge groups have proven elusive.  
 Our objective in this section is to provide a qualitative description of an alternative route to this end --- namely, we formulate a lattice construction of Chern-Simons theory which is naturally `topological', in the sense that it is independent of the lattice geometry and captures the topological character of the braiding of Wilson lines.   The interesting feature of these models is that they are independent of the lattice constant $a$, and thus there is no way in which the usual prescription of taking the continuum limit applies.   Rather, the correspondence to a continuum theory is achieved by means of a known mathematical equivalence \cite{WittenJones} between Chern-Simons theory and knot polynomials.  Hence our model does not encode the continuum theory as a long-wavelength limit, but rather encodes the continuum theory in a lattice representation.  This is possible because the topological model has, for any fixed number of excitations, a finite number of degrees of freedom.

\subsection{Chern-Simons theory}
\label{sub:CStheory}

We begin with a few important facts about Chern-Simons theory.  Our goal here is to sketch the relationship between Chern-Simons theory and knot theory first described by Witten\cite{WittenJones}, which is the cornerstone of our construction.  For a more pedagogical overview of Chern-Simons theory, see, for example, Ref.~\onlinecite{DunneCS}.  Readers who are relatively familiar with Chern-Simons theory may be able to skip to section \ref{subsub:omega} where we discuss the $\Omega$ string (which is likely to be less familiar, even to many experts).

To define a Chern-Simons theory, we pick a Lie group ${\cal G}$ and let the gauge field $A$ take values in its Lie algebra.  The Chern-Simons action is written
\begin{eqnarray}
& & \mathcal{S}_{CS}[A]= \label{eq:secondform} \\
&  &\frac{k}{4 \pi} \int_{{\cal M}}\, {\bf dx}\, \varepsilon^{lmn} \left [ A_l^a \partial_m A_n^a + \frac{2}{3}f_{abc} A_l^a A_m^b A_n^c \right ] \nonumber
\end{eqnarray}
where $k$ is an integer known as the level.   Here the integral is over the spacetime manifold ${\cal M}$ and we
write coefficients $A_\mu^a$ where $\mu$ is a spatial index and $a$ is the Lie algebra index where $f_{abc}$ are the structure constants of the algebra.   We denote this Chern-Simons theory  ${\cal G}_k$ (pronounced ``${\cal G}$ level $k$").

The topological character of Chern-Simons theory results from an unusual feature of the Chern-Simons action:  since all indices are contracted using the anti-symmetric tensor $\varepsilon^{ijk}$, the action is defined without a spacetime metric.  Hence the action must be invariant under deformations of space-time  -- in other words, it must be invariant under any continuous change in the geometry of the space-time manifold.
As a result the Chern-Simons partition function\footnote{ We note that one must be rather careful writing partition functions in this form, as strictly speaking such a functional integral is not a well defined mathematical quantity. For a discussion of the problems involved, see for example Ref.~\onlinecite{CSRefs}. Here we adopt the physicist's approach of working with these quantities in spite of these problems.  }
\begin{equation}
    Z_{CS}[{\cal M}] = \int {\cal D}[A] e^{i {\cal S}_{CS}[A]}
\end{equation}
is a topological invariant of the manifold ${\cal M}$.  Throughout this work, unless otherwise specified, all partition functions are to be understood as evaluated at $T=0$; in the presence of thermal excitations the partition function necessarily scales with the area, and hence is not a purely topological quantity.

One consequence of topological invariance is that the gauge-invariant physical observables in pure Chern-Simons theories must also be independent of the space-time metric.
This stringent condition leaves us with very limited possibilities for observables of the theory; in fact, the entire physics of Chern-Simons theory can be described as a theory of Wilson lines.  Explicitly, a Wilson line operator is defined as
\be
W_R(C) = {\rm Tr}_R P e^{\shs{i} \oint_C A_i dx^i }
\ee
where $C$ is a directed closed curve in the $3D$ space-time (a {\em knot}), $P$ denotes path ordering, and $R$ is a representation of the gauge group in which we take the trace.  Hence we integrate the component of the gauge field tangent to the curve $C$, to obtain some element of the gauge group.  To obtain an observable, we must stipulate a representation of the gauge group ${\cal G}$ in which to compute the trace.   For example, in the case of the gauge group $SU(2)$, the representations are labeled by their total spin $0,1/2,1, \ldots$.

There is one important subtlety in this approach, due to the impact of the constraints that make the action topological  on the quantized theory.  The up-shot of this technicality is that after quantizing the theory, only a finite
number of representations are allowed --- which we will label $0, \ldots, r$ with $0$ reserved to mean the trivial representation (which is equivalent to the absence of a Wilson line)\footnote{The allowed representations turn out to correspond to conformal blocks of a rational conformal field theory\cite{BigYellowBook} (a Wess-Zumino-Witten theory) whose current algebra is appropriate to the Chern-Simons theory in question.}.   For example, In the case of $SU(2)_k$ we have $r=k$ nontrivial particles.

Thus we may think of Chern-Simons theory, quantized in this way, as a theory of Wilson lines which are labeled by a finite set of allowed quantum numbers $0 \ldots r$.     All relevant information about the theory is then encoded in the expectation values of products of Wilson lines, which we can describe by a set of directed closed curves $C_j$  labeled with quantum numbers (representations) $i_j$ :  
\begin{eqnarray}  \nonumber
& &     \langle W_{i_1}(C_1) \ldots W_{i_n}(C_n) \rangle_{\mac{M}} = \\ & & \frac{1}{Z_{CS}(\mac{M})} \int_{\mac{M} } {\cal D}[A]  \,\, W_{i_1}(C_1) \ldots W_{i_n}(C_n) \,\, e^{i {\cal S}_{CS}} \label{eq:linkinv}
\end{eqnarray}
If $n>1$, we will call the collection of closed curves $C_j$ a {\it labeled link}, whose {\it components} consist of the individual curves $C_j$, and whose labels are determined by the representations carried by the corresponding Wilson line.  Denoting the resulting link $L$, we will use the notation
\be
\langle L \rangle_{\mac{M}} \equiv  \langle W_{i_1}(C_1) \ldots W_{i_n}(C_n) \rangle_{\mac{M}}
\ee 
as shorthand for such expectation values.  When the manifold $\mac{M}$ is not specified, we take it to be the $3$-sphere $S^3$. 

The expectation values (\ref{eq:linkinv}) are topological invariants of the labeled link -- and hence, we may evaluate expectation values of products of Wilson lines in the gauge theory by computing the relevant link invariants.  Indeed, as famously shown by 
Witten\cite{WittenJones}, in the case of $SU(2)$ gauge group, one obtains the colored Jones polynomial (with the 
``colors" being the different labels of the lines).   For more general gauge group one obtains other topological 
invariants including many previously known in the mathematical theory of knots.

In a more physical language, we can think of each of the labels as being a ``particle type" and the directed lines as being the worldlines traced out by these particles.  In fact, this picture can be generalized to include branched loops, where two particles come together at a vertex to form a third in a process known as {\em fusion}.  Roughly speaking, one can think of binding the incident pair of Wilson lines together such that the only physically relevant variable is the sum of their labels. In general, the end product of fusion will be a superposition of particle types-- analogous to combining two spin 1/2 particles to form a superposition of singlet and triplet states.   Thus we can think of Chern-Simons theory as more generally being able to assign a value to any graph of worldlines that has trivalent vertices as well as over- and under-crossings.

\subsubsection{Vacuum partition functions and the $\Omega$ String}
\label{subsub:omega}

Above, we motivated the claim that an expectation of an arbitrary product of Wilson lines in the Chern-Simons gauge theory at zero temperature can be evaluated by computing an appropriate link invariant.  But there is one key missing ingredient required to make this picture valid: we must know the correct normalization for Eq. \ref{eq:linkinv}, which requires evaluating the vacuum partition function.  At zero temperature, the topological theory contains a finite number of degrees of freedom; consequently, as Witten\cite{WittenJones} showed this, too, can be done in terms of link invariants-- specifically, by evaluating the invariant of a link (whose form is dictated by the space-time manifold) labeled by a special superposition of particle types which we call $\Omega$ (whose precise definition we defer to Sect.~\ref{sub:categories}).   In other words, for the correct choice of link components $C_i$, we have 
\ba \label{eq:WittenPart}
Z_{CS} (\mathcal{M} ) &=& \frac{1}{Z_{CS} (\mathcal{M} _0 ) }  \langle L(\mathcal{M} )  \rangle_{\mac{M}_0} \n
& \equiv & \frac{1}{Z_{CS} (\mathcal{M} _0 ) } \langle W_{\Omega}(C_1) \ldots W_{\Omega}(C_n) \rangle_{\mac{M}_0}  \ \ \ 
\ea
where $\mathcal{M} _0 $ is a `reference' space-time manifold (for our purposes, $\mathcal{S}^2 \times \mathcal{S}^1$), whose partition function we will choose to set to $1$.  The link thus encodes the topology of the manifold $\mac{M}$.  The precise relationship between $\mathcal{M}$ and $L(\mac{M})$ can be understood by a procedure known as {\it Dehn surgery}, which we will outline in more detail in Sect.~ \ref{sub:handleslide} and Appendix \ref{SurgApp}.

Because of its special relationship to the ground-state partition function, the Wilson line labeled by $\Omega$ (or $\Omega$ string) plays a pivotal role in our construction.   Its privileged status is related to the following useful property:

\begin{center}
\begin{figure}
\includegraphics{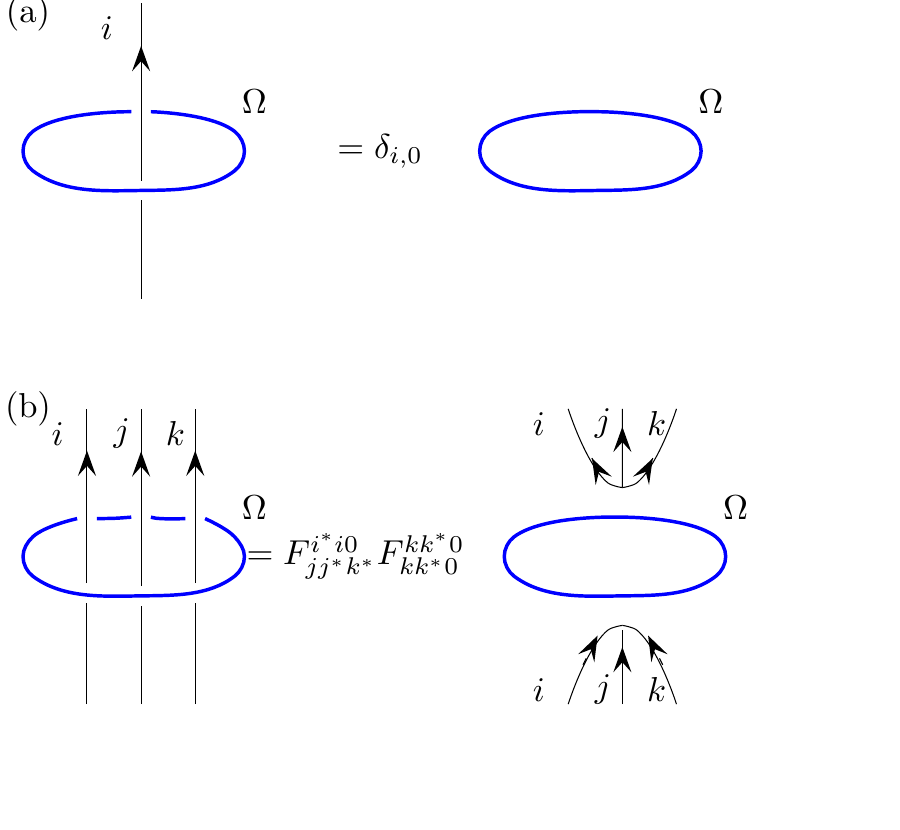}
\caption{\label{KillFig} The $\Omega$ string.  (a) The killing property.  (b) A useful consequence of (a). The meaning of $F$ will be explained in Sect.~\ref{sub:categories}.}
\end{figure}
\end{center}

{\it Killing property}  If any particle world line labeled $i$ passes through a  non-self-knotted $\Omega$ loop, the value of the evaluated link invariant Eq. \ref{eq:linkinv} will be zero unless that particle is the trivial particle $i=0$ (See Fig. \ref{KillFig}.a).   By adding an appropriate normalization, the $\Omega$ can thus be considered to be a projector that gives one if the vacuum particle passes through it and gives zero otherwise.   We emphasize that if multiple particle world lines pass through an $\Omega$ loop, the projector acts on the {\it combined} quantum number of all the particles.   Thus the
value of the link invariant will be nonzero if the quantum numbers of the multiple particles can be combined (or ``fused") to the vacuum quantum number.  (See Fig. \ref{KillFig}.b) Such fusions are described in detail in section \ref{sub:categories} below.

Hence we may use $\Omega$ to construct projection operators, which will be the constituents of our topological lattice Hamiltonian.  The importance of $\Omega$ in topological invariants has been discussed starting with the work of Ref.~\onlinecite{WittenJones,RT}.   More recently this type of projector has been used for analysing anyon models in Refs. \onlinecite{Gilsetal,KitaevVeryLongPaper,LW}.

\subsection{Lattice Chern-Simons theory via Wilson lines}
\label{sub:latticeCStheory}

Our initial question was how a pure Chern-Simons theory can arise from a lattice Hamiltonian.  At this point, let us turn the question around and ask how we may construct a lattice Hamiltonian based on a Chern-Simons theory.  To this end, let us choose a two dimensional lattice and give quantum numbers to the edges of the lattice chosen from the set of quantum numbers $ i\in \{ 0.. .r \}$ of a Chern-Simons theory.   Although, as we will describe below, such a starting point can indeed generate a lattice model that is equivalent to a continuum Chern-Simons theory, we will find that it is actually equivalent to the {\it double} of the Chern-Simons theory we started with --- that is, two copies with opposite chiralities\footnote{See also Ref.~\onlinecite{Gilsetal} for a different picture of this doubling}.

We will aim to make a direct correspondence between {\it partition functions} of our lattice model and those of a continuum Chern-Simons theory.  Hence we work directly with a three dimensional description, where the third direction is time.   We discretize time, so that we obtain a $3$D lattice consisting of multiple copies of our $2$D lattice separated by discrete time intervals $\delta t$.   We will refer to an edge or plaquette of this 3D lattice as being space-like if it is at fixed time, and time-like if it extends between neighboring time intervals.

Our objective is to construct a lattice Hamiltonian $H$ such that by using $H$ to propagate states in time, and tracing over intermediate states, we obtain a partition function of the form (\ref{eq:WittenPart}) -- in other words, a partition function which corresponds precisely to that of a topological quantum field theory.

Since Chern-Simons theory is essentially a theory of Wilson lines, it is quite reasonable to represent the quantum numbers of the 2d lattice edges (the space-like edges) in terms of Wilson lines.
We thus associate a quantum number on a space-like edge at a given time $t$ with a closed Wilson loop which runs along this edge at time $t$, up the two vertical edges at its endpoints, and back along the same edge at time $t + \delta t$ (Fig. \ref{Fig_Tloops}).  In other words an edge with quantum number $r_i$ at time $t$, is described in the continuum model as a Wilson loop just inside the perimeter of a time-like plaquette corresponding to a particular edge and that particular time.

The partition function will require us to sum over all possible edge variables at every time. Since $\Omega$ is actually a sum over all quantum numbers, it will turn out that labeling the Wilson loop around the perimeter of the time-like plaquettes with $\Omega$ will precisely effect the desired sum in the partition function.

\begin{center}
\begin{figure}[htp]
   \includegraphics[width=2.75in]{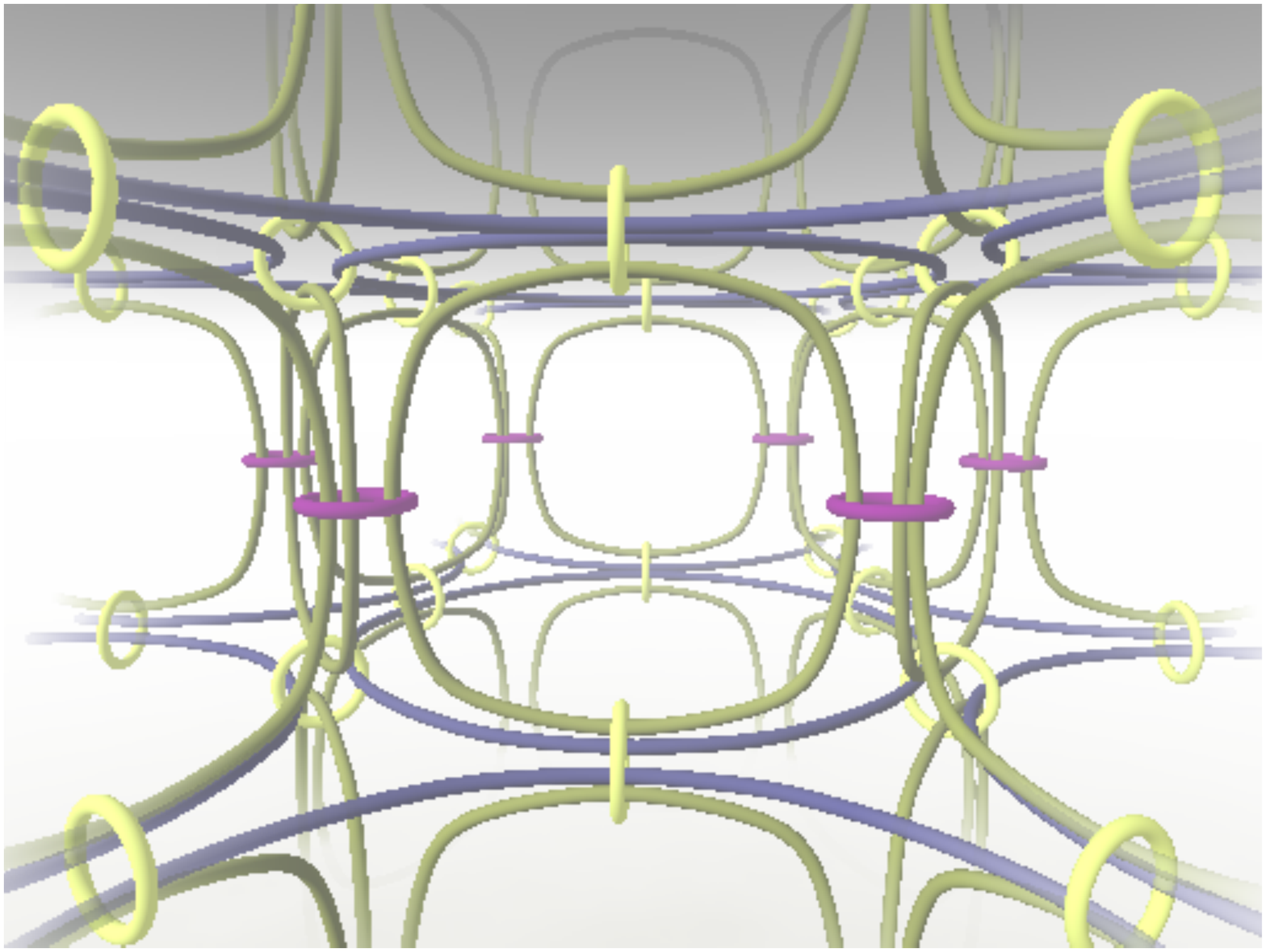}
   \label{Fig_FullCh}
\caption{ \label{CHFig} Wilson lines used to evaluate the partition function, drawn here on a lattice whose spatial slices are the honeycomb. 
Green strings, which encircle time-like plaquettes, represent the states at each time step.  Blue strings around the space-like plaquettes represent the plaquette projectors in the Hamiltonian; mauve strings encircling the time-like edges represent the vertex projectors.  The yellow loops along space-like edges are necessary to implement the action of the Hamiltonian, by forcing the plaquette projectors to fuse with the edge variables at each time step. 
All loops carry the label $\Omega$.
}
\end{figure}
\end{center}

Having extended the edge variables in 2d to Wilson loops in 3d,  we now attempt to  determine an appropriate lattice Hamiltonian which will yield a topological theory.  In existing constructions of topological lattice models such as that of the Levin and Wen\cite{LW}, and Kitaev's Toric Code\cite{KitaevToric}, the Hamiltonian is a sum of mutually commuting projectors  -- one class being applied to every vertex of the 2d lattice, and another to every plaquette.  As will be detailed below in section \ref{sub:LWmodels}, the vertex projectors assure that the quantum numbers incident on each vertex fuse to zero.  The plaquette projectors can be thought of similarly as a fusion condition (but in a dual basis, as discussed in section \ref{sub:ramifications}).
As mentioned above (Property 1 of section \ref{subsub:omega}), a loop labeled with $\Omega$ projects onto states where all lines passing through the $\Omega$ loop fuse to the vacuum particle.  Hence, up to normalization of the projectors, we may use the special Wilson line $\Omega$ to construct operators in the Chern-Simons theory that carry out the action of the Hamiltonian of a topological lattice model. Indeed, the vertex projector may be implemented by encircling time-like edges with an $\Omega$ loop and the plaquette projectors are implemented by including $\Omega$ loops on each space-like plaquette -- threading additional $\Omega$ loops which wrap space-like edges.

An example of the resulting link of Wilson loops is shown in Fig.~\ref{Fig_FullCh}.  This link is referred to as the ``\chainmail" link\cite{Roberts}, evoking its resemblance to \chainmail armor made of linked rings.   Evaluation of the associated knot invariant, known as the \chainmail invariant, is known to give an invariant of the three manifold\cite{Roberts} --- that is, it is independent of the lattice decomposition of the manifold.

Although all of the strings in the \chainmail picture in Fig.~~\ref{Fig_FullCh} are $\Omega$'s, we remind the reader that they have different meanings. The time-like plaquette loops represent the edge variables in the lattice models.   The remaining loops are part of the Hamiltonian which acts upon these variables at each time step to effect time propagation of the edge variables.  Thus, all of the pieces taken together,  this link is a explicit space-time representation of the partition function of a 2d lattice model.

Hence, we associate a link (the \chainmail link, $L_{CH}$, whose components are all Wilson lines labeled by $\Omega$), with the zero-temperature partition function of our lattice model.  Specifically, we have used the \chainmail link to engineer a lattice Hamiltonian with the interesting property that evaluating the link invariant gives, up to a constant, the partition function for the ground state sector of our lattice theory:
\be \label{eq:ChPart}
Z \sim \langle L_{CH} \rangle \ \ \ \ .
\ee 
  (This is nontrivial in topological models, since the ground state sector is degenerate).     As we will explain in detail in Sect.~\ref{LWSect}, the lattice Hamiltonians we construct in this way are precisely the set of Hamiltonians described by Levin and Wen (Ref.~\onlinecite{LW}) for which the spectrum of excitations corresponds to that of a doubled Chern-Simons theory.  
  
  Obtaining the partition function in the presence of quasi-particles turns out to be similar.  Since our formalism captures only the topological sector, we must work in the zero-temperature or infinitely-gapped limit, where these quasi-particles can appear only as topological defects, not as excitations. In the 3d picture, these defects are also Wilson lines (or external sources in the gauge theory) which trace out the quasi-particle worldlines' space-time paths.  Hence the partition function can again be evaluated by computing the appropriate link invariant:  
 \be
Z(W_{i_1}, ... W_{i_n})  \sim \langle L_{CH} \cup W_{i_1}(C_1) W_{i_n} (C_n)  \rangle 
\ee 
where the the link now includes both the \chainmail and the inserted quasi-particle world-lines.  
(This is illustrated in Fig. \ref{QPFig}, and explained in detail in Sect.~\ref{QPSect}).   Restricting this picture to a 2d slice at fixed time, the quasi-particle defects appear as string operators  --- meaning that defects must be created in pairs which appear at the end of unobservable strings.    

The true power of expressing the partition function in the form \ref{eq:ChPart} is that it allows us to exploit Eq. \ref{eq:WittenPart}, which relates the link invariant to the Chern-Simons vacuum partition function.   To evaluate the ground state partition function of the lattice model, we need to determine the Chern-Simons expectation value for the complicated ``\chainmail" link that is the space-time representation of the lattice model partition function.   However, we may invoke Witten's result to realize that this is equivalent to evaluating the Chern-Simons {\it vacuum} partition function in an appropriately chosen space-time manifold (whose topology is dictated by the precise linking pattern of the components of the \chainmail link).  
Thus, as promised, we make a connection between the lattice models and the continuum  Chern-Simons theory, not via the usual notion of taking the lattice spacing to zero, but by using Dehn surgery to see that the two partition functions are equivalent.  

As we will see in Sect.~\ref{CHSect}, it turns out that the space-time of the continuum theory arrived at in this way is always such that the vacuum partition function is {\em achiral}.  Specifically, if the (3d) space-time lattice can be `filled in' to obtain a manifold $\mathcal{M}$, then the corresponding continuum theory is a Chern-Simons theory on ${\cal M} \# \overline{\cal M}$, that is, ${\cal M}$ connected to its mirror image.   
In other words, though the lattice model is constructed entirely in terms of operators from a chiral theory, our unusual prescription for taking the continuum limit results in a continuum theory which is, in fact, doubled.  Hence we have made an explicit connection between the topological lattice models of Ref.~\onlinecite{LW} and the continuum doubled Chern-Simons theory.  

In the remainder of this paper, we will present the ideas outlined here in greater detail, focusing on the more general case of an arbitrary `doubled anyon' theory, which may or may not correspond in practice to a doubled Chern-Simons theory.  Though this will require a more technically involved treatment, the intuitive picture --based on the correspondence between partition functions of topological theories and link invariants -- is essentially the same as what we have outlined here.

One may also consider more complicated (or even irregular) lattice
geometries.  For example, the so-called Cairo-pentagon lattice tiling
and the prismatic pentagon lattice tiling both share the feature with
the honeycomb and square lattices that a LW model based  on these
lattices would couple only 12 edges at a time.

\section{Formalism of anyon theories}
\label{sub:categories}

In Sect.~\ref{OverSect}, we constructed a lattice Chern-Simons theory by placing projectors, comprised of a particular superposition of Wilson lines, on the edges and plaquettes of a lattice model.  To describe the
resultant microscopic action on the states, however, requires rules about how to evaluate the resulting link diagrams.  Here we briefly introduce the formalism of {\em anyon theories} (often called {\em modular tensor categories} in the literature) associated with these rules.  For a more comprehensive introduction to this subject, see Refs. \onlinecite{KitaevVeryLongPaper},\onlinecite{Bonderson},\onlinecite{Wang}.

Generally an anyon theory can be thought of as a generalization of a Chern-Simons theory.  Like a Chern-Simons theory  (Sect.~\ref{sub:CStheory}), the anyon theory specifies a topological invariant  of a manifold\cite{RT} (analogous to a vacuum partition function), and can be used to assign a value (the {\em link invariant}) to any link of world-lines within that manifold.
As with Chern-Simons theory, using the concept of fusion, the idea of a link invariant can be generalized to give a value to world line diagrams with branches as well as knots.

The formalism of anyon theories has three important elements.  First of all, fixing an anyon theory specifies a set of particle types allowed in the theory.  Second, it stipulates a set of {\em fusion rules}, which determine what happens when two particles combine at a trivalent vertex to form a third.  (Again, physically we can think of bringing the two particles so close together that at the length scale on which we examine the system they appear to be fused into one).   These fusion rules will be accompanied by so-called {\em $F$-matrices}, which describe how the order of fusing multiple particles together may be changed.   Finally, the anyon theory gives a set of {\em braiding rules}, which specify the statistics of each species of anyons by assigning a phase to any exchange process.

A familiar example of an anyon theory is the Chern-Simons theory described above.  (Note however that not all anyon theories are Chern-Simons theories!)  First, it has a finite set of particle types (the number of which depends on the group ${\cal G}$ and the value of $k$).  We represent a particle by its world-line, which carries a label to denote the particle type.  (When ${\cal G} = SU(2)$, this is the particle's angular momentum quantum number).   Second, if two particles are brought close together we can ``fuse" them to form a third by combining their labels appropriately.  (Again for ${\cal G} = SU(2)$, this corresponds to performing the appropriate addition of angular momenta, bearing in mind that for $SU(2)_k$ Chern-Simons theory there is a maximum possible angular momentum in the theory, and combinations which exceed this do not occur.)  And third, when two world-lines are
braided around each other, their anyonic nature ensures that the wave function acquires a complex phase.  The close connection between $SU(2)$ angular momentum addition (Yang-Mills) and $SU(2)_k$ Chern-Simons theory is shown in Table \ref{SU2Table}.  Table \ref{AnyTable} gives several examples of fusion rules in well known anyon theories.

\begin{widetext}
\begin{center}
\begin{table}
\begin{tabular}{l c c c}
  & $SU(2)$  &   &  $SU(2)_k$  \\
  \hline
 particle types:~~~~  &   $j \in  \{0,1/2,1, \ldots\}$    & $\Longleftrightarrow$ &   $j \in  \{0,1/2,1, \ldots k/2\}$ \\
 vacuum:  &   $j = 0$    & $\Longleftrightarrow$ &  $j =0$ \\
 fusion rules:  &   $ j \times l = \sum_{i=|j-l| }^{ j+l } i $    & $\Longleftrightarrow$ &  $ j \times l = \sum_{i=|j-l| }^{ \min \{j+l, k-j-l\}  } i$ \\
 $F$-matrices &  6-j symbols of $SU(2)$ & $\Longleftrightarrow$ & q-deformed 6-j symbols of $SU(2)_k$
 \end{tabular}
\caption{
 \label{SU2Table}Analogy between $SU(2)$ Yang-Mills theory (ie. conventional $SU(2)$ angular momentum addition) and $SU(2)_k$ Chern-Simons theory.
Note that the former does not  constitute an anyon theory in our sense because there are an infinite number of particle types.
Here each particle $j$ is its  own antiparticle, since $j$ can only fuse with $j$ to produce a singlet $0$ (the trivial particle).
In the Chern-Simons case, the fusion rules are similar to conventional $SU(2)$ except that they have been {\it deformed} so that no fusion ever produces $j > k/2$.  Note also that there is no analogue of $l_z$ in this theory: the particle types are uniquely specified by the total angular momentum $j$.
}
\end{table}
\begin{table}
\begin{tabular}{l l l}
 Anyon Theory & Particle Types &  Fusion Rules \\
 \hline
 $SU(2)_k$   & $j \in \{0,1/2,1, \ldots, k/2\}$ ~~~~& $ j \times l = \sum_{i=|j-l| }^{ \min \{j+l, k-j-l\}  } i$  \\
 Fibonacci = $(G_2)_1$ ~~~~ &  $\{0,1\}$  &   $1 \times 1 = 0 + 1$ \\
Toric Code & $\{0,e,m,em \}$ &  $e \times m = em$ ; $e\times e = m \times m = 0$ \\
& & $e \times e m = m$ ; $em\times m = e $ 
 \end{tabular}
\caption{
 \label{AnyTable}
 Examples of Well Known Anyon Theories, Their Particle Types, and Fusion Rules.  Note that it is always true that $0 \times y = y$ for any particle type $y$.}
\end{table}
\end{center}
\end{widetext}

Let us describe the formalism surrounding each of these three elements in turn.  First, we will graphically represent a particle as an arrow labeled with the name of the particle $i \in \{0,\ldots r\}$.  We may think of this arrow roughly as the world line of a particle. The identity (``trivial" or ``vacuum") particle (represented either by no line, or a dashed line when necessary) is always labeled $0$, and particle $i$ has a unique antiparticle type $i^* \in \{0,\ldots r\}$, which is equivalent to changing the direction of the arrow  (i.e, the particle $i$ going forward is equivalent to the particle $i^*$ going backward).  Note that a particle can be its own antiparticle (in which case, lines representing particles need not have arrows).

Each of these particle types has an associated {\em quantum dimension}, denoted $\Delta_i$.  This is the value of a circular world-line loop labeled by $i$ when nothing passes through the loop.   It is useful to define the ``total quantum dimension" given by the root sum of the squares of the quantum dimensions of all of the particles in the theory:
\be
\label{eq:totalquantumdimension}
 {\cal D} = + \sqrt{\sum_{a=0}^r \Delta_a^2} \ \ \ .
\ee
This quantity will appear in the normalizations of various quantities in our models.

\begin{center}
\begin{figure}[htp!]
\subfigure[] 
{
 \includegraphics{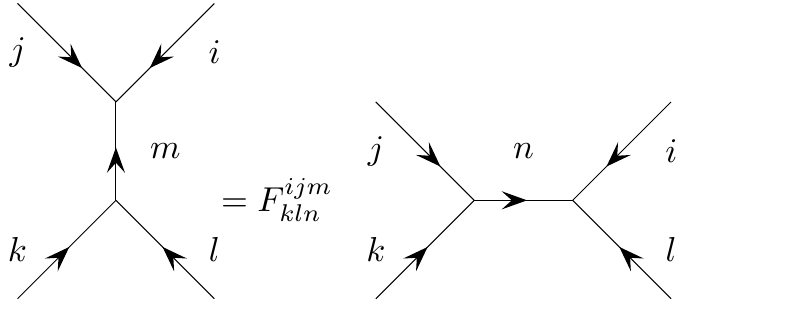}
  \label{FSym_Fig}
}
\subfigure[] 
{
\includegraphics{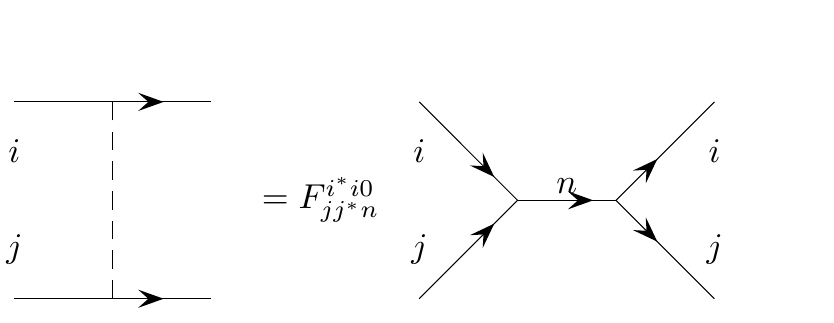}
    \label{CG_Fig}
}
\subfigure[] 
{
  \includegraphics{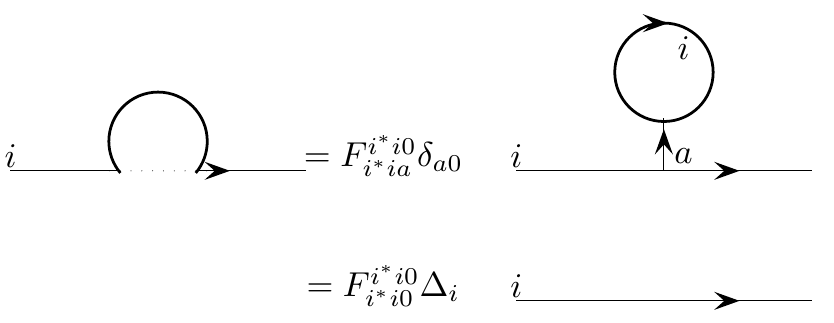}
    \label{Deltas_Fig}
} \hspace{1cm}
\subfigure[] 
{
  \includegraphics{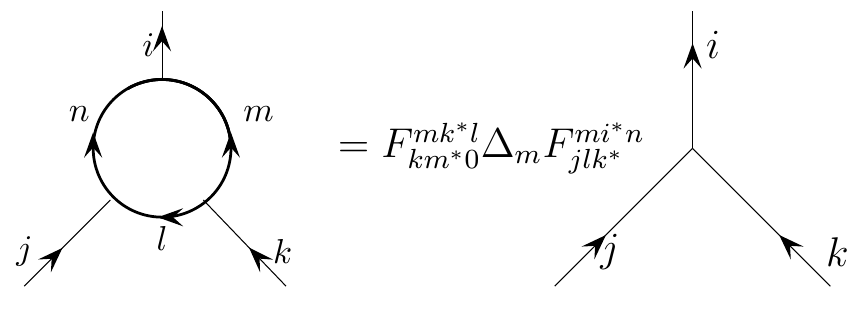}
  \label{Tetra_Fig}
}
\caption{Basic relations of tensor categories.  (a) shows the
defining relation for the $F$ matrices.  (b) shows how these
define fusion between two strings.  This prescribes a set of
allowed vertices.  (Throughout the figure, the dotted line represents the $0$ particle).  (c) shows the consequence of applying the rule
in (b) to a single string, defining the relationship between
fusion and the quantum dimension $\Delta_i$.  (d) shows a simple
consequence of fusion.} \label{LW_3}
\end{figure}
\end{center}

Second, we represent fusion with a tri-valent vertex,
which denotes combining the quantum numbers of two of the incident lines to give the quantum number carried by the third.  The rules for these fusions share certain properties with their more familiar analogues from the theory of angular momentum addition: (a) if the quantum number of any of the three lines is trivial (vacuum), then the other two lines carry the same quantum number.  (I.e, a trivalent vertex may have one incoming particle labeled $i$, one outgoing particle labeled $i$ and a third particle labeled $0$).   (b)  Generically the result of fusing two particles' lines is not unique -- the product may carry one of several possible quantum numbers.  This is analogous to two spin $\frac{1}{2}$ particles fusing to either a singlet or a triplet (c) Not all trivalent vertices represent allowed combinations.

As with conventional angular momentum addition, one can change bases (from $\frac{1}{2}+\frac{1}{2}$ basis to singlet and triplet basis, for example) and a wavfunction will appear different in the new basis.   The relation between two different bases is given by generalized 6j coefficients, called $F$ matrices or recoupling coefficients, and is shown in Fig. \ref{FSym_Fig}.    We can think of this equivalence between two sets of diagrams as being simply a basis change.
As shown in Fig. \ref{Deltas_Fig}, the $F$ matrices involving the trivial particle encode the values of the quantum dimensions of the theory via $\Delta_i = \left( F^{i^* i 0}_{i^* i 0} \right ) ^{-1}$.   Fig. \ref{Tetra_Fig} shows a relationship which will be useful for evaluating partition functions below.

\begin{figure}[htb]
\begin{center}
\subfigure{
 \includegraphics{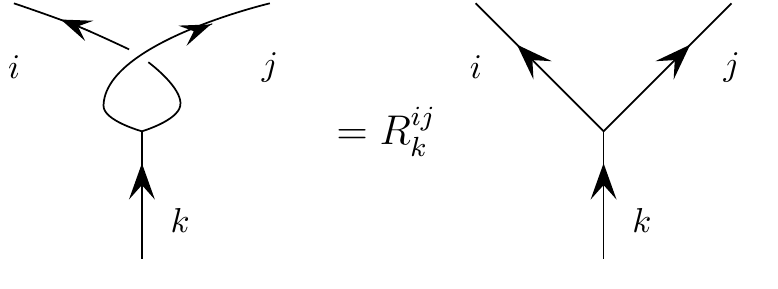}
\label{Rfiga}
}
\hspace{1cm}
\subfigure{
 \includegraphics[width=3.5in]{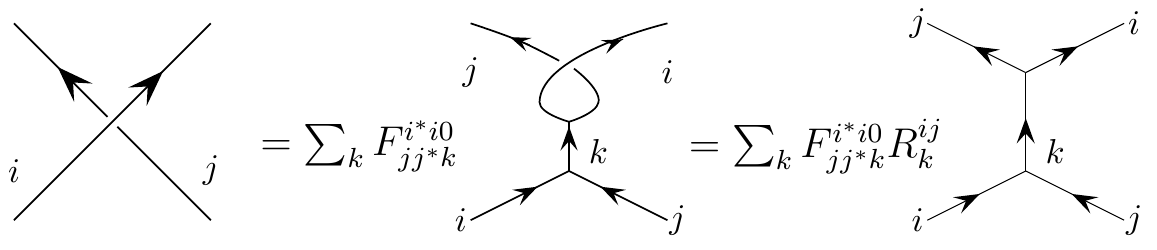}
\label{RFigb}
}
\caption{The un-crossing relations used to define the tensor $R$
which fixes a braiding structure on the modular tensor
category.}\label{RFig}
\end{center}
\end{figure}

Third, we must know how to describe braiding (or exchange statistics) in an anyon theory.  This information is encoded in a matrix $R$, as shown in Fig. \ref{RFig}.  As one might expect for consistency of anyonic statistics, $R$ is a pure phase, and obeys
\be
\left( R^{ij}_k \right )^* = R^{ji}_k \ \ \ .
\ee

While the world line diagrams are drawn in two-dimensions, they obviously represent a process  in three dimensional spacetime.   A crucial part of the structure of anyon theories (and link invariants in general) is that the value of the diagram will be independent of which two-dimensional projection is chosen. This independence of projection is guaranteed by certain consistency conditions on the $R$ and $F$ matrices which must be obeyed by any anyon theory (See Refs.~\onlinecite{KitaevVeryLongPaper},\onlinecite{Bonderson},\onlinecite{Wang}).

One important consequence of the rules described above is that in every anyon theory, we can define an $\Omega$ string, via
\be \label{Eq_Omega}
\Omega \equiv \frac{1}{\mathcal{D}} \sum_{i=0}^{r} \Delta_i \, | i \rangle
 \ee
which has the Killing property 
described above in section \ref{sub:CStheory}\footnote{ In fact, it is {\em always} possible to construct a projector onto the $0$-string, which can be used to generalize our construction of the ground-state partition function to the case of tensor categories which do not even have a braiding structure.}.  (Here we have slightly abused notation and written the right hand side as a ket vector to emphasize that the $\Omega$ string is just a linear sum over particle labels on that string. )  A string labeled with $\Omega$ does not require an orientation, as the sum over all particles necessarily includes all antiparticles; hence we represent $\Omega$ by an un-oriented string.

As noted in section \ref{subsub:omega}, the evaluation of a link of $\Omega$'s in a manifold $\mathcal{M}$ can be considered as a topological invariant of a different 3-manifold $\mathcal{M}'$, which is obtained from $\mathcal{M}$ by a process known as surgery.  (This connection will be described in detail in sections \ref{sub:handleslide} and Appendix \ref{SurgApp} below.)   The resulting manifold invariant, known as $Z_{WRT}(\mathcal{M}')$ (which stands for Witten\cite{WittenJones}, Reshitikhin and Turaev\cite{RT}), is precisely the Chern-Simons vacuum partition function $Z_{CS}(\mathcal{M}')$ in the case of a Chern-Simons theory.  It is this connection between links labeled with the element $\Omega$, and the topological invariant $Z_{WRT}$, which lies at the heart of the connection we establish between topological lattice models and doubled anyon theories.

Note that any anyon theory has a `mirror image', which is equivalent to the original anyon theory on the mirror image of the manifold or link under consideration.  Alternately, the mirror image can be obtained by complex conjugating all of the $R$ matrices and leaving the manifold and link unchanged.  Thus a mirror theory has conjugates value  $Z_{WRT}(\overline{\cal{M}})  = [Z_{WRT}({\cal M})]^*$ where $\overline {\cal M}$ means the mirror image of the manifold $\cal M$.    (For Chern-Simons theories, the mirror image theory can be obtained by taking taking the level $k$ to $-k$).  We will refer to these mirror image theories as {\em left-handed}-- and hence to the originals as {\em right-handed}.

We emphasize that the fusion rules in Fig. \ref{LW_3} do not describe the most general category (although they do describe most simple examples).  More generally two particles $i$ and $j$ may fuse to $k$ in multiple possible ways.  If one has such nontrivial ``fusion multiplicities", one needs to include an additional index at the fusion vertex indicating which of the possible ways the fusion takes place.  (See for example, the detailed discussion in Refs.  \onlinecite{KitaevVeryLongPaper},\onlinecite{Bonderson},\onlinecite{Wang}).  In this more general case, each $F$ matrix has four additional indices.  For notational simplicity, and following the lead of Levin and Wen, we do not include these explicitly here.

A second issue occurs when the anyon theory contains particles that are odd under time reversal.  In this case there are minus signs which enter into the translation between fusion diagrams and space-time world-line diagrams.  ( These factors are also explained in detail in Refs. \onlinecite{KitaevVeryLongPaper},\onlinecite{Bonderson},\onlinecite{Wang}.)   Again, for simplicity, we will not consider such models here.  We note, however, that these more general cases can easily be described by our constructions as well.

\section{The Ground State Partition Function} \label{LWSect}

In this section we will explain in detail a knot-theory based construction of the Levin-Wen partition function, which we will relate to the \chainmail invariant introduced by Roberts \cite{RobertsThesis} in Sect.~\ref{CHSect}.
This section will focus only on obtaining the zero-temperature vacuum partition function of the Levin-Wen models\footnote{the ground state is nontrivial in topological models, as it is generically degenerate}; quasi-particle defects will be considered in section \ref{QPSect} below.
Readers familiar with tensor categories should note that although our approach is most useful in cases where one starts with a valid anyon theory (a.k.a. a modular tensor category) and constructs the lattice model which gives the double of this input theory, the discussion of the current section is sufficiently general to apply to any Levin-Wen model.

\subsection{The Levin-Wen models}
\label{sub:LWmodels}

The models we will consider are a class of Levin-Wen models which describe doubled anyon theories.  That is, beginning with a valid anyon theory, we construct a lattice model which ends up being equivalent to the achiral {\it double} of that theory.  (We note that this is not the most general class of Levin-Wen model; we discuss a more general form of the approach presented here in Appendix \ref{NonModApp}).
    Here we briefly review the key features of these models;  we refer the reader to Ref.~\onlinecite{LW} for more details.

Analogous to the doubled Chern-Simons theory described above in section \ref{sub:latticeCStheory}, the ingredients for this construction are a Hilbert space consisting of a set of edge labels $0 \ldots r$, and two sets of projectors -- one acting at vertices, the other at plaquettes-- from which to build the Hamiltonian.
The essence of the Levin-Wen models is that 1) all terms in the Hamiltonian commute, but 2) vertex and plaquette projectors are not simultaneously diagonal in the basis of edge labels.  1) ensures that the model is exactly solvable, and 2) yields ground states that can be roughly described as a weighted superposition of all possible edge labelings that satisfy certain ``fusion" conditions at each vertex.
This is a natural generalization of the toric code \cite{KitaevToric} and RVB phase \cite{SondhiRVB} (in which ground states are a superposition of all possible closed loops on the lattice).
Ref.~\onlinecite{LW} demonstrates that the ground state degeneracy of these models depends only on the global topology of the 2d lattice (i.e, if the lattice forms a torus or a sphere etc), and that the low-lying excitations about this ground state have anyonic statistics -- in other words, they realize topological phases.

The detailed construction of these projectors has a natural interpretation in the language of tensor categories.
The vertex projector at vertex $V$ will be denoted $B_V^{ijk}$, where $i,j,k$ are the quantum numbers on the bonds incident to that vertex.  The projector $B_V$ gives one if the vertex is allowed by the fusion rules of the category\footnote{The Levin-Wen models require only fusion rules, and not the matrix $R$; such a mathematical structure is known as a tensor category and is more general than the anyon theories which we focus on here.  See section \ref{NonModApp}}, and gives zero otherwise.  A configuration of quantum numbers along the edge which satisfies all of these projectors (except possibly at the position of quasi-particles, if there are any in the system) is known as a ``string net".

\begin{figure}[htb]
\begin{center}
\begin{tabular}{cc}
\subfigure[] 
{
 \includegraphics[width=1.5in]{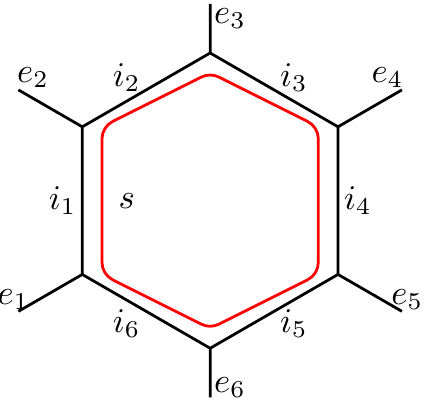}
  \label{LWF1a}
} \hspace{1cm}
&
\subfigure[] 
{
 \includegraphics[width=1.5in]{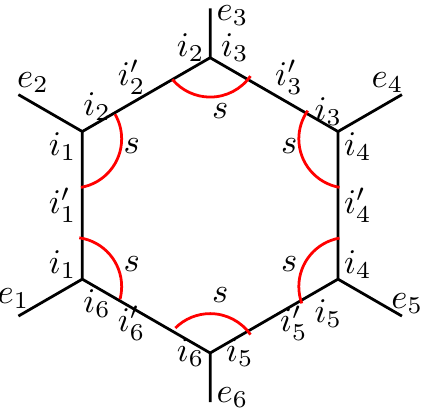}
  \label{LWF1b}
}
\end{tabular}
\hspace{1cm}
\subfigure[]{
  \includegraphics{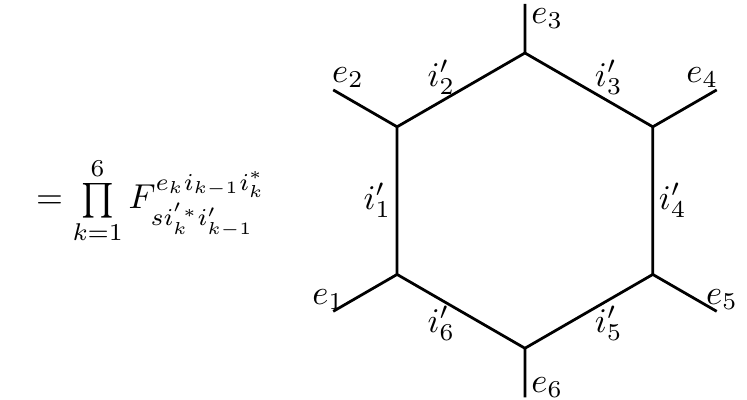}
  \label{LWF1c}
}
 \caption{String operator representation of the plaquette projector $\mathcal{{B}}_P(s)$.  (a) The projector is constructed by running a labeled string $s$ (shown in red) around the interior of a plaquette.  (b) Fusing this string into the edges of the plaquette.  (c) Evaluating the bubbles using the relation of Fig. \ref{Tetra_Fig} gives a  map between different labelings consistent with the vertex projectors.}
 \label{LWF_1}
 \end{center}
\end{figure}

The plaquette projectors, denoted $B_P$, act on a set of edges surrounding an individual plaquette $P$ to ``flip" the quantum numbers of these edges without violating  the vertex constraints, so that the ground state is indeed a superposition of string nets.
This is done by adding a `string operator' (with a particular linear combination of labels) around the plaquette
and fusing it into the existing configuration of edge labels.
The effect of such a string operator can
be described by a chain of  $F$-matrices describing the fusion processes involved.
A string operator labeled $s$ which runs around a single plaquette $P$
thus acts on the edge labels $i_1 ... i_6$ according to:
\be \label{Eq_PlStr}
{\mathcal{B}}_P(s) | i_1 ... i_6 \rangle = \left ( \prod_{k=1}^6
 F^{ e_k i_{k-1} i^*_{k}}_{s i^{' *}_{k} i'_{k-1} }\right )| i'_1 ... i'_6 \rangle
\ee
where $i_0 \equiv i_6$, and $e_k$ is the edge entering vertex $k$
from outside the plaquette, as show in
Fig. \ref{LWF_1}.
The plaquette projectors are a superposition of such string operators:
 \be \label{Eq_Bp}
{B}_P =\sum_{s=0}^{r} a_s {\mathcal{B}}_P(s)
 \ee
where $a_s$ are constants.  Ref.~\onlinecite{LW} shows that
choosing
\be \label{eq:as} a_s = \Delta_s / {\cal D}^2
\ee
ensures that $B_P$ is a projector onto states in which the plaquette $P$ contains no external sources, and hence
can be `filled in' without punctures.   (Other choices of $a_s$ project onto states in which an anyon world-line pierces $P$.  From the perspective of the lattice model, these anyons are not excitations of the theory, but rather topological defects -- specifically, punctures carrying flux -- on the plaquette.)

The Levin-Wen Hamiltonian consists of applying these projectors to each vertex and plaquette
in the honeycomb lattice:
 \be \label{Eq_LWHam}
  H= \epsilon_V \sum_V (1-{B}_V) + \epsilon_P \sum_P ( 1- {B}_P ) 
  \ee
where $\epsilon_P$ and $\epsilon_V$ are the mass gaps for vertex- and plaquette- violatIng excitations, respectively.  In the remainder of this section, we will show how to evalulate the partition function for the Hamiltonian, Eq.~(\ref{Eq_LWHam}) in the limit $\epsilon_P,\epsilon_V \rightarrow \infty$ , in a way that makes its relationship to the Chern-Simons formulation discussed in Sect.~\ref{OverSect} apparent.

The technically informed reader will notice that for simplicity we treat cases where there are no fusion multiplicities.  The more general case is a straightforward extension of this, in which the Hilbert space also includes variables at the vertices of the honeycomb lattice to indicate the relevant fusion channel, as explained in Appendix A of Ref.~\onlinecite{LW}.  In this case the $F$ matrices in the Hamiltonian have four extra indices to track the four vertex variables involved in such a change of basis.  All of the results derived here are equally valid in the presence of multiple fusion channels, provided that the lattice model is constructed from a valid anyon theory.

\subsection{Pictorial construction}  \label{LWPict}

  We now give a graphical construction of the partition function for the Levin-Wen models discussed in the previous section.
The essence of this construction is exactly that described for Chern-Simons theory in Sect.~\ref{OverSect}:  we represent the partition function as a graph -- or worldline diagram -- consisting of labeled loops encircling the perimeters of  plaquettes
in the $3$D lattice, and projector loops encircling its edges.  (The expert reader should note that this construction can also be made for tensor categories that do not admit a braiding structure.   We discuss this in more detail in Appendix \ref{NonModApp}).    We evaluate the
partition function using a Trotter decomposition approach.  That is, we discretize time in short steps and trace over each time slice separately, such that
\be
 \label{eq:Trotter}
Z = Tr \prod  e^{- H \, \delta \tau }
 \ee
where $\delta \tau$ is a small imaginary time step and the product is over many such small steps.  Hence we will visualize the theory as living on a $3$D lattice, with  $2$D honeycomb planes stacked in the time direction, referring to edges
within a honeycomb layer as space-like, and edges joining layers as time-like.
The reader should note that our diagrammatic approach calculates the partition function in {\em imaginary} time; one may view this either as a thermal description of a classical model, or as related to the quantum theory by analytic continuation.

There are two elements that make this construction of the partition function work.  First, in Sect.~\ref{Trotter}, we show that in the limit that the gap for spontaneously exciting quasi-particles is infinite , the operator $e^{ -\Delta \tau H_{LW}}$ is exactly equivalent to acting with a product of vertex and plaquette projectors on the state at time $\tau$.  Second, in Sect.~\ref{sec:detailedLW}, we establish that the action of the required product is given by evaluating the \chainmail diagram.   In particular, along the time-like edges, projecting onto the $0$ string effectively enforces the same constraint as $B_V$: the net quantum number entering each vertex in the lattice must be $0$ at each time step.  Similarly projecting onto the $0$ string along space-like edges ensures that $\Omega$ loops around spatial plaquettes give the action of ${B}_P$.

\subsubsection{Trotter Decomposition} \label{Trotter}

Before discussing the utility of this geometrical construction,
let us pause for a moment to understand the details of the
Trotter decomposition of the
partition function of Eq. (\ref{Eq_LWHam}).  We may express the
Levin-Wen partition function in imaginary time as
\ba
\label{LWPart_Eq}
 Z_{LW} &=& Tr \prod_{i=1}^N e^{ - H \, \delta \tau_i} \\
 &=& Tr \prod_{i=1}^N  e^{- \left [\epsilon_V \sum_V (1-B_V) + \epsilon_P \sum_P (1-B_P)
\right ] \delta \tau_i} \\ &=&  Tr \left[ \prod_{i=1}^N \prod_V e^{-(1-B_V)
\epsilon_V \delta \tau_i } \prod_P e^{-(1-B_P) \epsilon_P \delta \tau_i } \right] \n
 \ea
where the third
line follows because all projectors in the theory commute.  Here
we imagine computing the partition function by discretizing the
full imaginary time interval $\tau$ into $N$ time steps, each of duration $\delta \tau_i = \frac{\tau}{N}$.

Further, since $B_P$ and $B_V$ are projectors, we have $(1-B_\alpha)^n = 1 - B_\alpha$, and hence:
\ba
e^{-(1-B_\alpha) \delta \tau} &=& 1+\sum_{n=1}^\infty \frac{(-\epsilon_\alpha \delta \tau)^n }{n!} (1-B_\alpha ) \\ &=& e^{-\epsilon_\alpha \delta \tau}  - (e^{-\epsilon_\alpha \delta \tau} -1 )  B_\alpha
\ea
Plugging this into the partition function Eq.~(\ref{LWPart_Eq}) gives:
\ba \label{LWZ2_Eq}
Z_{LW} &=&Tr \left \{ \prod_V \left [\prod_{i=1}^N (e^{-\epsilon_V \delta \tau_i}  - (e^{-\epsilon_V \delta \tau_i } -1 )  B_V  ) \right ] \right .\n
&&\left. \prod_P \left [ \prod_{i=1}^N (e^{-\epsilon_P \delta \tau_i}  - (e^{-\epsilon_P \delta \tau_i} -1 )  B_P ) \right ] \right \}\\
 &=& Tr \left\{ \prod_V (e^{-\epsilon_V \tau}  - (e^{-\epsilon_V \tau } -1 )  B_V  )  \right. \n
 && \left. \prod_P (e^{-\epsilon_P \tau}  - (e^{-\epsilon_P \tau} -1 )  B_P ) \right\} \nonumber
 \ea
where the last equality follows from the fact that
 \ba
(e^{-\epsilon_\alpha \delta \tau_1} + (1- e^{-\epsilon_\alpha \delta \tau_1}) B_\alpha ) (e^{-\epsilon_\alpha \delta \tau_2} + (1- e^{-\epsilon_\alpha \delta \tau_2})
B_\alpha ) \n
 = (e^{-\epsilon_\alpha (\delta \tau_1+\delta \tau_2)} + (1- e^{-\epsilon_\alpha (\delta\tau_1+\delta \tau_2)}) B_\alpha ) \n
 \ea
In other words, the result is independent of the time slicing.

Now, if we take the limit $\epsilon_\alpha \tau \rightarrow \infty$, effectively
restricting the trace to the ground state sector, Eq.~(\ref{LWZ2_Eq}) reduces to
\be \label{eq:partfirst}
Z_{LW} |_{T=0} =Tr \prod_V B_V \prod_P B_P
 \ee
So, in fact, we may obtain the partition function by simply evaluating this operator over a single time slice.  However, since $B_V^n = B_V$ and $B_P^n = B_P$ and further all of the $B_V$ and $B_P$ are mutually commuting, we are free to apply these operators at every time slice and we will obtain the same partition function.  Thus we may write equally well
\be
    Z_{LW}|_{m_\alpha \tau \rightarrow \infty } = Tr \prod_{i=1}^N \left[ \prod_{V}B_{V(i)} \prod_{P} B_{P(i)} \right]  \label{eq:partsecond}
\ee
where by $B_{V(i)}$ and $B_{P(i)}$ we mean to apply $B_V$ and $B_P$ at time slice $i$ respectively.    In fact, this is the form of the partition function that we will actually use.

Our ability to manipulate the partition function in the above ways relies heavily on the fact the Hamiltonian is made of such simple mutually commuting projectors--  or equivalently, that the partition function is topological, which guarantees that $Z_{LW}$ is independent of the time slicing.  (Indeed, the typical Trotter decomposition is implemented precisely to deal with noncommuting operators, which we do not have).  Given this, it may seem to be overkill to calculate the partition function by using a full three-dimensional lattice as in Eq.~\ref{eq:partsecond} rather than only with a single time slice as in Eq.~\ref{eq:partfirst}.   However, such a representation will be crucial below.  First, we will be able to relate this three dimensional form to previously studied mathematical quantities such as the Turaev-Viro invariant\cite{TuraevViro} and the \chainmail invariant\cite{Roberts}.  Second, and more importantly, when we consider quasi-particles in our theory, we will want to consider quasi-particles following world lines in space time; collapsing the system down to a single time slice, as in Eq. \ref{eq:partsecond}, will be insufficient to describe this physics.

\subsubsection{Detailed Construction}
\label{sec:detailedLW}

It remains to show that the expression (Eq. \ref{eq:partsecond}) for the partition function can be obtained by applying the rules outlined in Sect.~\ref{sub:categories} to the appropriate \chainmail diagram of anyon world lines.  To do so, we first argue that time-like plaquette loops correctly propagate edge variables between time slices.  Second, we will show how the combination of $\Omega$ loops around the edges and plaquettes gives the product of projectors (\ref{eq:partsecond}).  The reader should note that $ \Omega$ is not strictly a projector; rather ${\cal D }^{-1} \Omega$ has eigenvalues $0$ and $1$.  To make the correspondence to the lattice Hamiltonian explicit here we use projectors, rather than $\Omega$ loops, to implement the Hamiltonian.  (We will show in Sect.~\ref{sub:combinatorial} that these factors of $\cal{D}$ cancel, and do not alter the normalization of the partition function.)

\begin{widetext}
\begin{figure}[htp]
\begin{center}
\begin{tabular}{c c}
\subfigure[]{
   \includegraphics[width=2.75in]{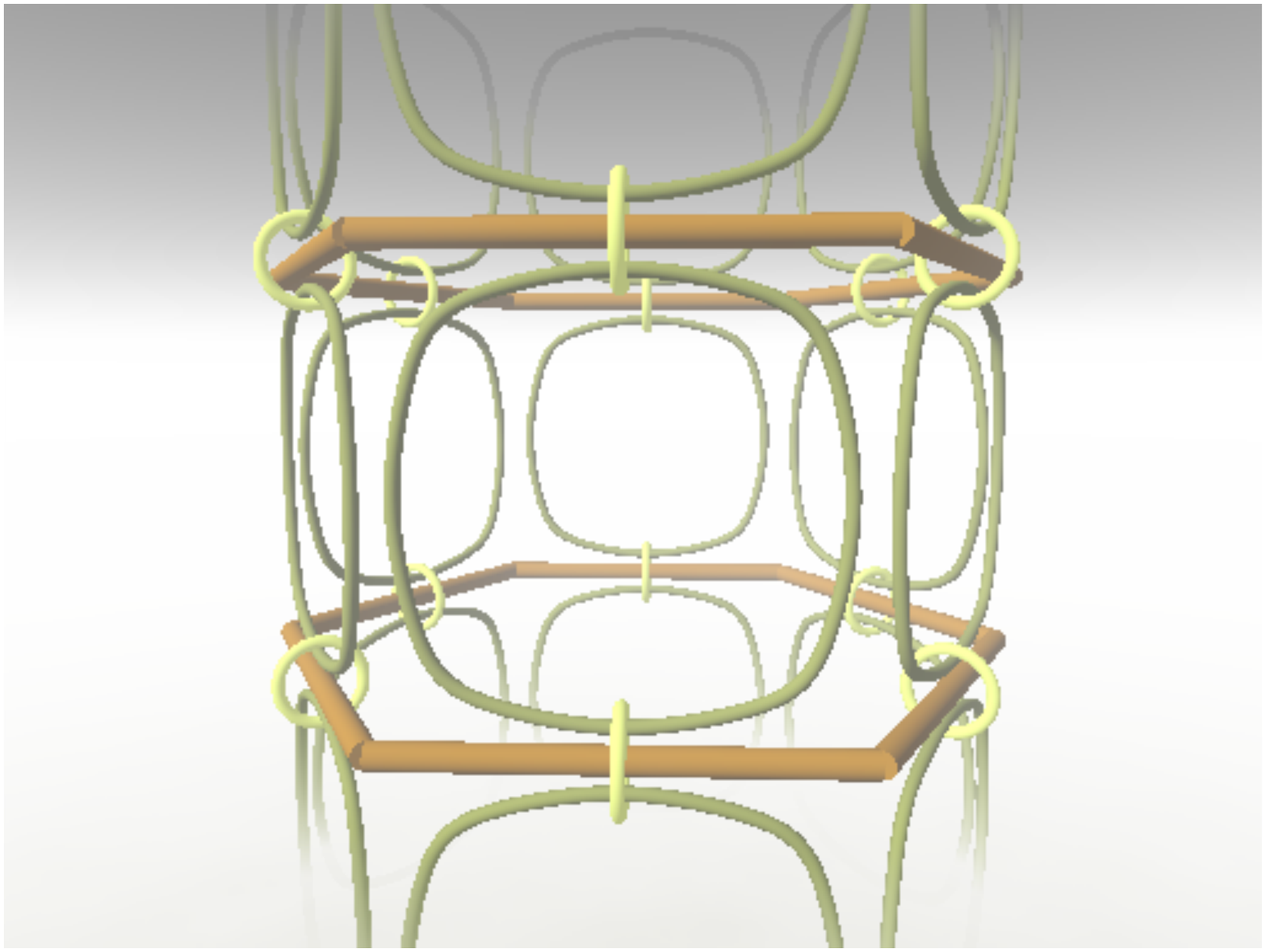}
   \label{Fig_Tloops}
} &
\hspace{1cm}
\subfigure[] 
{
    \includegraphics[width=2.75in]{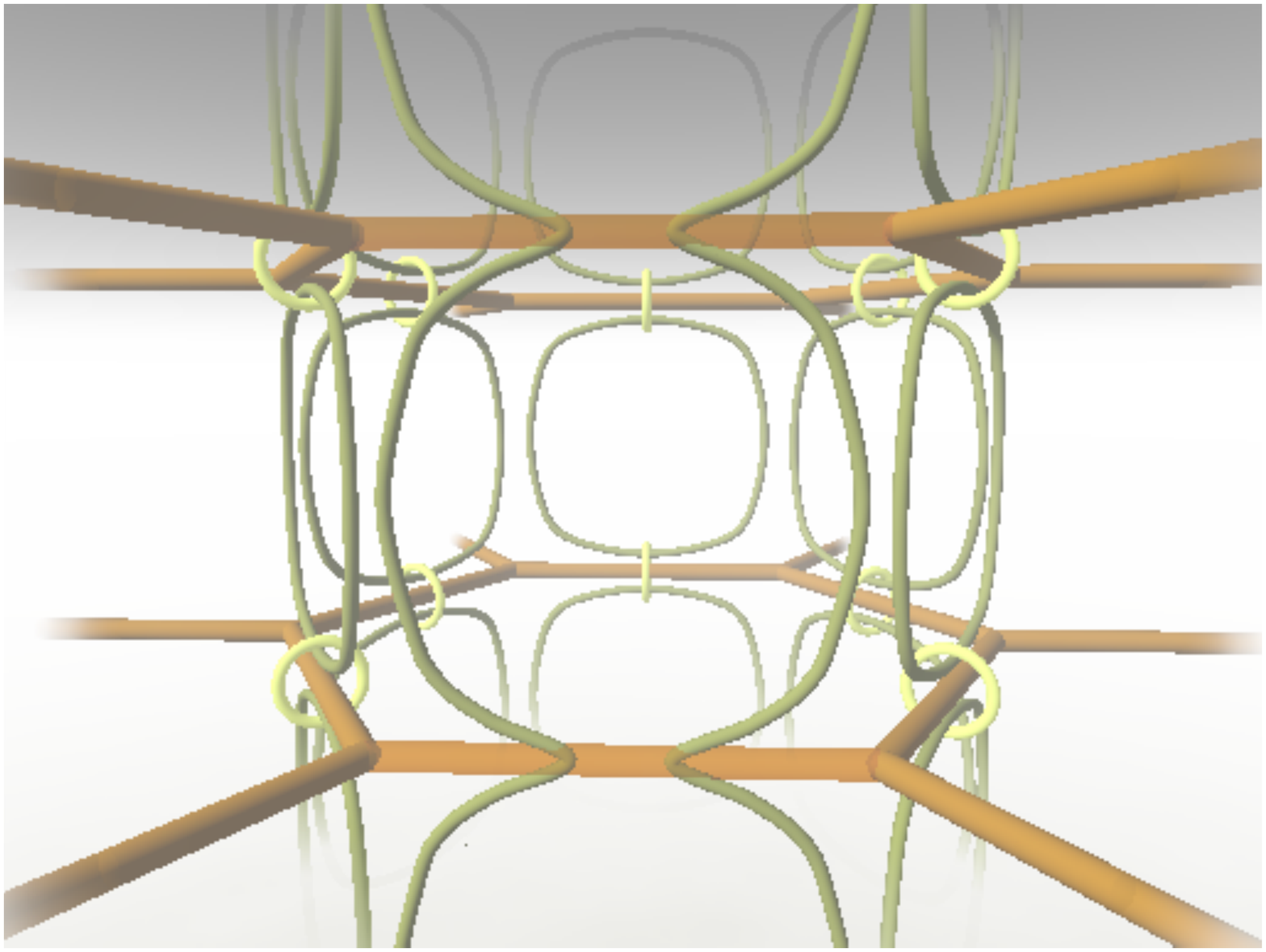}
} \\
\subfigure[] 
{
 \includegraphics[width=2.75in]{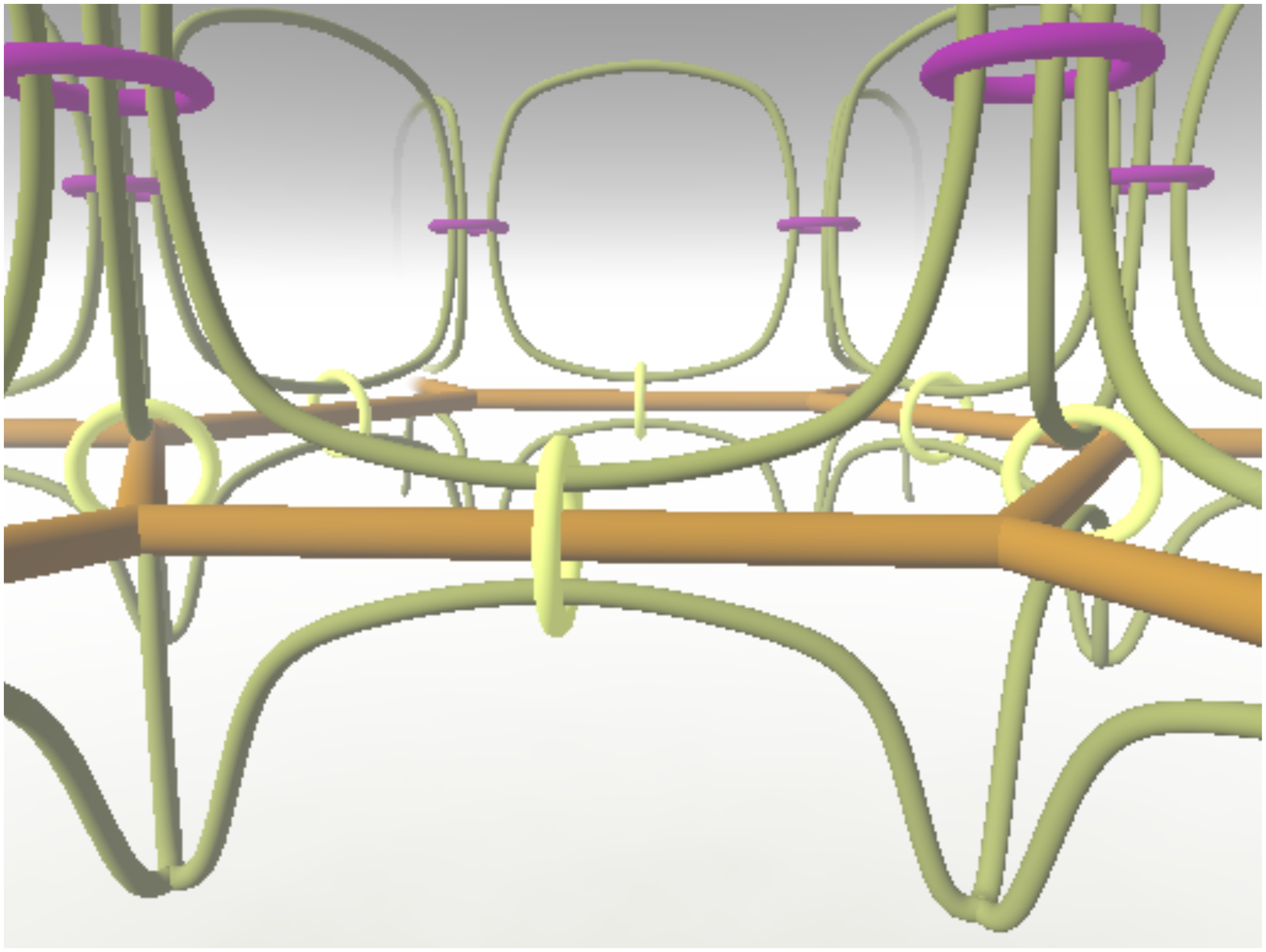}
} &   \\  
\end{tabular} 
\end{center}
\caption{The pictorial Levin-Wen model, drawn on
the honeycomb lattice tiled in time.  To construct the final picture
( the full \chainmail link with both vertex and plaquette projectors) shown in Fig. \ref{CHFig}, a closed loop is drawn
around every plaquette in the lattice, and a projector is applied on
each edge.  Here we show the intermediate steps in evaluating the diagram.  The thick golden lines show the plaquettes in the honeycomb lattice; 
strings encircling the time-like plaquettes, which serve to label the states in the model at 
each time slice, are show in green.  (a) and (b) Propagation
of edge labels without the Hamiltonian.   The yellow rings show the projector loops on space-like edges.  Applying these projectors propagates the label of each loop forward in time.  (c) Applying vertex
projectors (purple rings) between space-like layers forces the three edge labels incident at each vertex to fuse to the identity, satisfying the criteria for a string net ground state.      } \label{Fig_CH1}
\end{figure}
\end{widetext}

First, as in Sect.~\ref{OverSect}, we
represent states in this picture in terms of oriented closed curves representing anyon world lines.  Thus for each space-like edge labeled $i_t$ at time
$t$, we draw a closed loop above this edge and label it $i_t$, as
shown in Fig. \ref{Fig_CH1}.  We may imagine this string as
the world-line of a particle-antiparticle pair, which
has been created at time $t$, and annihilated at time $t+\delta
t$.  We therefore call such strings {\it particle loops}, to distinguish them
from the strings which comprise the Hamiltonian.

It is instructive to consider evaluating the diagram containing only these world-lines and the space-like edge projectors.  This describes particle world-lines propagating unaltered through time --- i.e., a Hamiltonian which is zero.  To see this, consider a space-like edge with a particle loop $i_{t - \delta t}$ below, and $i_t$ above (green loops in Fig. \ref{Fig_Tloops}).  Applying the projector (yellow rings in Fig. \ref{Fig_Tloops} along their shared edge at time $t$, as shown in Fig. \ref{Fig_CH1}, fuses the two strings $i_{t- \delta t}, i_{t}$ in a bi-valent vertex on each side of the projector.  The result can be interpreted as a world-line for the particle-anti-particle pair $i_{t- \delta t}$ propagating from time $t$ to time $t +  \delta t$.  In this way, adding projectors along all bonds lying in space-like directions results in a picture of an initial set of labels propagating unaltered through this time slice.  
In other words, the space-like edge projectors have the effect of taking an inner product between $i_{t - \delta t}$ and $i_t$ (which are orthogonal unless the labels $i_{t - \delta t} $ and $i_t$ are identical).

To be precise, we must normalize this inner product correctly.    To do so, we multiply each edge loop $i$ by $\Delta_i$.
This is because fusing $i_{t- \delta t}$ and $i_t$ gives a factor of
\be \label{Eq_EdgeNorm}
 \delta_{i_{t - \delta t}, i_t} F^{i_t^* i_t 0}_{i_t^* i_t 0} =\frac{1}{\Delta_{i_t}}
 \ee
 for each edge.  Hence we must give each state a weight of $\Delta_i$
to cancel these factors and ensure that the time evolution effected by the space-like edge projectors  is unitary.

To obtain the partition function Eq.~\ref{eq:partsecond}, we must add the Hamiltonian to this picture.   The vertex projectors $B_V$ correspond to adding edge projectors
to the time-like edges, as shown in Fig. \ref{Fig_CH1}c.  Since these projectors force the three lines at the vertex to fuse such
that their total quantum number at the vertex is $0$  (i.e., such that the three incident quantum numbers are an allowed fusion), they clearly reproduce the effect of
$B_V$.  To ensure that the correspondence is exact, one must keep track of the various coefficients induced by fusion.  This calculation is given in Appendix \ref{FusionApp}.

\begin{widetext}
\begin{center}
\begin{figure}[htp]
\begin{tabular}{c c}
\subfigure[]{
   \includegraphics{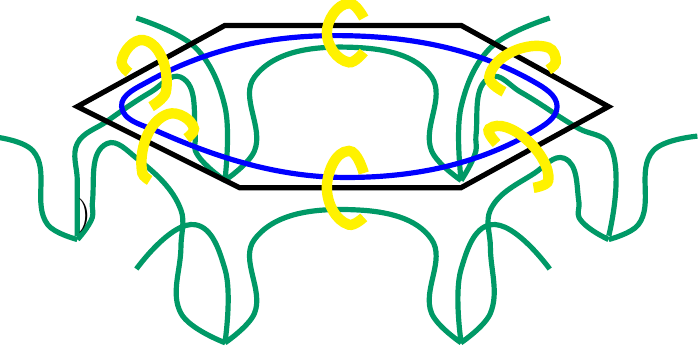}
} \hspace{1cm}
&
\subfigure[]{
   \includegraphics{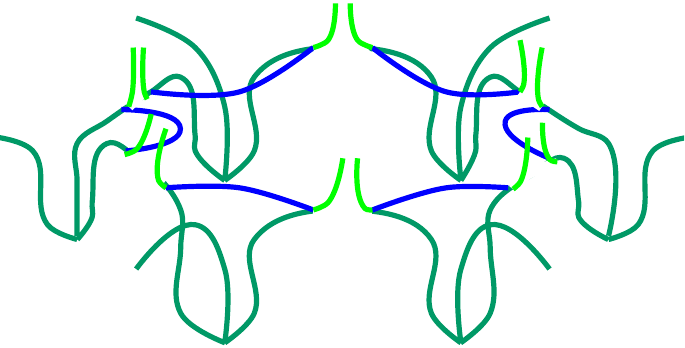}
}
\\
\subfigure[]{
  \includegraphics{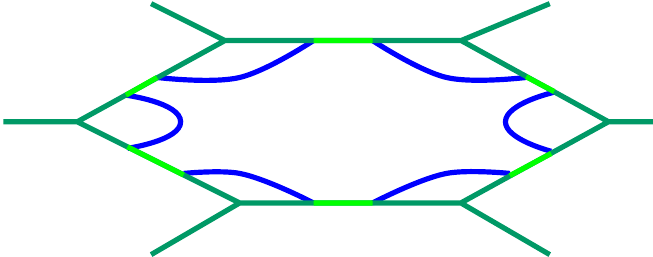}
}
&
\subfigure[]{\label{Fig7d}
 \includegraphics{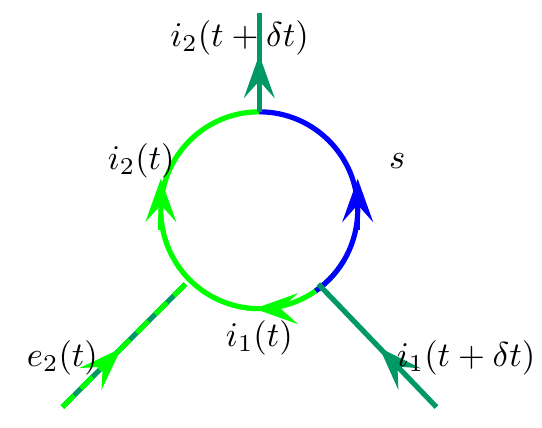}
}
\end{tabular}
 \caption{Applying edge projectors in a single time slice to implement the action
of the Hamiltonian on the edge labels of a single plaquette. (a) The plaquette with edge variables (green) at time $t$ fused beneath each vertex.  (b) Fusing the string corresponding to the plaquette projector  (shown in blue) to the edge variables at time $\tau$ (dark green) and $\tau+ \delta \tau$ (light green-- not shown in (a)) by implementing the space-like edge projectors (yellow rings in (a)).  Implementing the vertex projectors at time $\tau + \delta \tau$ requires fusing the open green strings to each other (not shown).  (c) Alternative picture of (b).  Here we have omitted the vertex projectors at $\tau + \delta \tau$, and pulled the  (light green) strings carrying these edge labels down into the plane.  The bubbles in this diagram can be collapsed to trivalent vertices using the fusion relations, just as in Fig. \ref{LWF_1}.  (d) The diagram which remains at each vertex can be evaluated using the identity of Fig. \ref{Tetra_Fig}.  The details of this calculation are given in Appendix \ref{FusionApp}.  Here  $e_2(t) = e_2(t+ \delta t)$. }
 \label{CHF_3}
\end{figure}
 \end{center}
\end{widetext}

To add the plaquette string operator $\mathcal{{B}}_P(s)$, we run a string labeled $s$ around the boundary of the hexagonal plaquette $P$,
passing it through all of the projectors on the edges of $P$.
The edge projectors force this string to fuse with the edge labels,
producing (after some diagrammatic algebra, given in Appendix \ref{FusionApp})
exactly the product of $F$-matrices  in Eq.~(\ref{Eq_PlStr}).  The action of the plaquette projectors 
$\mathcal{B}_P$ is thus included by labelling all space-like plaquettes by
 \be \label{ProjP}
{\cal D }^{-2}\sum_{i=0}^{r} \Delta_i \, | i \rangle \ \ \ .
 \ee

Fig. \ref{CHF_3} shows the effect of applying both Hamiltonian projectors to a single  plaquette at a given time slice.

Including edge projectors above each vertex, and plaquette projectors on every space-like plaquette,  we obtain precisely the expression given in Eq.~\ref{eq:partsecond} for the partition function.  To see this, it suffices to consider a single time slice. Applying the projectors above each vertex annihilates configurations with nonzero total quantum number entering or leaving a vertex; on the remaining (allowed) configurations, it gives:
\be \label{eq:CHPart}
\prod_{ \{P_l \}}  \left ( \sum_{s=0}^{r} \frac{\Delta_s}{ \mathcal{ D}^2} \prod_{k=1}^6
 F^{ e_k i_{k-1} i^*_{k}}_{s i^{'^*}_{k} i'_{k-1} }\right )
\ee
times the diagram for the remaining time steps.  Eq. (\ref{eq:CHPart}) is precisely the factor we expect from $\prod_{ \{V_i \} }B_{V_i} \prod_{ \{P_l \}} B_{P_l}$ at this time step, given a particular set of edge labels.  
To
obtain the zero temperature partition function, we then sum over all labelings $i$ of the
edges, with appropriate weights $\Delta_i$ as described above in Eq. \ref{Eq_EdgeNorm}.  
This is equivalent to labeling each edge loop with 
\be \label{Proj0}
\sum_{i=0}^{r} \Delta_i \, | i \rangle \equiv \mathcal{D} | \Omega \rangle \ \ \ .
 \ee

In summary, the pictorial representation of the full Trotter decomposition of Eq. \ref{eq:partsecond} is
as follows:  first we fill the manifold with layers of honeycomb
lattice at each (discrete) time step. Draw a labeled loop above
each edge in each honeycomb layer to specify the value of the edge variable in that time slice.     To operate with $B_V$ on a vertex, apply a projector to the time-like edge on that vertex.  To operate with $B_P$ on a plaquette,  draw the appropriate superposition of strings $\sum a_s s$ around that plaquette. Finally, add a projector on each spacelike edge, to take the inner product between the states represented by time-like loops above and below each time-slice, with the appropriate string operators sandwiched in between.  All projectors, as well as the time-like plaquette loops, are strings of the form (\ref{ProjP}).

Up to normalization, then, the diagram corresponding to the ground state partition function is thus
generated by 1) drawing a closed loop around each plaquette in the
$3D$ lattice, labeling each loop with the element $\Omega$,
 2) applying a projector onto $0$ quantum number to each edge of the $3D$
lattice by encircling it with a loop labeled $\Omega$ and 3) evaluating the coefficients given by the resultant fusions. The final resulting diagram, the \chainmail link, is shown in Fig.1b.

Physically, one way to think of the diagrams described here is rather complicated pictures detailing the action of a transfer matrix.  Imposing the edge projectors on a time-like edge gives the appropriate transfer matrix element for $B_V$ acting on the $3$ incident edges.  Similarly, imposing the edge projectors on all space-like edges touching a given plaquette yields the appropriate matrix elements of $B_P$ between the initial edge configurations-- represented by the labeled strings which emenate from the previous time slice, and run `backwards' along the edges of the plaquette-- and the final configurations, represented by strings which extend upwards towards the next time slice, and run `forwards' along the edges of the plaquette.  

\subsubsection{ Classical and quantum partition functions}

It is worth pausing to clarify one question which some readers may have at this point: if the lattice models we are interested in are truly quantum creatures, how is it that we can describe their partition functions in terms of state sums, such as the Turaev-Viro invariant, which are essentially classical entities?  Viewed as a gauge theory, our system is already quantized -- indeed, this quantization is essential to the relation between expectation values of Wilson lines and knot polynomials.  However, viewed as a theory of labeled strings, the link used to evaluate the partition function is essentially a classical object: each component of the link is labeled with the same element $| \Omega \rangle$; thus quantum superpositions are not required to define the \chainmail link (or, consequently, the partition function).  

In fact, quantum mechanics enters only if we wish to examine the state at a particular instant in time by cutting the manifold open at a particular time slice.  This is characteristic of Chern-Simons theories, in which the quantum mechanical behavior is captured by the appropriate conformal field theory at the boundary.  To expose this quantum theory in the lattice model, we must first pinch off the projectors above each vertex, so that the slicing will not cut through any of the strands of the \chainmail link.  In pinching off these vertices, however, we effectively limit the labels incident at each vertex such that only the combinations which fuse to $0$ flux remain.  At this point it is no longer sufficient to think of each edge as labelled by the single ket $| \Omega \rangle$, as this fusion treats each string label $|i \rangle$ in the superposition differently.  (In other words, the projectors are diagonal in the basis of string labels, rather than a basis in which $| \Omega \rangle$ is a basis vector).   Hence the sum over labels on the time-like loops is transformed into a sum over all possible intermediate {\it string net} configurations.

\section{Doubled Anyon Theories and Relation to the \chainmail Invariant} \label{CHSect}

Armed with the diagrammatic representation of the Levin-Wen partition function described in Sect.~\ref{LWPict}, we may now return to the main theme of this work -- namely, exploring and exploiting the connection of some of the Levin-Wen models to doubled anyon theories (including doubled Chern-Simons theories).  Since
such theories have been extensively studied both as field theories in physics, and as topological invariants in mathematics, this is a useful framework from which to study the lattice models.

To make the desired connection, we will begin by introducing  the \chainmail invariant $Z_{CH}$ of Ref.~\onlinecite{Roberts} described briefly in section \ref{OverSect} above.  Roughly speaking, $Z_{CH}$ is obtained by associating a link diagram (the \chainmail link, $L_{CH}$) with a space-time $3$-manifold, and evaluating the resulting link invariant (denoted $\langle L_{CH} \rangle$).  $\langle L_{CH} \rangle$ is evaluated using the diagrammatical rules of an anyon theory laid out in Sect.~\ref{sub:categories}.
Remarkably the \chainmail invariant $Z_{CH}$ is precisely the same as the invariant of Turaev-Viro\cite{TuraevViro} which itself is the same\cite{WalkerOld,Turaev} as the square of the Witten-Reshitikhin-Turaev invariant $Z_{CH} =
Z_{WRT} \overline{{Z}_{WRT}}$.  (Here the overline denotes either complex conjugation of the result, or, equivalently, evaluating the manifold invariant for the mirror image manifold.)

There are many equivalent links associated with a given space-time manifold (associated with different lattices which can be used to tile space-time -- or more technically, with different Heegard splittings of the space-time manifold);  we shall see that one of these produces exactly the same link diagram  which we associated with the Levin-Wen partition functions in Sect.~\ref{LWPict}.  We then explore what the other, equivalent, diagrams tell us about the lattice model.

\subsection{The \chainmail link and the lattice partition function}
\label{sub:combinatorial}

A general prescription\cite{Roberts} for constructing the \chainmail link $L_{CH}$ is as follows\footnote{Mathematically inclined readers should note that here we describe how to construct the link for a given lattice; the formal prescription associates a link to any Heegard splitting of the $3$-manifold}.
Given any 3D lattice, wrap a string just inside the perimeter of each plaquette; encircle the plaquette strings that  run adjacent to each edge by another string, and label all strings with the element $\Omega$.  On a cubic lattice, the result is a picture with $4$ strings running parallel to each edge, linked together by an $\Omega$ loop. In the case of the 3D lattice which is the space-time representation of the Levin-Wen model's honeycomb lattice, the result is exactly the link shown in Fig. \ref{Fig_CH1}.

Using this prescription to construct the \chainmail link, the \chainmail invariant is given by
\be  \label{Eq_ZCH}
Z_{CH}= {\cal D}^{-n_v - n_c} \langle L_{CH} \rangle \ \ \ .
\ee
(Here $\cal D$ is the total quantum dimension from Eq. \ref{eq:totalquantumdimension} and $n_v$ is the number of vertices of the lattice and $n_c$ is the number of 3d cells of the lattice\footnote{Strictly speaking $n_c$ is the number of 3-handles of the Heegard splitting that defines the manifold\cite{Roberts}.})  Here, $\langle L_{CH} \rangle$ means that one should evaluate link invariant of the \chainmail link using the combinatorial rules of the given anyon theory.

The \chainmail diagram defined here, and the link diagram presented in Sect.~\ref{LWPict} to represent the lattice partition function, are identical links.   Hence in evaluating the two diagrams using the combinatorial rules outlined in Sect.~\ref{sub:categories}, the only  difference is in the normalization: the strings are labeled  as specified in Eqs.
(\ref{Proj0}) and (\ref{ProjP}) for the lattice partition function, and by $\Omega$ for the \chainmail link.  To check that the normalizations agree, we must
count the factors of ${\cal D} = \sqrt{ \sum_{i=1}^r
\Delta_i^2}$ in the two descriptions.    Compared to the projectors we used in the construction of the Levin-Wen partition function in section \ref{sec:detailedLW}, the \chainmail link obtains one factor of $\cal D$ for every edge (since we apply projectors along both space-like and time-like edges) and for each space-like plaquette of the lattice.
On the other hand, to obtain the Levin-Wen construction as in section \ref{sec:detailedLW} we found in Eq.~\ref{Eq_EdgeNorm} that we should normalize the sum over edge variables by multiplying all edge variables with a factor of $\Delta_i$.  In the \chainmail link, the time-like plaquette $\Omega$ loops correspond to the edge variables, but are normalized to include a factor of $\Delta_i/{\cal D}$.  Thus, the \chainmail link obtains a factor of ${\cal D}^{-1}$ for every time-like plaquette, cancelling the factor of $\cal D$ we obtained for the associated space-like edge.  This leaves one extra factor of $\cal D$ for every time-like edge and for every space-like plaquette.  However, the number  of time-like edges is equal to the number of vertices $n_v$, and the number of 3d cells $n_c$ equals the number of space-like plaquettes.  (This counting is valid provided we take periodic boundary conditions in imaginary time, and the topology of space remains fixed during the evolution).  Hence the prefactor of the \chainmail invariant ${\cal D}^{-n_v - n_c}$ precisely cancels these factors and we end up with the same normalization in both cases.

In fact, there is some arbitrariness in the normalization of the topological invariant, which is evaluated relative to a reference manifold whose partition function we choose to be $1$, as explained in Sect.~\ref{subsub:omega}.  Physically, however, this is not the case: we expect the zero-temperature partition function to count the ground-state degeneracy.  The simplest spatial manifold with ground-state degeneracy $0$ is the sphere $S^{(2)}$; if we use periodic boundary conditions in imaginary time, we conclude that  $Z_{LW} (S^{(2)} \times S^{(1)}, T =0) = 1$.  The two agree in this case because the \chainmail invariant is also normalized such that $Z(S^{(2)}\times S^{(1)} ) =1$.

\subsection{Surgery, Handle-Slides, Topological Invariants and Invariance}
\label{sub:handleslide}

In the previous subsection, we showed that the \chainmail invariant is
none other than the partition function of a Levin-Wen Hamiltonian
constructed from an anyon theory.
Here we will discuss how topological invariance can
be understood in the context of these link invariants.  This will give us a convenient framework for
discussing topological invariants of quasi-particle world-lines,
which we will exploit in Sect.~\ref{QPSect}.

We will focus on two observations about the partition function of the lattice model which are apparent from the topological properties of the link invariant.
First, we can relate the partition
function of the lattice model on a space-time manifold ${\cal M}$ to that of
a theory with no $\Omega$ loops on a space-time manifold ${\cal M}\# {\cal \overline{M}}$, through
a process called {\em surgery} which we will outline below.
This is the generalization of the connection pointed out in Sect.~\ref{OverSect} between
these lattice models and doubled continuum Chern-Simons theories.
Second, the \chainmail invariant has the property, known as the {\em handleslide} property,
which implies that  certain re-arrangements of the link components do not alter the partition function.
Such re-arrangements are an interesting tool for visualizing the topological nature of the partition function.

\subsubsection{Surgery and Handleslides}
\label{subsub:handleslide}

We begin with a description of the two properties which we will use in the remainder of this section. (For a pedagogical introduction to the subject, see Ref.~\onlinecite{Gompf}).

For the purposes of this work, a technical description of surgery will not be necessary; the interested reader may consult Appendix \ref{SurgApp}.  The essential point is that surgery gives a way of establishing an equivalence between partition functions of anyon theories on pairs of space-time manifolds $ \mac{M}$ and $\mac{M}'$ with different topologies.   The manifolds are related by `performing surgery' on one or more closed loops in $\mac{M}$ to produce $\mac{M}'$.   In fact, surgery is an essential component of the interesting correspondence between links in the $3$-sphere $S^3$ and the topological classification of $3$-manifolds.
Specifically, it is a well-known, but nontrivial result\footnote{Amusingly, in Ref.~\onlinecite{Witten}, Witten refers to the Lickorish-Wallace theorem\cite{LickorishOld1,LickorishOld2,Wallace} as a `not too deep result'.}, that {\it any} closed $3$-manifold can be described by starting with $S^3$ and performing surgery on an appropriate link\cite{LickorishOld1,LickorishOld2,Wallace,Gompf}.   This is the root of the correspondence between topological invariants of  $3$-manifolds, and invariants of the corresponding links.

\begin{figure}[htp]
\begin{center}
  \includegraphics{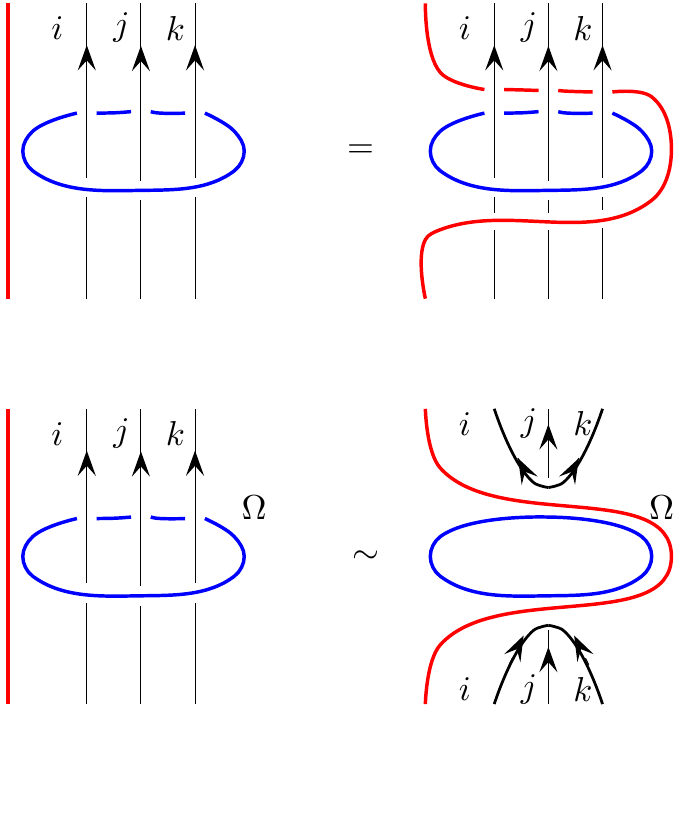}
   \caption{\label{Handleslide} The handle-slide property.  A loop (blue) is said to have the handle-slide property if any string can be deformed freely around it in the manner shown in (a) without changing the value of the link invariant associated with the diagram.   Fig.~(b) shows the intuitive reason why $\Omega$ loops have the handle-slide property: Using the Killing property of $\Omega$ we forces all strings which pass through the $\Omega$ loop
 to fuse to the identity, allowing the external string to slide freely across the loop as shown.  Once the string has been slid across the loop on the right-hand side, the strings ${i,j,k}$ may be reattached through the middle of the $\Omega$ loop again (exactly cancelling the coefficient $ F^{i^* i 0}_{j j^* k^*} F^{k k^* 0}_{k k^* 0}$ incurred due to fusion) to give the right hand side of Fig.~(a). } 
\end{center}
\end{figure}

Another feature of surgery which has interesting consequences for the lattice model is that certain re-arrangements can be made to the link in $\mac{M}$ without affecting the topology of $\mac{M'}$\footnote{Such re-arrangements are known as Kirby moves; see e.g. Ref.~\onlinecite{Gompf} for a detailed discussion.}. Thus, in order for a link invariant to be a topological invariant of a 3-manifold, it must also remain invariant under these same re-arrangements.   For our purposes, the most important of these is the {\it handle-slide} property\footnote{This move has a natural interpretation in the surgery picture, as one handle (attached during surgery) sliding over another -- hence the name} shown in Fig.~\ref{Handleslide}.

The handle-slide property is a statement about how loops in a link can `slide' over one another, whilst leaving the value of the link invariant unchanged. The handle-slide is shown in Fig. \ref{Handleslide} a: quite literally, we slide the (red) string on the left over the (blue) loop, irrespective of what other strings pass through the loop's center.   
One way to construct such a link invariant which is unchanged by handle-slides, is to label all  strings with $\Omega$.  An intuitive justification\footnote{Strictly speaking handle-slide is a more general property which must also apply to $\Omega$ loops that have been self-knotted; a proof that $\Omega$ satisfies the general handle-slide property can be found in Refs.~\onlinecite{Lickorish}, \onlinecite{Wang}.  In this paper, however, we will not need to consider this more general move.} for this is shown in Fig.~\ref{Handleslide} b:  Using the Killing property of $\Omega$, we may replace any strings threaded through $\Omega$ by the identity or vacuum string.  Since there are now no strings linked with the $\Omega$ loop, we may now freely slide any string around this loop without obstruction.  The fusion of encircled strands onto the vacuum can be un-done once the string has been pulled around the $\Omega$ loop, giving the original diagram with a handle-slide.

\subsubsection{Insights on topological invariance}
\label{sub:ramifications}

The surgery and handleslide properties give numerous insights into the nature and 
properties of the topological lattice models described with the \chainmail link.  Here 
we describe the most interesting among these.

{\it Lattice theory and continuum gauge theory:} Through surgery, we can make an explicit connection between the lattice model and (when a Chern-Simons anyon theory is used) doubled Chern-Simons theory.  The latter can be obtained not by taking the
naive continuum limit, but rather by identifying both the edge variables (appropriately summed over to give the ground-state partition function) and the terms in the Hamiltonian with anyon world-line loops, and performing surgery on the resulting link.

{\it Doubling:} Surgery gives one route to understanding how the lattice theory, which is in general constructed from labels in a chiral anyon theory, gets doubled.  Ref.~\onlinecite{RobertsThesis} showed that performing surgery on the  \chainmail link of a manifold ${\cal M}$ produces ${\cal M}\# {\cal \overline{M}}$.  Here  ${\cal \overline{M}}$ is the mirror image of ${\cal M}$, and the $ \#$ (or connected sum) indicates that the two copies are joined together by cutting a solid ball out of each and gluing the two holes together along there $2$-sphere boundary.  Since the partition function of the anyon theory obeys
\be   \label{Eq_CHSurg}
Z_{WRT}({\cal M}\# {\cal \overline{M}} ) = Z_{WRT}({\cal M}) Z_{WRT}({\cal \overline{M}}) \ \ \ ,
\ee
this shows that the lattice model, whose partition function is given by the \chainmail invariant $Z_{CH} = Z_{WRT}({\cal M}\# {\cal \overline{M}} ) $, is a doubled anyon theory.  
In appendix \ref{SurgApp} we elaborate further on this connection, examining how the rigorous mapping works even when ``quasi-particle" defects are inserted into the \chainmail link.

{\it Turaev-Viro invariant}: Roberts \cite{RobertsThesis} showed
that the \chainmail invariant is rigorously equivalent to
the Turaev-Viro invariant, which has been understood\cite{LW}
to describe the Levin-Wen ground state partition function.  
(We give some details of the argument orignially presented in Ref.~\onlinecite{RobertsThesis} in appendix \ref{SurgApp}).
Hence we can both establish a rigorous correspondence between
the two, and understand how the models of Levin and Wen differ
from the Turaev-Viro invariant in the presence of quasi-particle defects (which we will explore below in section \ref{QPSect}).

{\it Independence of Lattice}: the \chainmail construction can be  carried
out on any lattice, and gives the same result -- that is, it is a topological invariant of the space-time manifold and does not depend on the particular lattice discretization.  We will discuss
the implications of this for possible alternative Hamiltonians, on lattices with different geometries, in
Sect.~\ref{NTriSect}.

Further, once a lattice is selected, it is possible to coarse-grain the \chainmail description without altering the partition function.  Specifically, as emphasized by Ref.~\onlinecite{RobertsThesis}, in the absence of excitations it is always possible to eliminate all but a finite number of the plaquettes in the lattice through a series of handle-slides.  (The specific number will depend on the topology of the space-time manifold, and in particular the minimal number of $1$ and $2$-handles required to construct it).  This procedure demonstrates an exact equivalence between the partition function of the lattice model with an {\em arbitrary} choice of lattice constant, and a simple product of a finite number of $F$-symbols.   Indeed, it gives a  geometrical understanding of the result found by Ref.~\onlinecite{LevinNave, WenGu}, who construct an algebraic coarse-graining procedure which is exact for these topological phases.

{\it Commutativity of operators in $H$:}  The handle-slide of $\Omega$ also gives a convenient picture of various manifestations
of topological invariance in our picture of the Levin-Wen Hamiltonian.
For example, with some effort one can show algebraically that the Hamiltonian is comprised
of commuting projectors.  In the \chainmail picture, this has a simple geometrical
meaning:   $\Omega$-loops on adjacent plaquettes can handle-slide past each other,
changing the order in which they are to be evaluated in time.  Similarly, an
$\Omega$ plaquette loop can handle-slide (along a time-like plaquette loop)
past an edge projector.   Hence
handle-slide invariance {\it requires} that all operators in $H$ commute.
Ref.~\onlinecite{LW} motivates the choice of commuting projectors by arguing that 
the Hamiltonian of a purely topological theory should play the role of imposing 
an appropriate set of constraints on the wave-functions; 
the \chainmail picture gives an interesting alternative route to understanding why commuting 
projectors are the appropriate building blocks for the Hamiltonian of a 
topological lattice model.

{\it Independence of time-slicing}:  At the end of Sect.
\ref{Trotter}, we argued that as the Hamiltonian is comprised of
commuting projectors, the partition function does not depend on
how many time steps are used in the Trotter decomposition, or on
whether we evaluate all plaquette operators simultaneously or at different times.
(Here we assume that the topology of space is fixed during the evolution).
From the point
of view of the partition function, this freedom is a consequence
of the fact that all operators in $H$ commute.  Hence a time step
may be `subdivided' into separate applications of $B_P$ -- or
collapsed, so that all $B_P$'s act simultaneously.  In practice, we can do this in
the \chainmail picture by handle-sliding an $\Omega$-loop
which effectuates an application of $B_P$ from one time slice to the next --
in a manner similar to that used to establish commutativity above.
Interestingly,
from the point of view of the \chainmail link, this freedom is
required for topological invariance: it corresponds to changing
the lattice by simultaneously adding or removing a
plaquette (2-handle) and a 3-cell (3-handle --in this case, a solid ball
filling in the cells of the lattice).  This transformation does
not alter the topology of the space-time manifold, and hence
leaves the partition function unaltered.  Thus again, we find 
a fundamental connection between commuting projectors 
and Hamiltonians with topological ground states.

To summarize, by using a pictorial construction of the Levin-Wen partition function at zero temperature, we arrive at a natural correspondence to the \chainmail invariant.  This makes explicit the relation of Levin-Wen models to doubled Chern-Simons theory.   Further, it underlines the relationship between topological invariance and exactly solvable Hamiltonians written in terms of commuting projectors.  Finally, in this language we are naturally lead to consider some of the flexibilities of the Levin-Wen models, such as analogous construction on arbitrary lattices.

\section{Quasi-Particles in the Pictorial model}  \label{QPSect}

We now turn to the question of understanding the quasi-particle defects of
the theory.   We note that, whereas the topological invariants discussed in the above sections (Chain-Mail, Turaev-Viro), which correspond to the ground state of the Levin-Wen Hamiltonian, have been well studied in the mathematical literature, not all of the situations with violations of terms in the Hamiltonian which we discuss here have, to the best of our knowledge, yet been studied (but see Refs. \onlinecite{Martins1, Martins2}).

Since there are two types of terms in the Hamiltonian, the vertex terms and the plaquette terms, we should think about quasi-particle defects that violate one or both, of these terms.  A useful intuition for these two types of quasi-particles, which is exact in the case of Abelian models \cite{SondhiSC}, is the idea of electric and magnetic defects in analogy to gauge theories: electric defects violate vertex projectors, and magnetic defects violate the plaquette projectors.  More generally, though, we should refer to these types of defects as vertex or plaquette defects.
Here we will explain how to construct each type of violation in the pictorial model, by inserting extra strings (quasi-particle world lines) into the \chainmail link.  This gives geometrical insight into the difference between the two types of quasi-particles, which is reflected, via surgery, in
their relationship to Chern-Simons theory.

\subsection{Partition Functions with quasi-particles}
\label{sub:partitionqps}

Before delving into the details of how quasi-particle world-lines are included in the \chainmail link,  let us understand what the \chainmail invariant computes outside of the ground-state sector -- i.e., in the presence of quasi-particle world lines.  As emphasized in Sect.~\ref{OverSect}, the partition function that we compute is a topological invariant of the space-time -- which is possible only if it is evaluated at $T=0$.  
Thus strictly speaking, the quasi-particle world-lines that we insert are perhaps best conceived of as sources created in the system along some space-time path by an external field, rather than as particles generated by thermal fluctuations.  The new ground state contains quasi-particles following the world-lines of these sources; 
the \chainmail invariant then computes the ground-state partition function in the presence of these quasi-particles.  
Alternatively, we can view the \chainmail link as capturing only the topological portion $Z_{top}$ of the partition function at finite temperature-- that is, it captures the physics of the linkings of quasi-particle world lines around non-contractible curves in the spacetime, or around each other, but is insensitive to the  $e^{- E_{g}\tau}$ contribution of the quasi-particle creation energy $E_g$.  This is apparent in the discussion of Sect.~\ref{Trotter}, where we see that the correspondence between the partition function and the evaluation of the pictorial model is exact only at $0$ temperature.  

What is this topological contribution from the quasi-particle
world-lines to the partition function?  
In the simple case where
world-lines do not enclose non-contractible curves on the manifold, we will show presently that they
may be detached completely from the rest of the link by
handle-sliding.  In this case $Z_{top}$ factors into the contribution of the ground state partition function, times a contribution from the world line link.  More generally world lines may link around non-trivial topology in the manifold, in which case they can also mix different ground state sectors in the theory.  Hence, $Z_{top}$ tracks the topology of the spacetime manifold, and the linking of quasi-particle world lines with each other and around non-contractible loops in the space-time.

\subsection{Brief Review of Quasi-particles in the Levin-Wen models}
\label{sub:QPreview}

Ref.~\onlinecite{LW} describes all possible excitations by constructing operators which act on all edge states $i_k$ along some continuous path in the lattice.  These string operators are constructed from products of ``simple'' string operators, which are given by:
\ba \label{LWSOps}
\mathcal{O}_s &=& \prod_k F_s^k  \omega_s^k \\
F_s^k &\equiv &\begin{cases} F^{e_k i^*_k i_{k-1} }_{s i'_{k-1} i^{'*}_k} & \mbox{ if $s$ turns left at } V_k \\
F^{e_k i^*_{k-1} i_k}_{s i^{'}_k i^{'*}_{k-1} } & \mbox{ if $s$ turns right at } V_k
\end{cases} \n
\omega_s^k &\equiv &\begin{cases} \omega^{i_k i^{'}_k}_s &  \mbox{ if $s$ turns right then left at }
 V_k, V_{k+1} \\
\overline{\omega}^{i_k i^{'}_k}_s & \mbox{ if $s$ turns left then right at } V_k, V_{k+1} \\
 1 &  \mbox{ otherwise} \nonumber
\end{cases}
\ea where $V_k$ are the vertices, $i_k, i_k^{'}$ are the states
along the path of the string before and after fusion with the
string $s$, and $e_k$ is the state on the external leg (not
traversed by the string) at $V_k$. The $F_s^k$ are the $F$-matrices of the constituent anyon theory that the Levin-Wen model is built on.  These are precisely what we obtain by running a string labeled $s$ along the chosen path,
and applying a projector on each edge to force it to fuse
trivially with the two states $i_k, i'_k$.   Note that factors $\omega_s^k$ introduce phases each time the string crosses from one plaquette to another.  As we will see below, these $\omega$ matrices are closely related to the $R$ matrices of the constituent anyon theory.

The operators for closed strings can be shown to commute with the Hamiltonian, Eq.~(\ref{Eq_LWHam}), whereas operators for open strings connect a particle-anti-particle pair at the opposing ends of the string.    If the phases $\omega_s^k$ are trivial, then the particle-anti-particle pairs at the ends of the strings violate the vertex projectors only.   In order to create violations of the plaquette projectors at the ends of the strings, it is also necessary to introduce nontrivial values of the phases $\omega_s^k$.

\subsection{Quasi-particles in the \chainmail picture}
\label{sub:QPMTC}

We now turn to the task of constructing these quasi-particles in the
pictorial model.  
The basic idea is that
passing an additional string (labeled $s$) through an $\Omega$ loop  (projector) forces all other
strings passing through this loop to fuse to $s^*$, rather than to
$0$   (since $s^*$ must fuse with $s$ to give $0$).   For example, threading the projector loop above vertex $V$ with a string labeled $s$ fuses the string labels of the edges entering $V$ to $s^*$, rather than $0$, creating a defect in the ground state which violates the vertex
constraint $B_V$.  This suggests that quasi-particles should be added as 
labeled strings, representing quasi-particle world-lines, linked appropriately through the various projectors of the \chainmail link.  Here we will first describe how these world lines thread the space-time link, and then explain how they are interpreted in the Hamiltonian language.  We will see in Sect.~\ref{subsub:QPSTats} that depending on whether we link this new world line with vertex projectors or with both plaquette and and vertex projectors, we will get either the right- or left-
handed quasi-particle sectors respectively.  This gives an interesting explanation and proof of conclusions about quasi-particle statistics reached by Ref.~\onlinecite{Gilsetal}.  Further, as expected for the doubled theory,  these two sectors will be completely decoupled. (We refer to the original anyon theory as ``right" handed, and its mirror image as ``left" handed).

\subsubsection{Threading quasi-particle strings into the \chainmail link}
\label{subsub:threads}

\begin{figure}[htb]
 \begin{center}
\subfigure[]
{
    \includegraphics[width=2.75in]{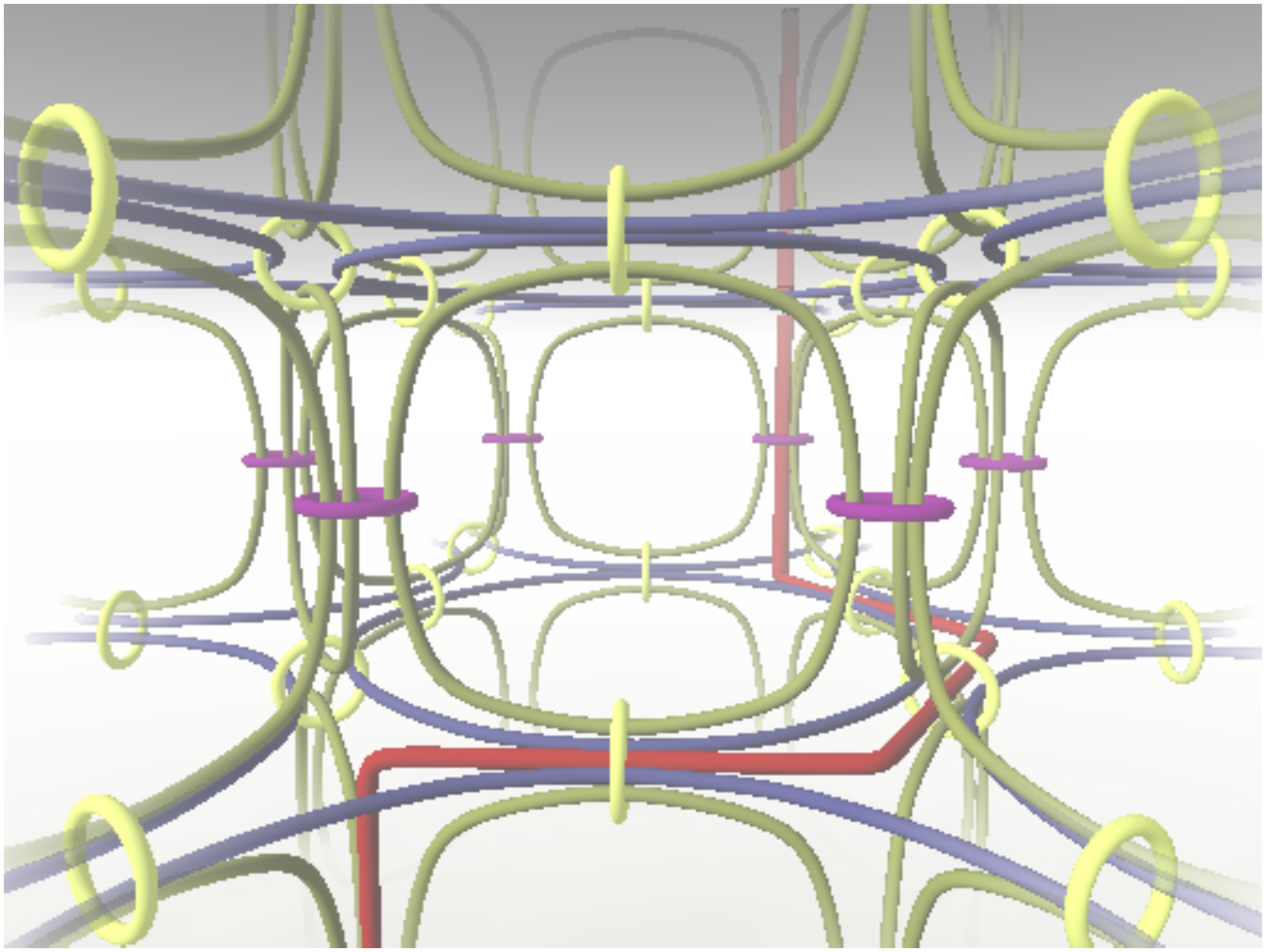}
    \label{QPFig_El}
} \hspace{1cm}
\subfigure[]
{
    \includegraphics[width=2.75in]{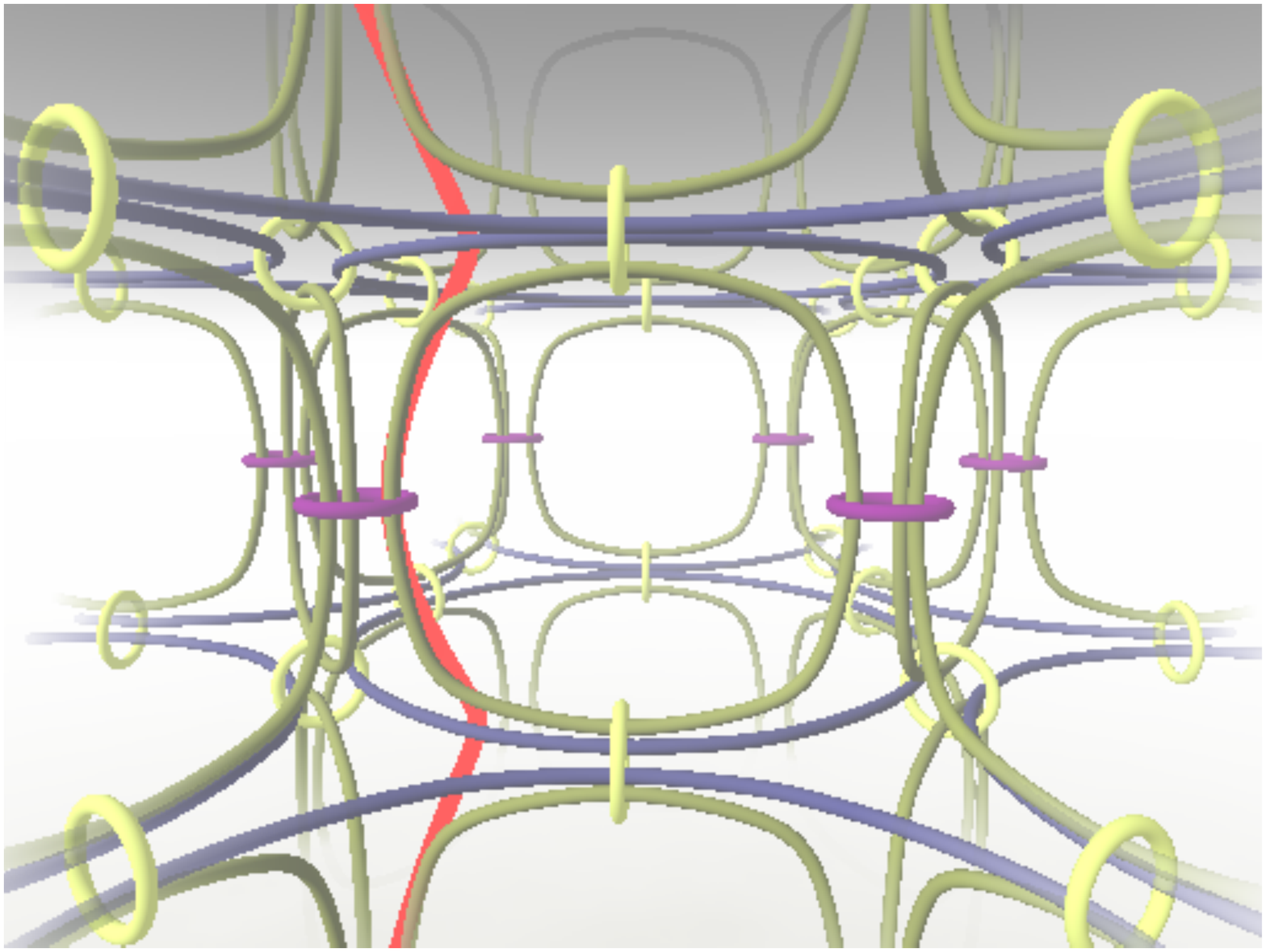}
    \label{QPFig_Mag}
} \hspace{1cm}
\caption{Quasi-particle trajectories.  Quasi-particle strings are shown in bright red, and oriented according to 
the direction of positive imaginary time. All other loops are labeled by $\Omega$. (a) Right-handed
quasi-particles run along the centers of the thickened edges,
passing through the edge projectors (yellow and purple rings) only.  (b) Left-handed
quasi-particles thread alternately through edge (yellow and purple) and plaquette (blue) 
projectors, and can be thought of as living on the plaquettes.}
\label{QPFig}
 \end{center}
\end{figure}

Let us begin by describing how to construct the two types of quasi-particles in the \chainmail picture.  To insert vertex violating quasi-particles, we run a string (representing the quasi-particle's world line) labeled $s$ along the edges of the lattice, linking it through the projector
for every edge it runs along (Fig. \ref{QPFig_El}).  We show shortly that such strings are associated with right-handed excitations, and violate only the vertex terms $B_V$ of the Hamiltonian.  We call these quasi-particles ``R-particles" for short.

Violations of the plaquette operators (Fig. \ref{QPFig_Mag}) are constructed in a
similar fashion ---  by running a labeled quasi-particle world-line along some set of
edges in the lattice.  However, to create a violation of the plaquette operators $B_P$,
this string must pass through the loops encircling the space-like
plaquettes.  We also stipulate that such a
string  links through a time-like plaquette every time it
crosses into a new plaquette in space\footnote{The reason for this
stems from requiring that the left-handed strings be able to
handle-slide freely about the lattice.}.  In other words, whenever a string passes between cells of the 3D lattice, it must cross through a plaquette (thus becoming linked with the corresponding plaquette string), rather than an edge.  Such strings are associated with left-handed excitations (landing in $\overline{{\cal M}}$ after surgery, as we will explain shortly) in
the doubled theory; we will call this type of quasi-particle ``L-particles".   Note that because they pass through both edge and plaquette loops, L-particles
violate {\it both} types of projectors.  To construct what we would normally associate with a ``magnetic" defect (violating only plaquette terms) in fact requires a combination of L- and R- particles.

To avoid ambiguity, it is possible to formalize the rules for choosing trajectories corresponding to the two quasi-particle types.  The trajectory for an R particle consists of a closed curve along the edges of the (3D) lattice.  This fixes a sequence of edge projectors through which the string is linked, and hence fully describes its action on the states in the theory.  The trajectory for an L particle consists similarly of a closed curve on the edges -- but in this case we must also specify a continuous trajectory along the $3$-cells bordering these edges, in order to fix which plaquettes the string is linked with.  The rules for selecting these $3$ cells are simply that for each edge on the path, there must be one $3$ cell which contains this edge, and that only $3$-cells containing at east one vertex on the path of edges can be included.  When passing between adjacent $3$-cells, the string links with the plaquette string on the face shared by the two $3$-cells.  This fixes a series of plaquettes `adjacent' to the edges on the world-line trajectory through which the string is to be linked.

To understand these strings in the Hamiltonian viewpoint, we must consider
the effect they have on the evaluated link invariant.  As both types of strings thread
through the edge projectors, applying these projectors forces them to fuse with the variables on
each edge that they traverse.
 If the edge in question lies in a
spatial plane, the string is linked only through
space-like edge projectors and, in the case of L-strings,  time-like plaquette loops.  Since
neither of these loops comprise terms in the Hamiltonian, such a
string has no associated energetic cost.   Rather, they
can be thought of as operators acting on the quantum state at this
time step.  (Indeed, evaluating the projector along this edge
gives exactly the operator $F^s_{k}$ of Eq.~(\ref{LWSOps}), and hence gives precisely the string operators of Levin and Wen).

Excitations in the Hamiltonian picture occur when a string follows an
edge that is oriented in the time direction.  In this case the string
travels through a vertex projector, creating an excitation at this
vertex by forcing the quantum numbers of the edges to no longer fuse to the vacuum.
In the case of L-particles, the world-line links alternately with vertex
projectors and plaquette
projectors, creating violations of both.

From the point of view of the state of the system in this
time slice, the $2D$ string operator
terminates at the last vertex (or vertex and plaquette, for L-particles) that it enters,
leaving a source at this vertex (vertex and plaquette).  This mimics an open string in the Levin-Wen formulation 
described above.
In the $3D$ pictorial representation, which is sensitive only to the topological part of the partition function, this distinction between space and time disappears, as the topological partition function is indifferent to the energy gap.  The $3D$ world-lines are simply closed curves, and  $Z_{top}$ depends only on how they are linked with each other and with the spacetime manifold.

\begin{center}
\begin{figure}\subfigure[]
{\begin{tabular}{cc}
    \includegraphics[width=1.7in]{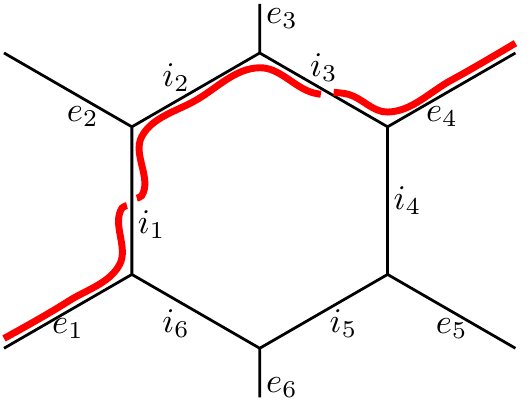} &
      \includegraphics[width=1.7in]{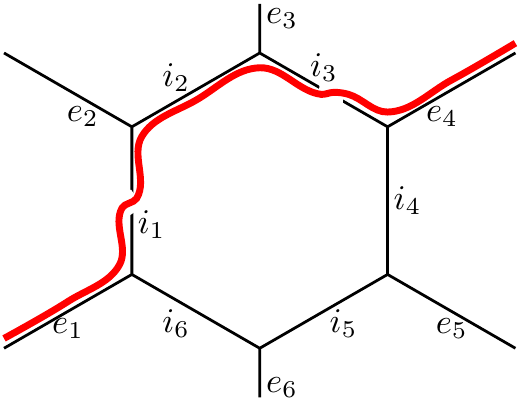}
    \label{QPFuse1}
\end{tabular}
} \hspace{1cm}
\subfigure[]
{
    \includegraphics{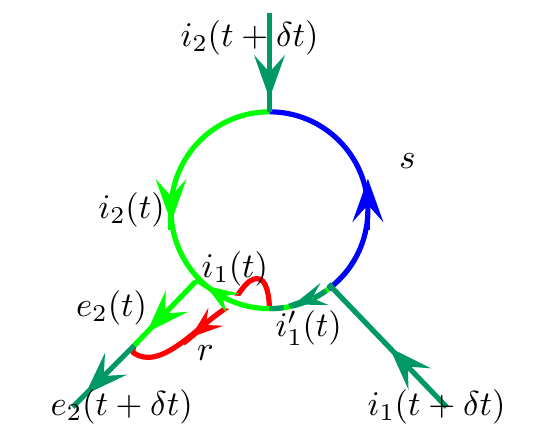}
    \label{QPFuse2}
} \hspace{1cm}
\caption{ \label{QPFusionFig}  Evaluation of diagrams involving quasi-particle world-lines.  (a) The projection used to reproduce the Levin-Wen string operators.  Here we look down at edge variables $i_t$.  L particles (at left) link with the plaquette loops which extend downwards in time below the hexagon shown, and hence pass under the strings $i_t$.  R strings (right) are not linked, and pass over $i_t$.  (b) The result of contracting all projectors about a vertex containing a quasi-particle world-line.  (This is the equivalent of Fig. \ref{Fig7d} in the case where a quasi-particle world-line enters the vertex).  As shown in Appendix \ref{FusionApp}, the diagram evaluates to $F^{ i_2(t) e_2^*(t) i_1(t) }_{r^* i_1' e_2^*(t+ \delta t)} R^{r i_1(t)}_{i'_1( t)}$ times the result of the vertex with no quasi-particle string.  If the string turns in the other direction, passing under the edge $i_2$ instead of $i_1$, this factor is changed to $F^{ i_1(t)  i_2(t) e^*_2(t)}_{r  e_2^*(t+ \delta t)i_2'} \left( R^{r i_2(t)}_{i'_2( t)}\right )^*$. This agrees with Eq. (\ref{LWSOps}) once we account for the differences in the string orientations.}
\end{figure}
\end{center}

\subsubsection{ Equivalence to quasi-particles in the Levin-Wen models}

The \chainmail link with quasi-particle world lines inserted as described above is shown in Fig. \ref{QPFig}.  To show that evaluation of the link invariant associated with this modified \chainmail link gives $Z_{top}$ for the Levin-Wen models, we must show that strings lying along spatial edges should reproduce the effect of string operators in the Hamiltonian theory.

The link invariant is evaluated by choosing a $2$D projection in which to draw the ($3$D) link, and applying the rules (described above in section \ref{sub:categories}) for fusion and un-crossing which are specified by the anyon theory.
The anyon theory satisfies certain consistency conditions ensuring that the result does not depend on the projection chosen.
It will prove convenient to choose a projection from the positive $t$ direction, looking down into the $x-y$ plane.  Let us consider a
configuration with a single layer, together with time-like
plaquettes below this layer to represent the state of the system
at this time slice,  as shown in Fig. \ref{QPFusionFig}.   In the projection that we have chosen,
L- strings always cross under the strings from the
time-like plaquettes, since they are linked with these before
projection; the R- strings, conversely, always pass over
these time-like strings.  Hence L- and R- strings
are assigned opposite phases at each crossing, as one would
expect.  Finally all edge projectors are applied;
 Fig. \ref{QPFuse2} shows diagramatically that the result is simply a factor of $R$ for each crossing, together with a factor of $F^s$ as expected for a Levin-Wen string operator.
The details of the diagrammatic evaluation are explained in Appendix \ref{FusionApp}.

To see that this gives the same prescription in terms of left- and
right- turns as Eq.~(\ref{LWSOps}), notice that if the R-
(L-) string turns right and then left, it crosses over
(under) a timelike plaquette string from right to left.
Conversely, if it turns left and then right, it crosses over
(under) a plaquette string from left to right.  Hence by applying
the un-crossing rules of Fig. \ref{RFig}, we obtain the expected
prescription for assigning phases to right- and left- turns, provided we choose our uncrossing tensor $R$ to be the same as $\omega$ in Eq.~(\ref{LWSOps}).

An entertaining consequence of the \chainmail formulation is that
the choice of phases in Eq.~(\ref{LWSOps}) is not unique.  The notion
of strings crossing depends on the projection chosen to evaluate
the link invariant -- though the end result does not.  When the
projection is into the $x-y$ plane, as described above, the
$R$- (vertex) particles appear to cross over, and the $L$ (vertex and plaquette)
particles under,  the time-like plaquette strings.  If the plane
of projection is orthogonal to one of the space-like axes,
however, right-handed strings cross no other strings after the
edge projectors are applied.  From this angle, left-handed strings
cross both over and under the time-like plaquette below each edge
shared by a pair of plaquettes on the string's trajectory.
Fortunately the algebraic structure of the modular tensor category
guarantees that the end result will be independent of the angle of
view, provided that the relative phases acquired by
left- and right- handed string operators $j$ when crossing an edge
labeled $i_t$ is $ (R_{i_t j}^{i_{t+\delta t} } )^2$.

\begin{figure}[htb]
\begin{center}
\begin{tabular}{cc}
 \includegraphics[width=1.75in]{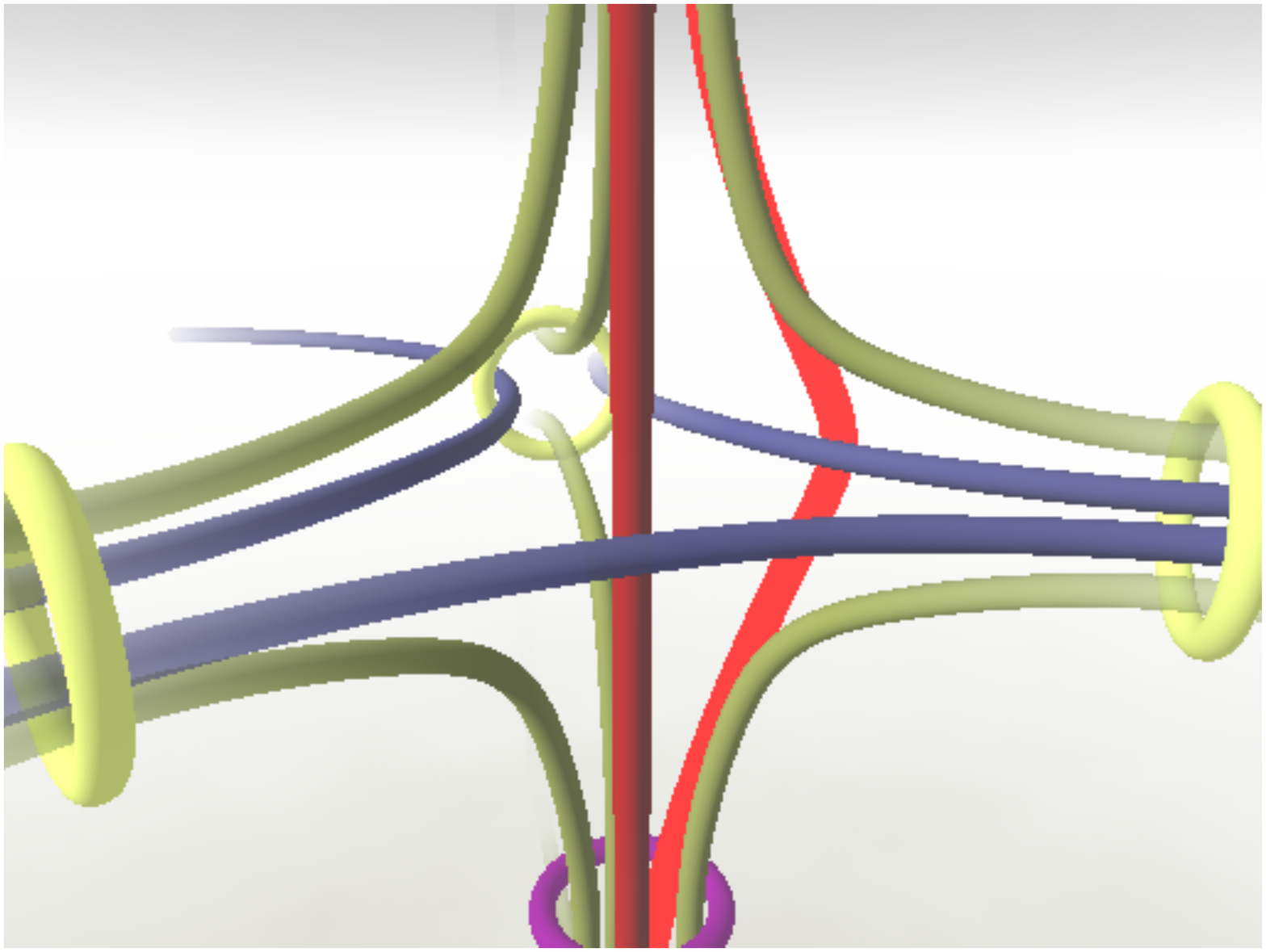} &
\includegraphics[width=1.75in]{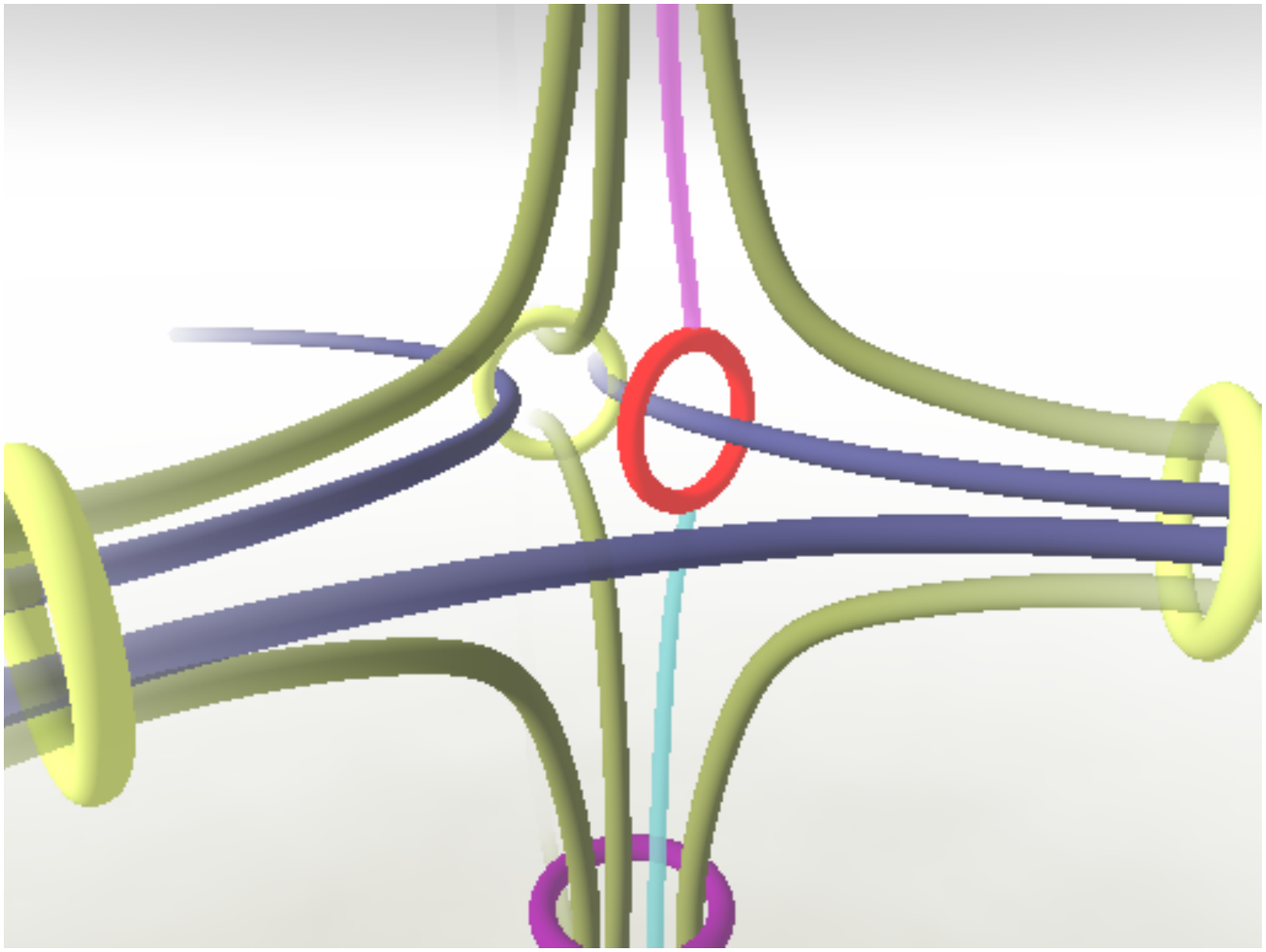}
\end{tabular}
\caption{Combining right- (dark red) and left- (light red) handed quasi-particle strings
to create a magnetic excitation.  In general fusing the two strings does not produce the $0$ string along the edges, but rather a linear superposition of the allowed fusion products (pale purple and pale blue strings in the right-hand Figure).} \label{MagPartFig}
 \end{center}
\end{figure}

As noted above, though our R- particles ( or ``electric" particles in Abelian gauge theory) violate only vertex projectors, the L- particles violate {\it both} vertex and plaquette constraints.  Thus the L- particles are not strictly ``magnetic"
excitations (violating plaquettes only) in the language of Abelian gauge theory , but rather a combination of magnetic and electric.  A purely `magnetic' excitation  can only be constructed by taking both
a right- and a left- handed string of the same label $s$ (See also Ref.~\onlinecite{Gilsetal}).  The
left-handed string crosses under, and the right-handed string
over, every plaquette loop which the left-handed string links.  We
may fuse the two strings together along each edge, as shown in
Fig. \ref{MagPartFig}.  This fusion will result in a superposition
of labels for the resulting strings along the edge, as shown in
the Figure.  As both strings carry the same label $s$, one element
of this superposition is the $0$ string.  When the two strings
fuse to $0$, the result is an isolated loop labeled $s$ encircling
each plaquette string with which the left-handed quasi-particle
was linked.  This violates these plaquettes, but no vertices,
giving a purely magnetic excitation.  (In general the fusion will
generate a superposition of excitations.  Only the component of this superposition corresponding to the $0$ string corresponds to a purely `magnetic' excitation which affects only plaquette variables. )  Interestingly, this shows that the purely magnetic
quasi-particle (when it exists) is achiral\footnote{This type of modification to the \chainmail invariant has also been discussed by Ref.~\onlinecite{Martins2}, who provides a proof, via surgery, of this achirality.}.

\subsection{Statistics and topological invariance of quasi-particle strings}

Thus far, we have found that we can reproduce the operators [Eq.~(\ref{LWSOps})] by creating R- strings, which violate vertex projectors and are linked through edge loops, and L- strings, which violate both plaquette and edge projectors and are linked through both edge and plaquette loops.  It is instructive, however, to ask how the topological characteristics expected of quasi-particles in the doubled theory are manifest in this construction.  There are several issues to address here.  First, the result should be invariant under local geometric deformations of the quasi-particle world lines.  Second, we must convince ourselves that the right- and left- handed sectors of the theory have the expected mutual statistics --- in other words, that they do not interact with each other. Finally, we have judiciously named the two types of quasi-particles right-  and left- handed --- we must  show that these names are indeed justified. 

It turns out that all three of these are relatively easy to show using the handle-slide property.  Handle-sliding allows any string $s$ to slide over a loop labeled by $\Omega$ without changing the value of the link invariant, irrespective of what strings pass through the $\Omega$ loop.  This allows us to slide quasi-particle strings around on the \chainmail link witout affecting the partition function-- guaranteeing that only the topology of the quasi-particle world-lines is relevant.  Further, one important consequence of the handle-slide is that it changes the linking of $s$ with any string that passes through the $\Omega$ loop.  This plays an important role in understanding the statistics of the quasi-particles, as we shall see shortly.

\subsubsection{Handle-slide prescriptions for quasi-particle strings}
\label{subsub:handle}

We will begin by describing the effect of handle-sliding (Sect.~\ref{subsub:handleslide}) on both types of quasi-particle strings.  Though arbitrary handle-slides of each string type are allowed, we use a convention in which handle-sliding preserves the linking conventions of R- and L- particles,  as specified in Sect.~\ref{subsub:threads}, respectively.

The simplest such prescription for handle-sliding the two types of
quasi-particles is as follows. R- strings need only to slide
over plaquette loops: since they are never
linked with the plaquette strings, there is no obstruction to
sliding over them and the strings can be maneuvered freely
from plaquette to plaquette in this way  (See Fig. \ref{QPSlide_1}).  L- strings must slide over both edge loops and plaquette loops in sequence.
The first slide unlinks the quasi-particle string from the
plaquette loop (in the process, linking it with another plaquette
loops), and the second slide moves it across the plaquette  (See Fig. \ref{QPSlide_1}).  These
maneuvers have the advantage that they do not introduce additional twists into the strings, and also do not alter the prescription that L- strings
must link with a plaquette loop each time they move between $3$-cells, while R strings are linked only through edge projectors.  Analogous to the R- strings, handle-sliding L- strings in this way allows the
string to move freely among the plaquettes in the $3D$ lattice.
These two types of slides are illustrated in Fig. \ref{QPSlide_1}.

\begin{widetext}
 \begin{center}
\begin{figure}[h!]
\subfigure[]{
  \begin{tabular}{c c}
    \includegraphics[width=2.75in]{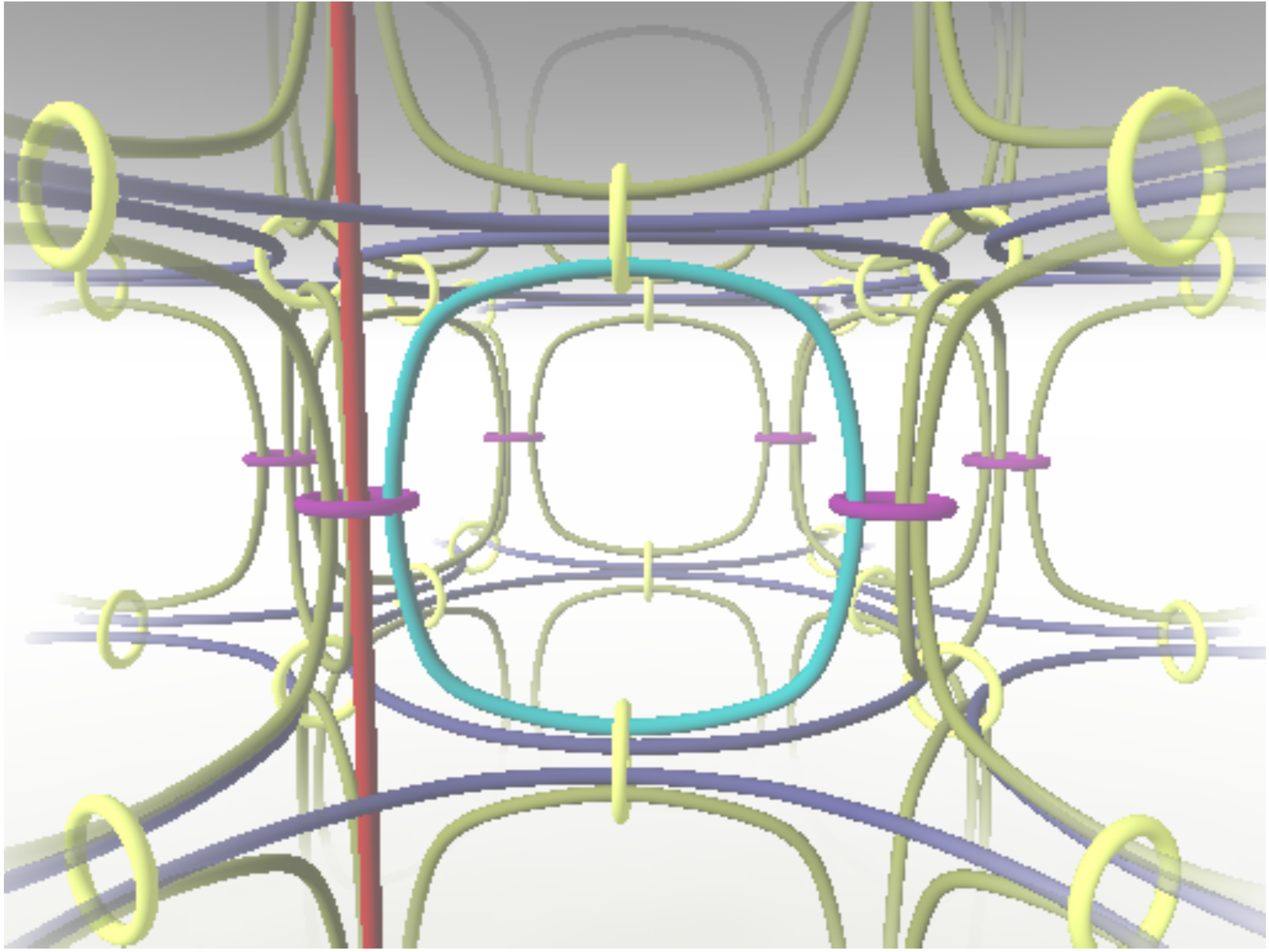}
  &
    \includegraphics[width=2.75in]{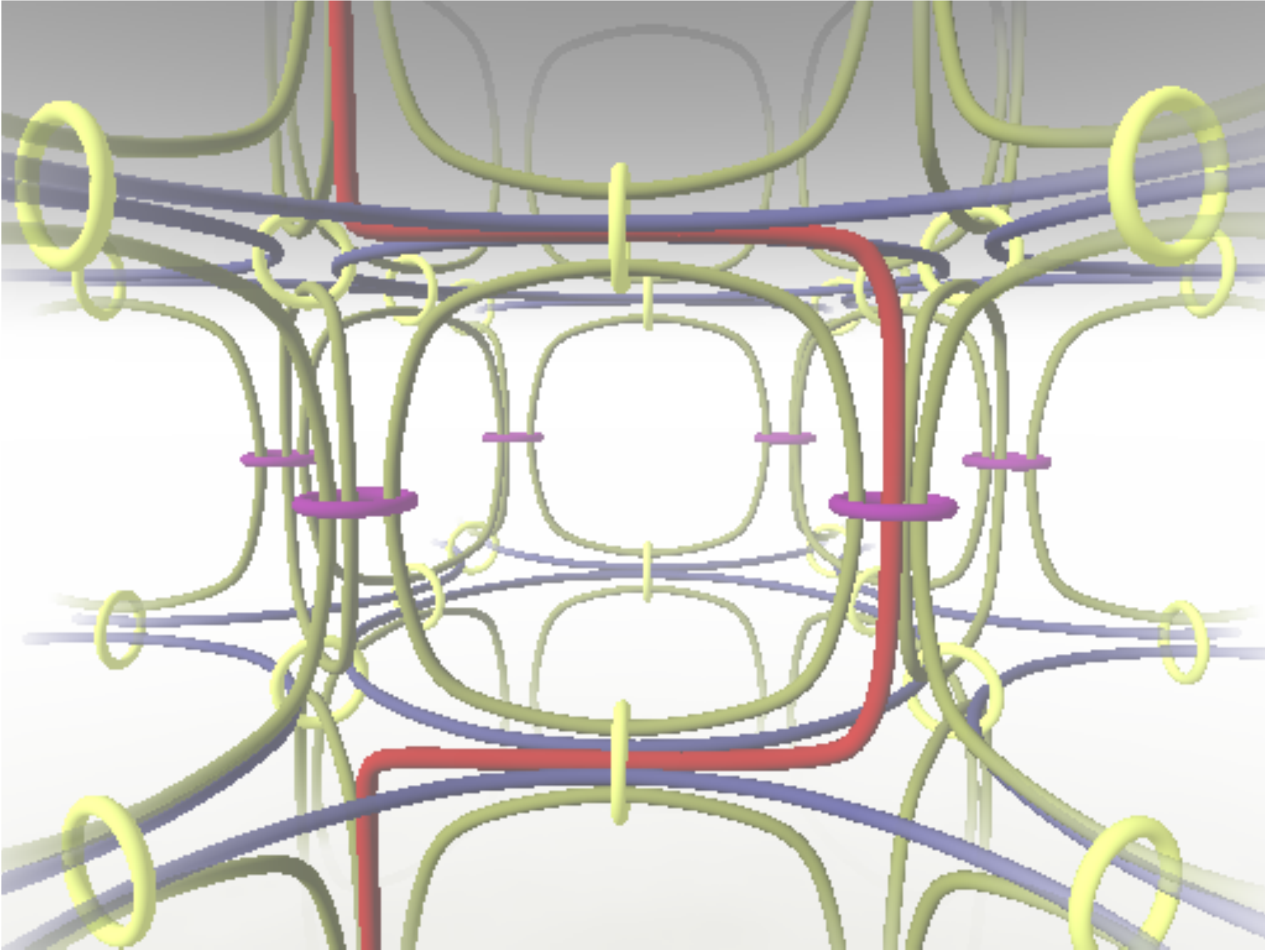}
  \end{tabular}
} \hspace{1cm}
\subfigure[] {
  \begin{tabular}{c c}
     \includegraphics[width=2.75in]{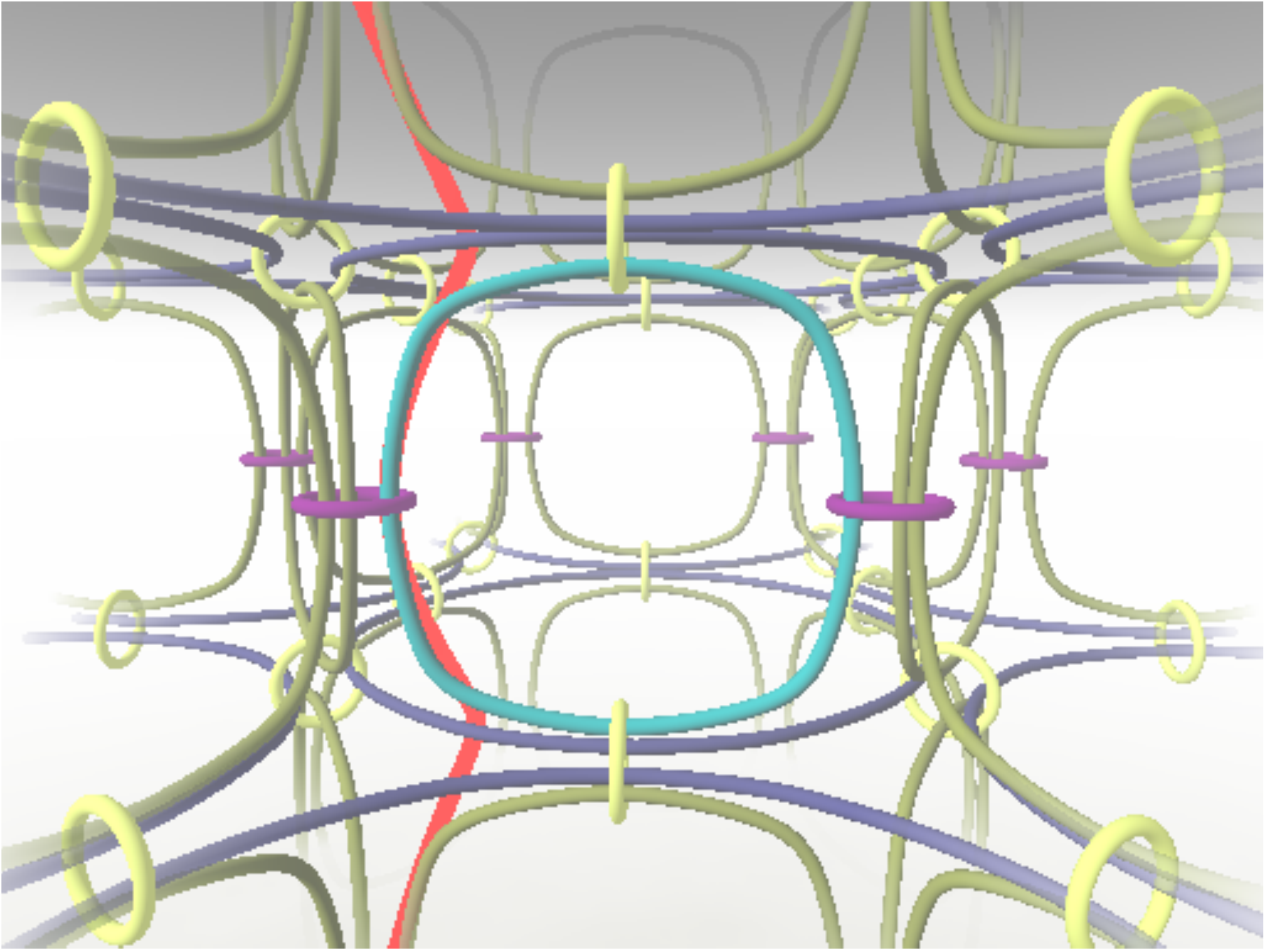}
  &
     \includegraphics[width=2.75in]{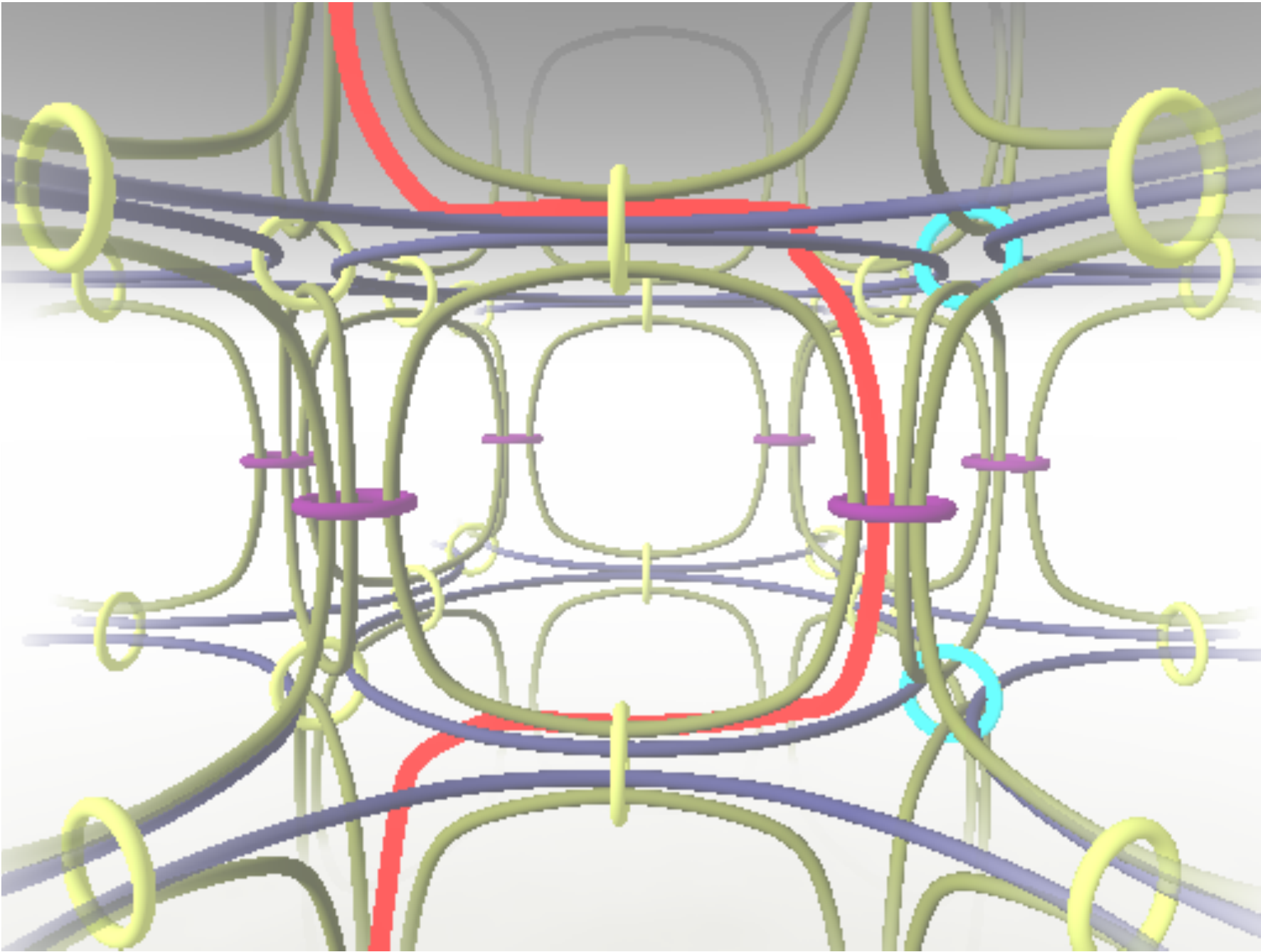}
  \end{tabular}
}
\hspace{1cm}
\subfigure{
     \includegraphics[width=2.75in]{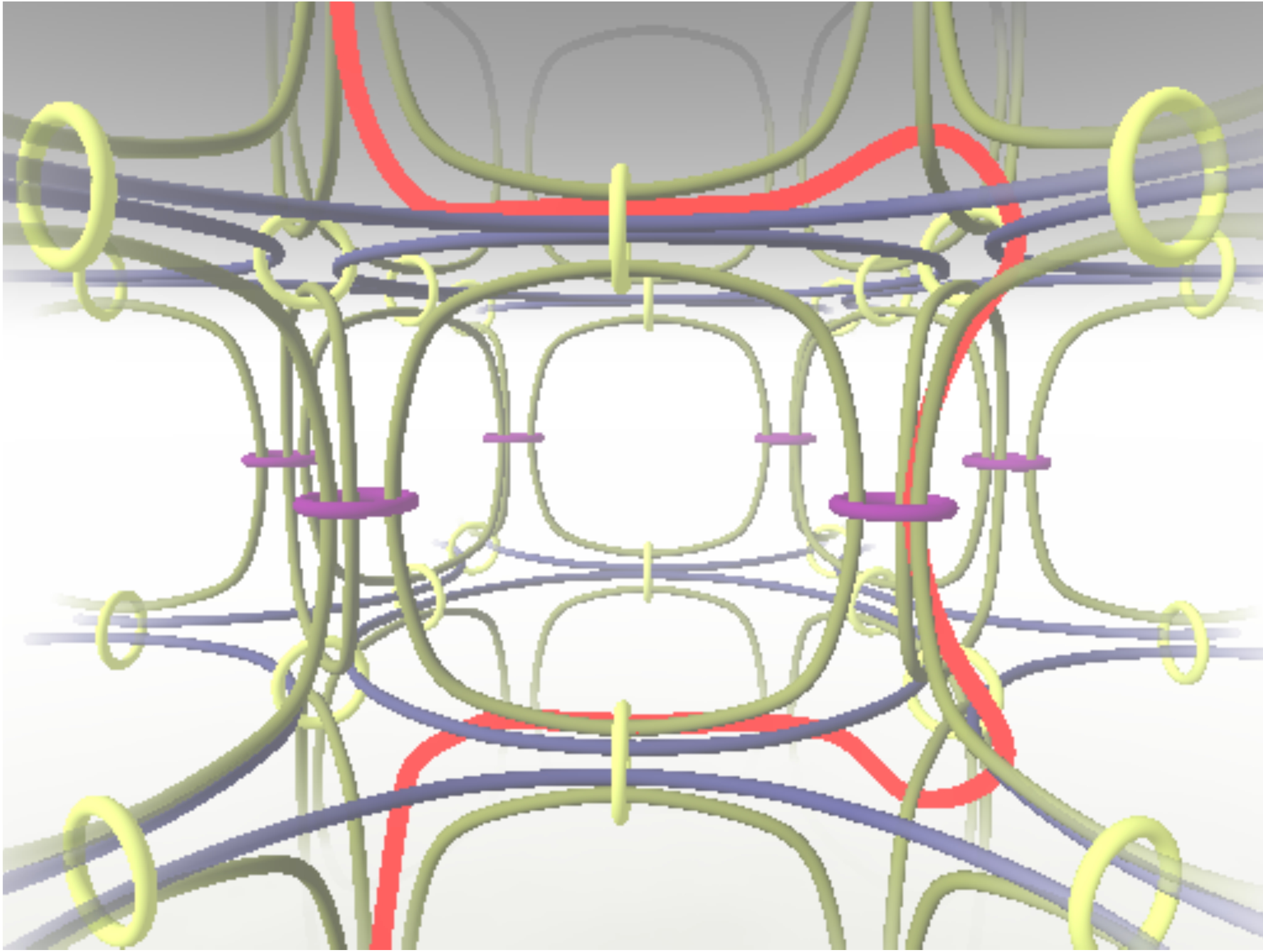}
} \hspace{1cm}
\caption{Handle-slide prescriptions for right- and left- handed
quasi-particles.  Plaquettes to be slid over are highlighted in turquoise.  (a) right- handed quasi-particles move around
the diagram by handle-sliding over plaquette loops.  (b)
left-handed quasi-particles slide first across a plaquette loop, and then
through edge loops, to
un-link from the plaquette.}
\label{QPSlide_1}
\end{figure}
 \end{center}
\end{widetext}

It is not hard to convince oneself that any deformation of the quasi-particle world-lines that does not change their linking with each other, or their winding around topological features of the space-time manifold, can be achieved by an appropriate series of handle-slides.  Hence, as promised, invariance of the value of the corresponding link evaluation under handle-slides guarantees that the partition function is completely independent of the local geometry of the quasi-particle strings.

\subsubsection{Statistics of quasi-particle strings}
\label{subsub:QPSTats}

We may now consider evaluating the partition function in the presence of quasi-particle world lines, with the goal of understanding their statistics.   We will find that 1) R-strings are ``right-handed", in the sense that they obey the same statistics as the original anyon theory; 2) L-strings are ``left-handed", having the statistics of the mirror anyon theory; and 3) R- and L- strings have trivial mutual statistics.    Further, if no quasi-particle world lines  encircle non-contractible loops of the space-time manifold, the partition function factors into separate ground-state and quasi-particle contributions; we will show this by un-linking the quasi-particle world-lines from the \chainmail link entirely. 

To understand the effect of adding quasi-particles to the theory without resorting to a brute-force evaluation of the partition function, we may use two powerful tools:  surgery and handleslides.  
 First, we may perform surgery on the \chainmail link in the presence of quasi-particles, and track the locus of the quasi-particle world-lines.  Second, we can use handle-slides to visualize more directly which quasi-particle world lines can pass through each other on the lattice, and which cannot.  Though less general than the first approach, the second is more straightforward, and will be the focus of this section.   The more technical surgery approach is discussed in Appendix \ref{SurgApp}. 

\begin{figure}[h!]
 \begin{center}
\subfigure[]{
    \includegraphics[width=3in]{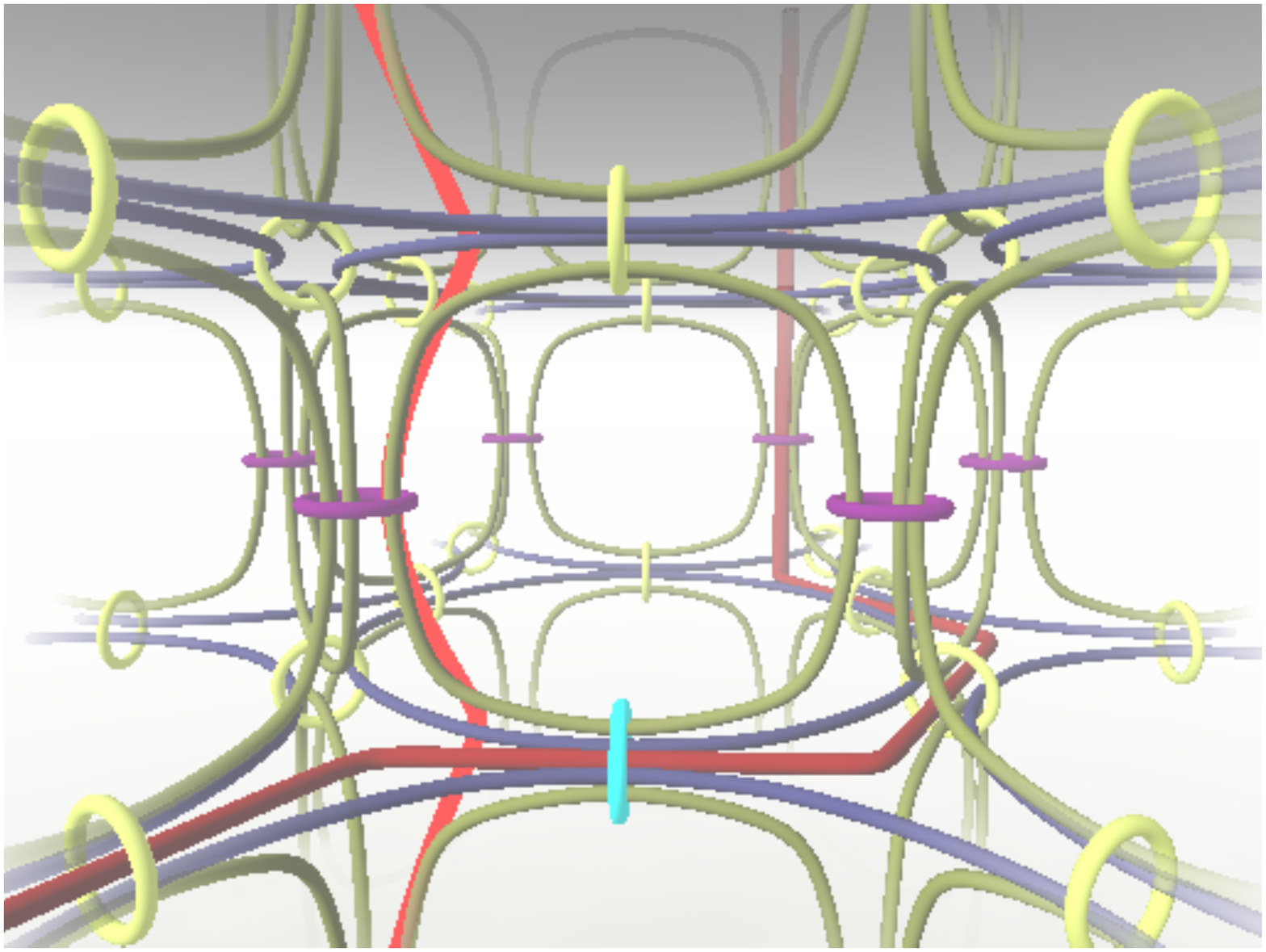}
} \hspace{1cm}
\subfigure[] {
  \begin{tabular}{c c}
     \includegraphics[width=2.2in]{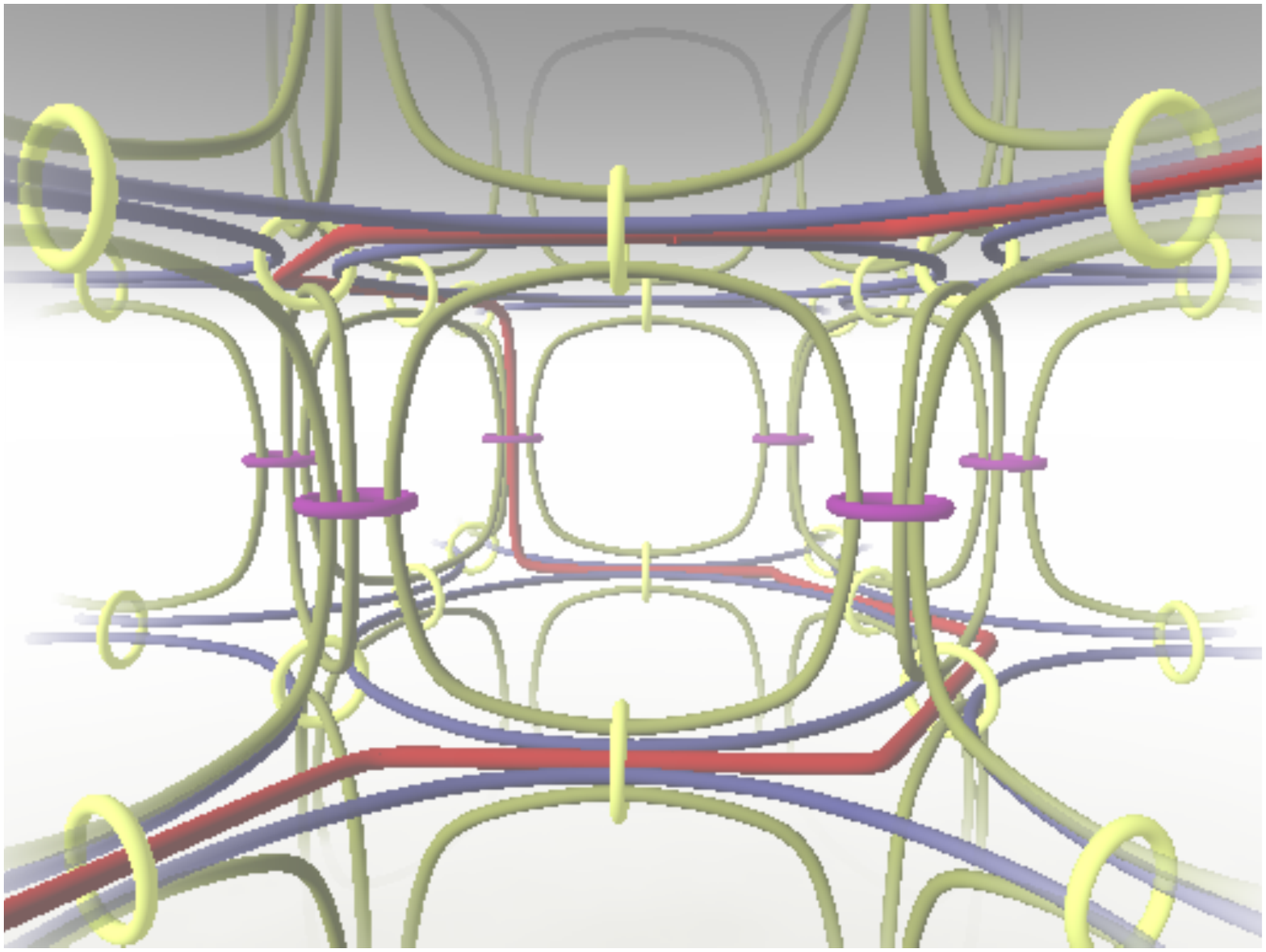}
  &
     \includegraphics[width=.75in]{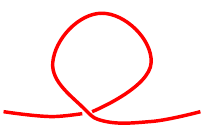} \end{tabular}
} \hspace{1cm}
\subfigure[] {
  \begin{tabular}{c c}
     \includegraphics[width=2.2in]{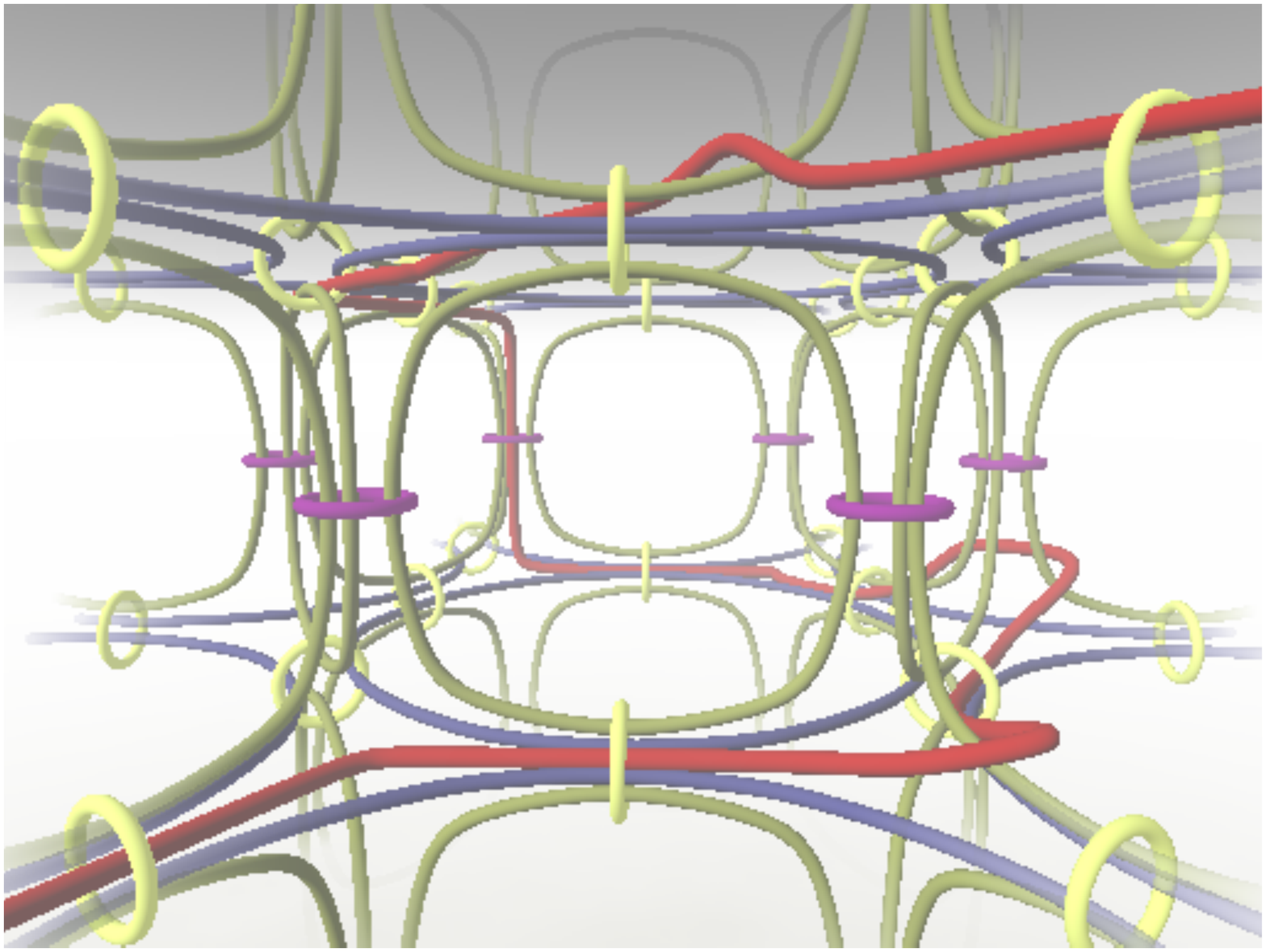}
  &
     \includegraphics[width=.75in]{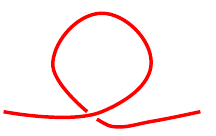}
  \end{tabular}
} \hspace{1cm}
\caption{Evaluating quasi-particle statistics using handle-slides.  Quasi-particle strings are shown in red; all other loops are labeled with $\Omega$.  (a) Passing an $L$ string through an $R$ string can be accomplished by handle-sliding the (vertical) $L$ string over a space-like edge projector (shown here in turquoise). (b) An $R$ string link in the lattice model, and the resulting link, seen here from above, after handle-sliding off all $\Omega$ loops.  (c) As in (b), but for an $L$ string.  Note that the direction of the crossing is reversed in the process of handle-sliding.}
\label{QPSlide_2}
 \end{center}
\end{figure}

First, we verify that there are two non-interacting sectors.
That is, any link of $R$ and $L$ strings can be reduced, via handle-slides, to separate $R$-particle and $L$-particle links.  To un-link $R$ and $L$ strings using handle-slides, we may slide the $L$ string over an edge projector through which the
$R$ string passes.  Since the $R$ string is not
linked to any of the adjacent plaquettes, the $L$ string
may now be `pulled through' without further affecting the
crossing, thereby unlinking the two strings (See Figure \ref{QPSlide_2}a).  This tells us immediately that the right and left handed particles form independent sectors of the resulting theory  (i.e, that the two sectors have trivial mutual statistics).

Careful inspection of the handle-slide rules in  Fig. \ref{QPSlide_1}
shows that strings of like chirality cannot be un-linked by
handle-slides.  For the R- strings, the reason is clear:
these strings slide only over plaquette loops, with which they are
never linked.  Hence the linking of R- strings with each other is never  altered by handle-slides.

For the L- strings, the situation is slightly more complicated: if two L- strings $s$ and $t$ cross, in attempting to slide $s$ past $t$ we must slide it over both an edge projector and a plaquette projector
with which $t$ is linked.  The strings appear un-linked after the
first slide -- if $t$ initially crossed over $s$, it now crosses
under -- but cannot be separated without re-linking, and restoring
the over-crossing of the initial configuration.   Hence two
left-handed strings also cannot be un-linked by handle-slides.

Thus by handle-sliding, we see that R- and L- strings constitute two sectors with mutually trivial statistics.  With this knowledge in hand, let us understand the statistics of the two sectors.
To do so, we will factor the partition function: that is, using handle-slides, we reduce the evaluation of the \chainmail invariant with quasi-particle strings (provided these do not encircle non-trivial topology in ${\cal M}$) to the separate evaluations of the \chainmail link (given by the \chainmail invaraint with no sources) and the evaluation of the link invariant of the quasi-particle world lines:
\be
Z(W_{C_1}, ...W_{C_N} ) = \langle L_{CH} \rangle \langle L_{QP} \rangle
\ee
where $W_{C_i}$ is the Wilson lines -- or more generally anyon world-lines -- of the external source on the curve $C_i$.   $L_{QP}$ denotes the link of these anyon world lines after they have been separated from the \chainmail link.   Once the partition function is factored, it is easy to evaluate the quasi-particle statistics.

To factor $Z$, we begin by using handle-slides to un-link all of the L- particles from all of the R- particles (since they can freely slide through each other).  We next perform a series of handle-slides to un-link all remaining world-lines from the $\Omega$ loops of the \chainmail link.  The result is illustrated in Fig. \ref{QPSlide_2}.
In the case of the R- quasi-particles, the resulting link, after being separated from the \chainmail scaffolding, has the same self-linking as when it was attached to the scaffolding (since as mentioned above, handle sliding never changes the self-linking of the R- particles).   A rather more involved argument (see Appendix \ref{HSApp}) shows that L-particles can also slide entirely off the scaffolding, but that in the process all over-crossings become under-crossings, and vice versa --- giving the mirror image of the original world-line link.
Hence by examining the world line links after they have been separated from the \chainmail link in this way, we conclude that the statistics of the R-particles are precisely given by the statistics of the particles from the consituent anyon theory, and those of the L-particles, by its mirror image.

Hence handle-sliding provides a convenient visualization of quasi-particle statistics.  However, if the quasi-particle world-lines do encircle non-contractible curves in the spacetime (for example, if they wind around periodic boundaries in space, should our manifold be the torus cross time), one must resort to the more rigorous approach of tracking the position of each type of quasi-particle world-line after surgery.
Recall that surgery on the \chainmail link gives two connected copies of the original space-time, with opposite orientations.  (That is, surgery on the \chainmail link in $\cal{M}$ produces $\cal{M} \# \cal{\overline{ M}}$).  As show in Appendix \ref{SurgApp}, we find that after surgery, R strings land in the right-handed copy of the manifold, and L strings land in the left-handed copy.  This gives a more general proof that the two types of quasi-particle world lines must have opposite chirality.

In summary, for doubled anyon theories, we find that violating vertex projectors or the combination of vertex and plaquette projectors creates independent right- and left- handed quasi-particle sectors.  By handle-sliding over the \chainmail link which describes the ground-state partition function, we see that the picture is invariant under geometrical deformations of the world-lines, and that the right and left handed sectors have precisely the statistics that we expect of the doubled theory.  Mathematically speaking, this follows from the fact that the two quasi-particle types land in opposite handed copies of the manifold after surgery.

\section{Alternative formulations of the Hamiltonian} \label{NTriSect}

From the point of view of topological quantum computation, or more
generally condensed matter physics, the most pressing question is
how to design experimentally realizable systems which are likely
to exhibit topological phases.  One significant obstruction to
doing this in the Levin-Wen models is the apparent complexity of
the interactions.  The plaquette projectors act on all 6 edges of
a plaquette in the honeycomb lattice, and the result in general
also depends on the states of the 6 external legs entering the
vertices at this plaquette.  An obvious question, then, is whether
the pictorial description allows us to re-express the model
Hamiltonians in a simpler form.

\subsection{Non-trivalent geometries}

One is tempted, at first, to try to utilize the fact that the correspondence to the \chainmail invariant guarantees that the partition function is independent of the lattice chosen.  The Hamiltonian will still be comprised of (local) vertex projectors, and plaquette projectors.  However, one might hope that  plaquettes with fewer edges might lead to simpler, more local, plaquette projectors.

There are in fact two complications here.  First, we must understand what happens to ${B}_V$ when the valence of a vertex is changed.  The vertex projector can always be implemented as a single operator, enforcing the constraint that the net flux entering the vertex is $0$.  From this perspective, the vertex projector is equally physical for any valence of vertex --- in the \chainmail picture we simply wrap an $\Omega$ around all the strings entering that vertex.
However, since the rules of the fusion category specify only how to fuse three strings at a point, to evaluate the link invariant, one ultimately must evaluate such a projector in terms of string configurations with only trivalent vertices\cite{LW,Roberts}. In other words, greater than three strings can be fused successively to determine their joint quantum number.   However, the intermediate quantum numbers may take multiple values and these values must be summed over after the projection.     As noted Sect.~\ref{sub:categories}, such additional quantum numbers (multiple fusion channels) associated with vertices are actually something that can generically occur even at trivalent vertices; for simplicity we have so far chosen not to consider this case.   However, when we consider vertices with valence greater than three, additional quantum numbers at vertices always occur, except for in a very few trivial (Abelian) theories.  Thus, while the higher valence case is in principle similar, in practice it becomes more complicated.

The second complication in attempting geometrical simplification is with the plaquettes themselves.  The plaquette projector always acts on all edges in a plaquette, and is sensitive to the state of all external legs.  In general, if the valency of each vertex is $v$ and there are $n_V$ vertices per plaquette, the final state on the $n_V$ edges in the plaquette depends on the initial state of all $n_V (v-1)$ edges which enter the vertices on this plaquette.  Thus, by reducing $n_V$, we decrease the number of states which ${B}_P$ alters -- but the total number of edges involved in constructing ${B}_P$ is still $n_V(v-1)$.  It is thus not obvious how to construct a $2$D lattice for which ${B}_P$ appears significantly more local, in this sense: for example, on the hexagonal, square and triangular lattices, we have $n_V(v-1) = 12$, $12$, and $15$, respectively.  One may also consider more complicated (or even irregular) lattice
geometries.  For example, the so-called Cairo-pentagon lattice tiling
and the prismatic pentagon lattice tiling both share the feature with
the honeycomb and square lattices that a LW model based  on these
lattices would couple only 12 edges at a time. 

\subsection{Duality transformations}
\label{sub:duality}

 A more promising approach to render the Hamiltonian more plausible is to attempt to construct models in which it is natural to impose the constraint of $0$ flux through edges on both the lattice and the dual lattice.  As is apparent in the \chainmail construction, for doubled anyon theories the plaquette projectors serve merely to implement the constraint of $0$ flux entering each vertex of the dual lattice; hence these projectors are perfectly local in the dual lattice description.
This yields a picture quite similar to that proposed by Ref.~\onlinecite{FendleyTopological}, in which a quantum loop gas model was constructed with a Hamiltonian consisting of two sets of vertex projectors -- one  on the lattice, and one dual lattice.  (The difference here is that the mapping between lattice and dual-lattice projectors involves all $n_V(v-1)$ edges required to construct ${B}_P$; the construction of Ref.~\onlinecite{FendleyTopological} is slightly different and renders the final result somewhat simpler.)

The seeming non-locality of ${B}_P$, then, can be seen as induced by the change of basis from the dual lattice picture, in which ${B}_P$ is diagonal in the edge variables, to the lattice picture (in which it is ${B}_V$ that is diagonal).  The most obvious way to leverage this fact is to construct a model in which ${B}_V =0$ is a constraint on the Hilbert space in the absence of quasi-particle sources; in this case the Hamiltonian consists of a single, maximally local, term.  This would essentially be a generalization of Yang-Mills theory in the strongly deconfined limit: in the lattice version of this theory, ${B}_V=0$ is the constraint that the electric field be divergenceless; ${B}_P$ gives the lattice action for the magnetic field.  In the deconfined limit the coupling of the $E^2$ term is $0$, and the Hamiltonian may be expressed in the magnetic basis as a term acting at a single vertex on the dual lattice.  Conceptually at least, this picture is quite natural; indeed, one way to motivate the Levin-Wen models is as generalizations of lattice gauge theories\cite{LW}.

\section{Conclusions} \label{Conclusions}

We have seen how, for Levin-Wen models constructed from modular tensor categories, the relationship between the partition function -- {\it both for ground state and excited state sectors} -- and the \chainmail invariant (decorated with quasi-particle world lines, where appropriate) allows us to connect these models to doubled Chern-Simons theory, and more generally to doubled anyon theories.  In doing so, we uncovered several interesting facts about the physical models.  First, the particular combination of terms in the lattice Hamiltonian of Ref.~\onlinecite{LW} can be understood as arising from a link (the `picture' of the model's partition function) which in fact corresponds after surgery to a right- and a left- handed copy of the original space-time manifold.  There are therefore two non-interacting sectors in the excitation spectrum -- one associated with each copy -- and the resulting model is achiral.  When we translate these excitations back into the more familiar language of vertex (electric) and plaquette (magnetic) violations, we find that electric excitations are chiral, while magnetic excitations are not.  Further, when the quasi-particle strings do not enclose non-contractible curves, we can easily evaluate the topological part of the partition function, as it factors into ground-state partition function and a piece that represents the quasi-particle excitations only.

We would also like to point out that in the context of the \chainmail and 
Turaev-Viro invariants, the construction of the
left-handed particle is not, to the best of our knowledge,
discussed in the existing literature-- though previous
authors have considered introducing right-handed \cite{Martins1}
and achiral \cite{Martins2} quasi-particle strings.  In this sense
our discussion completes the connection between the work of
Roberts \cite{RobertsThesis}, which introduces the \chainmail
invariant for the ground state, and that of Witten
\cite{WittenJones}, and Reshetikhin and Turaev \cite{RT}, which
describes both the ground state partition function and
quasi-particles in the un-doubled theory. By constructing both
left- and right- handed quasi-particles, we are able to describe
the full doubled Chern-Simons theory by adding lines of gauge flux
to the \chainmail invariant.

Besides making the above connections between Levin-Wen models, Chern-Simons theory, and the \chainmail (or equivalently, Turaev-Viro) invariant, our work opens several interesting new directions.  From the point of view of physics, the main challenge is to utilize the flexibility inherent in the \chainmail construction to connect these rather abstract model Hamiltonians to more physically motivated systems.  In a future publication, we will also address the question of whether the above construction can be replicated on $3$D lattices to generate a topologically non-trivial theory.

Another interesting direction is whether this construction can be usefully extended beyond the case of doubled anyon theories.  Indeed, the lattice models described by Ref.~\onlinecite{LW} encompass topological phases which are {\it not} described by doubled anyon theories, and hence do not fit into the framework outlined in this work.  (For example, perhaps the simplest topological theory that can be constructed by the Levin-Wen prescription, the Toric-code\cite{KitaevToric}, is not a doubled anyon theory.)   By replacing the encircling edge loops with generalized projectors, the ground state partition function can be constructed as described in Sect.~\ref{LWPict} without requiring any information about braiding\cite{ThanksParsa}.    However, in this case the formulation is far less instructive, as we know of no analogue of surgery in this case to relate these models to a continuum description.

The ultimate goal of understanding how to engineer a topological qubit remains beyond the scope of our present understanding.  Nevertheless the search for topological phases has inspired a new approach to finding topological theories in physics -- namely, through the construction of topological lattice models.  The deep connections between these lattice constructions, combinatorial topological invariants as they are studied in mathematics, and the topological field theories familiar from high-energy physics, reveal an elegant unity between these seemingly disparate approaches which, we hope, unveils interesting questions in all three areas.

\begin{appendix}

\section{Fusion coefficients in the ground state partition function} \label{FusionApp}

Here we will track the details of the fusion coefficients to prove that our partition function is exactly equivalent to that of the corresponding Levin-Wen Hamiltonian.

For many of the results derived here, the following symmetry relations of the $F$-symbols will prove useful\cite{LW}:
\be \label{Eq_LWFsyms}
F^{ijm}_{kln} = F^{lkm^*}_{jin} = F^{jim}_{lkn^*} = F^{klm^*}_{ijn^*}
\ee

We begin by adding the vertex projectors $B_V$, which fuse the $3$ strings $i_t, j_t$ and $e_t$ entering each vertical
edge at a trivalent vertex.  That is, we first fuse $i_t$ and $j_t$ to give $k_t$, pushing one of the two ensuing trivalent vertices through the projector loop.  Next we fuse $k_t$, with $e_t$, again pushing one of the two trivalent vertices through the projector.  The line through the projector is now the fusion product of $k_t$ with $e_t$, and must be the identity string.  Hence we conclude that $k_t= e^*_t$, and 
the fusion incurs a factor of
 \be
\frac{1}{ \Delta_{e_t}} F^{j_t^* j_t 0}_{i_t i^*_t  e_t^*}
\ee
at each time step.

\begin{center}
\begin{figure}[h!]
 \includegraphics{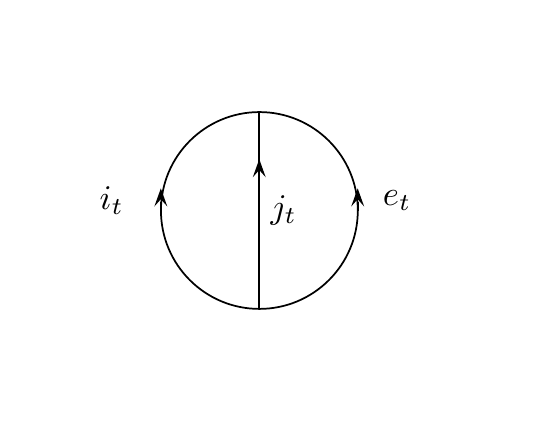}
\caption{ \label{Thetas} }
\end{figure}
\end{center}

\begin{center}
\begin{figure}[h!]
 \includegraphics[width=3in]{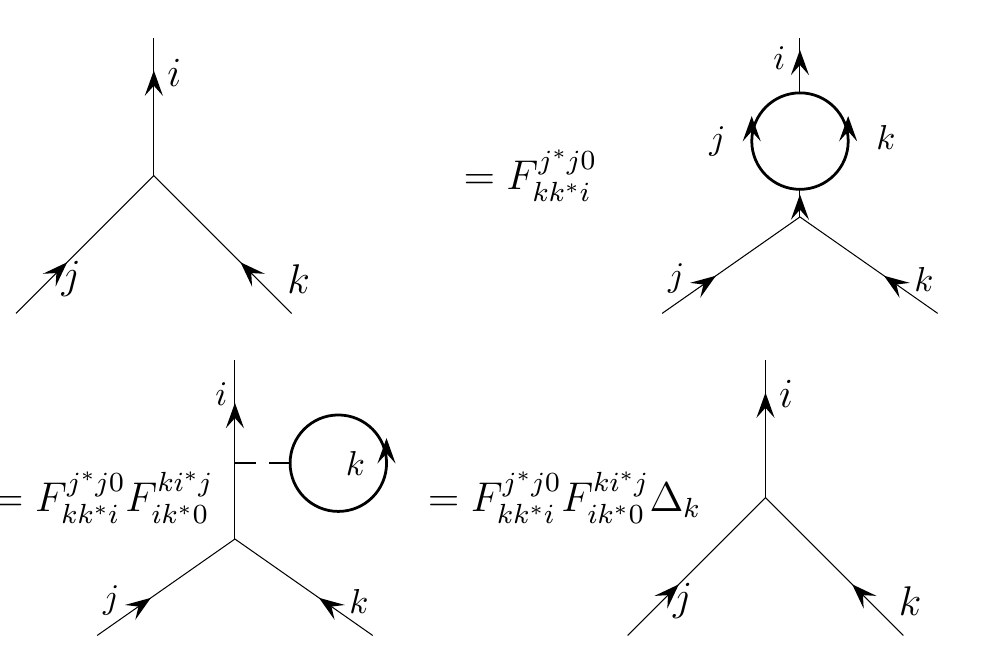}
\caption{ \label{FuseId} Graphical justification for the identity $F^{e_t i_t j_t}_{i^*_t e_t^* 0}  F^{j_t^* j_t 0}_{i_t i^*_t  e_t^*} \Delta_{i_t}  =1$.}
\end{figure}
\end{center}

If there are no space-like plaquette strings present, applying projectors along the space-like edge labeled $i$ results in a bivalent vertex between $i_t$ and $i_{t+
\delta t}$, etc, forcing $i_t = i_{t+ \delta_t} \equiv i$ along each edge.  The coefficient for this fusion process is $F ^{i^*_t i_t 0}_{i^*_t i_t 0} = \frac{1}{\Delta_{i_t}}$, which precisely cancels  the factor of $\Delta_{i_t}$ included in the partition function.  
Applying the vertex projector in the next time slice results in the diagram shown in Fig. \ref{Thetas}.
This diagram evaluates to
$ F^{e_t i_t j_t}_{i^*_t e^*_t 0} \Delta_{e_t} \Delta_{i_t} $.   Combining the two factors (fusion below the
vertex due to the projector, and evaluation of the diagram) gives:
\be
F^{e_t i_t j_t}_{i^*_t e_t^* 0}  F^{j_t^* j_t 0}_{i_t i^*_t  e_t^*} \Delta_{i_t} 
\ee
which is in fact unity, (as justified in Fig. \ref{FuseId}) 
as expected.
This holds in general: acting with $B_V$ at a vertex to fuse the $3$ strings together will induce a coefficient which is exactly cancelled by the coefficient obtained by evaluating the closed diagram obtained after imposing $B_V$ both above and below the vertex at the preceding time slice.

Next, we must add the plaquette projectors.  To
see that these gives the correct form for ${\mathcal{B}}_P(s)$, we will show that the net effect of adding a plaqutte string $s$ to the picture is to induce a factor of $F^{e  i_t j^*_t}_{s  j^*_{t + \delta t} i_{t+\delta t} }$ at each vertex, and flip the labels from $i_t, j_t$ to $i_{t+\delta t}, j_{t + \delta t}$.  Again,  at each vertex $V$
the strings $i_t, j_{t}$, and $e_t$  are fused at a
trivalent vertex by the  projector on the time-like edge emanating from $V$.  
As  illustrated in Fig. \ref{CHF_3}, the space-like edge projectors  around the plaquette now fuse the string $s$ to the strings $i_{t}$ and $i_{t+ \delta t}$ along each edge, to produce the diagram of Fig. \ref{Fig7d} at each vertex (with $i_t \equiv i_1(t), j_t \equiv i_2(t), e_t \equiv e_2(t)$).  This fusion incurs a factor of $\frac{1}{\Delta_{i_{t+ \delta t} } } F^{i_t^* i_t 0}_{s s^* i_{t+ \delta t} }$ along each edge.  The factor of $\frac{1}{\Delta_{i_{t+ \delta t} } }$ cancels the factor of $\Delta_{i_{t+ \delta t} }$ which we have explicitly included in the partition function.  
Applying the fusion rules in Fig. \ref{LW_3}, we may collapse the bubble at each vertex, leaving a trivalent vertex between the $3$ strings $i_{t + \delta t}, j_{t + \delta t}$, and $e_t$.  Including the factors incurred by fusing $s$ into the edges, this diagram comes with a coefficient:
\be
\label{VertEval}
 \left( \Delta_s  F^{s i^*_{t+\delta t} i_t}_{i_{t+\delta t} s^* 0} F^{i_t^* i_t 0}_{s s^* i_{t+ \delta t} } \right ) F^{e_t i_{t} j^*_ t}_{s j^*_{t+\delta t} i_ {t + \delta t} } 
\ee
where we have used the symmetry relation of Eq. (\ref{Eq_LWFsyms})
\be
F^{s j^*_{t + \delta t} j_t }_{e_t i_t i^*_{t + \delta t} } =    F^{e_t i_{t} j^*_ t}_{s j^*_{t+\delta t} i_ {t + \delta t} } \ \ \ .
\ee
 The term in parentheses is unity, due to the relation shown in Fig. \ref{FuseId}.  As noted above, the coefficients due to fusion along the time-like edges will always give unity once all vertex projectors are imposed.  Hence, as promised, the effect of the plaquette string is to induce a coefficient $F^{e_t i_{t} j^*_ t}_{s j^*_{t+\delta t} i_ {t + \delta t} } $ at each vertex, and interchange the labels $i_t, j_t$ with the labels $i_{t+ \delta t} j_{t + \delta t}$.  
After imposing the space-like edge projectors, then, the plaquette string $s$ acts by fusion on the labels around a plaquette to give exactly the product of $F$-matrices  used to define ${\mathcal{B}}_P(s)$.

\subsubsection{Multiple strings and quasi-particles}

In fact, it is easy to generalize this to the case of several extra strings acting on the vertex.  Again, we may apply all edge projectors except the one above the vertex, to obtain a diagram with $3$ external legs.  These digarams can always be reduced to the $3$-vertex by a series of applications of the identity shown in Fig. \ref{LW_3} d and, for quasi-particle strings, the un-twisting move of Fig. \ref{RFig} a.  The extra $\Delta$ coefficients cancel, as do the $F$ factors for fusing the strings together, such that the effect of passing this extra string through is always to multiply by a factor of $F$ (or $R F$ for quasi-particle strings which cross the edge variables).  In general these diagrams will contain internal edges, whose labels correspond to those neither of the initial variables $i_t, j_t, e_t$ nor the final variables $i_{t+\delta t}, j_{t+\delta t}, e_{t+\delta t}$.  These can be consistently assigned labels $i_{t_1} .. i_{t_n}$, etc. -- in other words, we may equally well subdivide the action of all string operators at a vertex such that each operates at a different intermediate time step $t_k$.

\begin{center}
\begin{figure}[h!]
 \includegraphics{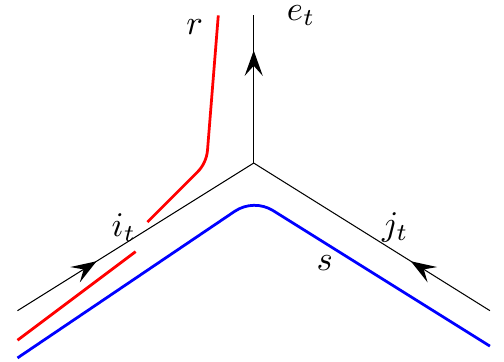}
\caption{ \label{FuseApFig3} }
\end{figure}
\end{center}

To clarify the above description, let us consider adding a single quasi-particle string $r$ to the vertex $i_t, j_t, e_t$, as shown in Fig. \ref{FuseApFig3}.  

The digram shows the $3$ strings $i_t, j_t, e_t$, which have been fused to a trivalent vertex by the application of $B_V$ at time $t$, as well as a plaquette string $s$ and the quasi-particle string $r$.  Note that as $r$ crosses under $i_t$, it represents an L-particle; R particle strings at time $t$ cross over $i_t, j_t$, and $e_t$ (but under $i_{t+\delta t}, j_{t+\delta t},$ and $e_{t+\delta t}$).

\begin{center}
\begin{figure}[h!]
 \includegraphics{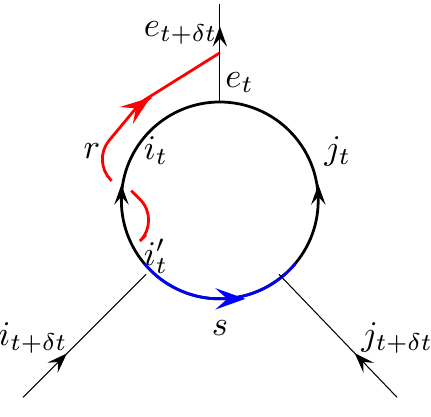}
\caption{ \label{FuseApFig4} }
\end{figure}
\end{center}

To evaluate this diagram, we first fuse $r$ to the edges $i_t$ and $e_t$ which it traverses.  Next, fuse the string $s$ associated with ${B}_{P}(s)$ to the edges $i_t, j_t$ on which it acts.  Finally, apply the in-plane edge projectors, which fuse the $3$ strings $i_t, j_t, e_t$ shown above to $i_{t + \delta t}, j_{t + \delta t}, e_{t + \delta t}$.  This results in the diagram show in Fig. \ref{FuseApFig4}, with the coefficient:
\be
F^{r^* r 0}_{e_t e^*_t e_{t+\delta t} } F^{i^*_t i_t 0}_{r r^* i'_t} F^{i'^*_t i'_t 0}_{s s^* i_{t+ \delta t} }  F^{j_t j^*_t 0}_{s s^* j^*_{t+ \delta t} } \ \ \ .
\ee

The arrows on $e_t$ and $i'_t$ are taken to have the same orientation as those on $e_{t+ \delta t}$ and $i_t$, respectively.

The diagram is evaluated by first un-twisting $r$ and $i_t $ using the identity shown in Fig. \ref{RFig}a, and then using the identity of Fig. \ref{LW_3} d twice to collapse the resulting diagram to a trivalent vertex.  The result is a trivalent vertex $i_{t+\delta t}, j_{t+\delta t}, e_{t+\delta t}$ multiplied by the coefficient 
\ba
\left ( \Delta_{r} F^{r^* e_{t+\delta t} e^*_t}_{e^*_{t+\delta t} r 0} F^{r^* r 0}_{e_t e^*_t e_{t+\delta t} } \right ) 
 \left ( \Delta_s F^{s i^*_{t+\delta t} i'_t}_{i_{t+\delta t} s^* 0}  F^{i'^*_t i'_t 0}_{s s^* i_{t+ \delta t} } \right ) \n
\left( F^{i^*_t i_t 0}_{r r^* i'_t}   F^{j_t j^*_t 0}_{s s^* j^*_{t+ \delta t} }  \right )
R^{r i_t}_{i'_t}  F^{r^* i'_t i^*_t} _{j_t e^*_t e_{t + \delta t} }
 F^{s j_{t + \delta t} j^*_t} _{e^*_{t+ \delta t} i'_t i^*_{t + \delta t} } 
\ea
The factors on the first line give unity (see Fig. \ref{FuseId}), once we account for the fact that the symmetries of the $F$ symbols guarantee that
\ba
 F^{r^* r 0}_{e_t e^*_t e_{t+\delta t} } = F^{e_t e^*_t 0}_{r^* r e_{t+\delta t}^* } \n
 F^{s i^*_{t+\delta t} i'_t}_{i_{t+\delta t} s^* 0}  = F^{ i^*_{t+\delta t} s i'_t}_{s^* i_{t+\delta t}  0}  \ \ \ .
 \ea
 Similarly, the two $F$ factors in parenthesis in the second line will cancel corresponding factors at the adjacent vertices traversed by the strings $r$ and $s$, respectively.  This leaves a net factor of:
 \be
 R^{r i_t}_{i'_t}  F^{r^* i'_t i^*_t} _{j_t e^*_t e_{t + \delta t} }
 F^{s j_{t + \delta t} j^*_t} _{e^*_{t+ \delta t} i'_t i^*_{t + \delta t} }  = R^{r i_t}_{i'_t}  F^{j_t e^*_t  i_t} _{r^* i'_t e^*_{t + \delta t} }
 F^{e^*_{t+ \delta t} i'_t  j_t} _{s j_{t + \delta t} i_{t + \delta t} } 
 \ee
 at each vertex (where we have again used symmetries of the $F$ symbols to obtain the equality).  
  Taking $\omega^{i j}_s \equiv R^{is}_j$ for R-particles, and $\omega^{ij}_s =\left(R^{is}_j \right )^* = R^{si}_j$ for L-particles (which cross over, rather than under, the edge $i_t$), we see that the net effect is equivalent to first applying the quasi-particle string operator (c.f. Eq. (\ref{LWSOps})) running from edge $i_t$ to edge $e_t$, which gives  the matrix element $\omega^{i_t i'_t}_r F^{j_t e^*_t  i_t} _{r^* i'_t e^*_{t + \delta t} }$, and then applying $\mac{B}_P(s)$ to the resulting state, giving the matrix element $F^{e^*_{t+ \delta t} i'_t  j_t} _{s j_{t + \delta t} i_{t + \delta t} } $.   (Note that the orientations of the edge labels $j$ and $e$ in Fig. \ref{FuseApFig4} differ from those of Fig. \ref{LWF1c}, resulting in different labels appearing in their conjugate representations).
Hence adding quasi-particle strings to the chain-mail link has precisely the same effect on the partition function as acting with the quasi-particle string operator of Eq. (\ref{LWSOps}).

The derivation here can be extended inductively to show that any number of strings can be added to the diagram, and will contribute to the partition function a product of string operators.  One useful consequence of this is that the partition function does not differentiate between adding all plaquette projector strings in the same time slice, and adding each at a separate time.

\section{Handle-slides of right- and left- handed quasi-particles} \label{HSApp}

Here we describe in more detail the handle-slide argument for the statistics of the quasi-particles.  As explained in the main text, R-particle strings that do not encircle non-contractible curves in the spacetime can be detached from the scaffolding by a a series of handle-slides over plaquette loops only.  Since they are not linked with these, it follows that the link has the same chirality after removal from the scaffoding as it did when threaded through the \chainmail link.  Hence the statistics of R particles are exactly those of the original theory.

For the L particles, however, the situation is more complicated.  To separate any section of a link from the \chainmail requires a series of handle-slides which necessarily include slides over both plaquette and edge $\Omega$-loops.  The handle-slide prescription described in Sect.~\ref{Handleslide} ensures that, where possible, these slides occur only over $\Omega$-loops from which the string has already been un-linked.  Hence,  we first slide over an edge loop to un-link with a particular plaquette, and only then handle-slide over the plaquette.  However, if the L string is non-trivially knotted, it is not possible to slide it off the lattice without handle-sliding over $\Omega$ loops with which the string is linked.  This changes the self-linking of the L string, and ultimately as we shall see the chirality of the knot.  More generally, if two L strings are linked, they cannot be separated from the \chainmail link using handle-slides without handle-sliding one string over $\Omega$ loops with which the other is linked, thereby changing the chirality of their linking.

It is helpful to consider the case of a single crossing of $L$ strings.  In order to bring this crossing `off' the scaffolding, we must first perform handle-slides until both strands are in the same horizontal slice.  This process always involves handle-sliding one strand over a plaquette loop with which the other is linked, changing the crossing from an over-crossing to an under-crossing.  As pointed out in the text, if we wish to continue separating the two strands in this direction, additional slides over edge loops must be performed which re-link the two strands with the same chirality as before, so that the two L stands have non-trivial mutual statistics.
However, if we merely wish to slide the L-string link off the \chainmail link, then we stop at this intermediate stage.

To be more precise, we may slide the L-string off the \chainmail link by first performing handle-slides so that the entirety of the link sits in a single spatial plane $\mathcal{S}$, so that its strands are linked only through plaquettes, and through edge projectors in this plane.  This can be done in such a way that the only plaquettes with which the string is linked are time-like plaquettes above $\mathcal{S}$.  Once the link has been positioned in this way, the handle-slides required to detach it from the \chainmail involve sliding over edge projectors sitting above $\mathcal{S}$, and plaquette loops which sit in $\mathcal{S}$.  As the world-lines comprising the link are unlinked from both of these types of loop, once the link has been maneuvered into the plane in this way it slides off the lattice without further alteration to its linking structure.

But, as noted above, when sliding the two strands at a crossing such that the crossing lies in, and linked only to time-like plaquettes above, a single spatial plane, all over-crossings and under-crossings are reversed.  Hence the link, when separated from the scaffolding, has the opposite chirality as it does when drawn as quasi-particle world lines in the lattice model.

\section{Surgery considerations}

Here we review in more detail the effect of surgery on the \chainmail link itself, and track the location of the two types of quasi-particle world-lines.  Our ultimate goal is to provide an alternative, more mathematically satisfying explanation of why right- and left- handed quasi-particle strings are in fact right- and left- handed.  To achieve this, however, requires a digression into the process of surgery itself.

\subsection{Surgery: a brief description}
Let us start with a brief review of the concept of surgery.   Dehn surgery is a prescription for constructing any closed 3-manifold from a reference closed 3-manifold  (usually taken to be the 3-sphere $S^3$) and a (framed) link within that manifold.    The prescription for performing surgery on a link is as follows:   Consider a single strand of the link.  Topologically, this strand is a circle $S^1$, although it may be embedded in the manifold nontrivially -- i.e, it may be knotted with itself or with other strands of the link.   Now thicken this strand into a solid torus, $S^1 \times D^2$ with $S^1$ being the strand of the link, and $D^2$ being a small disk cross section where we thickened the strand.  (If the strand is knotted or linked with other strands, the torus is non-trivially embedded in $S^3$).   Now cut this solid torus from the manifold, leaving a boundary that is the surface of the solid torus, $T^2 = S^1 \times S^1$ where the first $S^1$ is the original strand of the link, and the second $S^1$ is the boundary of the thickening disk $D^2$.   (Again this may be nontrivially embedded but topologically this $T^2$ is just a standard torus surface).    To obtain a new manifold without  boundary, we can sew onto this resulting $T^2$ boundary any other manifold whose boundary is also $T^2$ such that the two boundaries meet and ``cancel" leaving a new manifold without boundary.  While there are many manifolds that might have such a $T^2$ surface, the simplest would just be a solid torus.   One trivial possibility is to sew back in the same torus $S^1 \times D^2$ which we removed in the first place --- in which case we get exactly the same manifold we started with.     We can think of this as ``filling in" the second $S^1$ of the torus surface $S^1 \times S^1$.    The next simplest possibility (and the one we will be interested in) is to instead sew back in a different solid torus $D^2 \times S^1$ which would be ``filling in" the {\em first} $S^1$ of the torus surface $S^1 \times S^1$. (It is hard to visualize such a thing since either the initial manifold or final manifold cannot be embedded in 3-dimensions).

\subsection{Surgery, Chern-Simons theory,  and knot polynomials} \label{SurgIntro}

For the interested reader, we will briefly describe how surgery leads to a correspondence between knot polynomials and the partition function of Chern-Simons theory, as first described by Ref.~\onlinecite{WittenJones}.   (We note that for a first reading this section may not be crucial for understanding Sect.~\ref{SurgApp} below).   There are three essential ingredients to this connection.  First, the Hilbert space at a fixed instant in time is finite-dimensional; its dimension is fixed by the topology of the space-like surface $\Sigma$ and the number of Wilson lines piercing this surface.  For example, if there are no Wilson lines piercing $\Sigma$, and $\Sigma$ has genus $0$, then the Hilbert space is one-dimensional.   When there are multiple Wilson lines piercing $\Sigma$ the Hilbert space is typically multidimensional, as should be expected for a topological (non-Abelian) system with quasi-particle defects.  Second, certain geometrical transformations on $\Sigma$ result in a change of basis in the Hilbert space.  Third, the effects of such transformations on the Hilbert space can equally well be carried out by adding certain Wilson lines in parts of the spacetime manifold which do not intersect $\Sigma$.  For example, if $\Sigma$ is a torus, then linear transformations in the Hilbert space on $\Sigma$ can be carried out by threading Wilson lines through the solid torus bounded by $\Sigma$.
These three crucial properties all have roots in the connection between Chern-Simons theory and rational conformal field theories.  We will not attempt to explain their origins here, but merely briefly describe their interesting consequences.

First, consider the effect of performing a modular transformation on the surface of the torus.  The effect of this modular transformation is to interchange the meridian and the longitude of the torus, which is carried out via the action of the modular $S$ matrix: $|a \rangle  \rightarrow S_{ab} |b \rangle$.  Ref.~\onlinecite{WittenJones} showed that such a transformation on the surface of the torus can be obtained by threading a Wilson line $\Omega \equiv \sum_{i=0}^k S_{0i} |i\rangle$ around the solid torus enclosed by the $T^2$ space-like surface upon which the modular transformation is to act. (It turns out that, for the cases of interest where the associated CFT is rational and $S$ is unitary,  this gives precisely the same definition of $\Omega$ as  in Eq. (\ref{Eq_Omega}) ).  Of course, we could equally well add no Wilson lines to the theory but instead excise the torus, perform the modular transformation, and then glue it back into the space-time manifold.  Hence adding a Wilson line labeled by $\Omega$ to the theory is equivalent to computing the partition function on a different manifold, related to the first via surgery.

\begin{center}
\begin{figure}[h!]
 \includegraphics{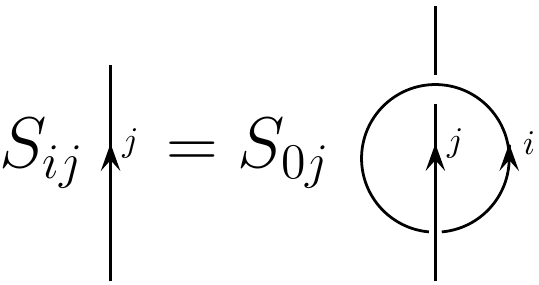}
\caption{ \label{SurgApFig} The $S$ matrix can be used to unlink loops from lines.}
\end{figure}
\end{center}

Incidentally, the connection between the element $\Omega$ and the modular $S$ matrix of a rational CFT furnishes an easy proof that $\Omega$ projects onto $0$ flux\cite{Gilsetal}.  The precise correspondence which we will need is shown in Fig. \ref{SurgApFig}.
Using Fig. \ref{SurgApFig}, we see that passing a string labeled $j$ through the $\Omega$ loop gives a factor of 
  \be
\frac{1}{S_{0j}}\sum_{i=0}^k S_{0i} S_{ij}
 \ee
 times the diagram with only the string labeled $j$.  
Hence since the $S$ matrix is unitary and symmetric (and $S_{i0}$ is real), the partition function of any Wilson line linked with $\Omega$ {\it vanishes} unless the Wilson line carries trivial flux.  Thus a Wilson line $\mathcal{L}$ labeled with $\Omega$ is an operator that projects onto states with $0$ flux passing through $\mathcal{L}$.

Thus, through its connection to modular transformations in the corresponding rational CFT, we see that a loop labeled with $\Omega$ projects onto $0$ flux in the lattice model, and also represents surgery on the space-time.

\subsection{Surgery arguments for the chirality of quasi-particles} \label{SurgApp}

We are now almost ready to track the positions of the quasi-particles after surgery.  In order to do so simply, however, we will first present an alternative description of surgery, in which one obtains the $3$-manifold as the boundary of an appropriately constructed $4$-manifold.  In this case we can give explicit instructions for constructing the $4$ manifold by gluing handles to the boundary of the $4$D ball (a.k.a. the $3$-sphere) along the strands in a link.  The $3$ manifold is then given by taking the boundary of this $4$-manifold.

The reason that this can be done at all is that all of the interesting topology of the boundary of the $4$-manifold can be captured by attaching {\it $2$ handles} alone to the $S^3$ boundary of the $4$ ball.  A $2$-handle is, by definition, something that gets glued onto a manifold $M$ along the boundary (edge) of a disk.  In $2$ dimensions, attaching a $2$-handle to a manifold $M$ means precisely gluing a disk to $M$ along its boundary.  In higher dimensions the process is essentially the same, except that the object to be glued is a disk thickened in the appropriate number of dimensions.  The crucial difference, however, is that in higher dimensions the (thickened) circle along which the disk is glued in can be twisted and knotted-- hence this gluing can, in fact, produce any $3$ manifold as the boundary of a $4$ manifold\cite{Gompf}.

To give the flavor of how this produces the same end-product of surgery as described above, let us consider the simplest possible
link-- a single, un-knotted circle.  The recipe, then, is to
thicken this circle into a solid tube, and remove this tube from
$S^3$.  Next, glue a $4$D $2$-handle onto $S^3$ by attaching it
along the excised region.  The relevant feature of the $2$-handle
is that for every longitude of the excised tube, there is an
associated disk in the $2$ handle which is attached along this
longitude.  (This is simply the higher-dimensional analogue of
saying that gluing in a $2$-handle amounts to sewing on a disk
along its $S^1$ boundary -- here we sew in a (meridinal) circle's
worth of disks along their (longitudinal) $S^1$ boundaries).  The
boundary of the finished product consists of a $3$-sphere from
which a solid tube has been removed, and glued back in {\it with
the meridian and longitude of its $2$-torus boundary
interchanged}.  In fact, this description is valid for surgery on
any link: surgery can be performed by thickening each strand of
the link to a hollow tube in $S^3$, excising all of the tubes, and
then gluing them back in with meridian and longitude interchanged.

Armed with this understanding of surgery, we may now consider the
fate of the \chainmail link and its quasi-particles. Let us first
introduce some convenient terminology.  A handle decomposition of
a $3$-manifold is made up of a manifold  $H^{+}$ of $0$ and $1$
handles (thickened vertices and edges, in our lattice
construction), a manifold $H^{-}$ of $2$ and $3$ handles
(thickened plaquettes, together with solid balls filling in the
cells on the lattice), and a surface $\Sigma$ which bounds both $H^+$ and $H^-$ equipped with
information about how these two are glued together.   Thus the 3-manifold is written $M=H^+ \cup \Sigma \cup  H^-$.   (This is known as a Heegard splitting.)  A convenient
way to describe surgery on the \chainmail link is given by Barrett
et. al. \cite{Martins2}, and is based on the following fact:
performing surgery on the plaquette (resp. edge) strings in
$\Sigma \times I$ gives $H^- \#_{h_3} \overline{H}^-$ (resp. $H^+ \#_{h_0}
\overline{H}^+$).  (The subscripts $h_0$ and $h_3$ denote a connect sum for each vertex or $3$-handle, respectively, in the lattice, as in Ref.~\onlinecite{Martins2}.)  Basically, this is because surgery attaches
each of the required $3$-dimensional $2$-handles, crossed with an
interval in the $4^{th}$ dimension, to $\Sigma$.  Taking the
boundary of this object, we obtain a manifold which has $2$
oppositely oriented copies of the $2$-handles -- but only one copy
of each of the $3$-handles which fill these in.  This is equivalent to
taking $2$ oppositely oriented copies of $H^-$, cutting out all of
the $3$-handles from one copy, and gluing the other copy in along
these excised balls.  The same construction holds for surgery on
the edge loops, except that in this case there is one connect sum
for each vertex in the lattice.

Now, imagine decomposing $S^3$ in the following way: start with $H^-$, the ensemble of $2$ and $3$ handles.  Glue onto this a copy of $\Sigma \times I$ which contains the edge strings of the \chainmail link.  (Since $\Sigma = \partial H^-$, we can glue these together on $\Sigma \times \{0\}$).  Glue in a second copy of $\Sigma \times I$, containing the plaquette strings.  (Here we join $\Sigma \times \{1\}$ in the first copy to $\Sigma \times \{0 \}$ in the second).  Finally, glue this ensemble into the $3$ sphere with $H^-$ removed  (which we write as $S^3 / \mbox{ Int} (H^-) $ .  This gives:
\be
\label{eq:divideS3}
S^3 = H^- \mycup (\Sigma \times I)_+ \mycup (\Sigma \times I)_- \mycup \left (S^3 / \mbox{ Int} (H^-) \right)
\ee
where the subscripts $+$ and $-$ remind us that we will do surgery on the plaquette strings in the first copy of $\Sigma \times I$, and on the edge strings in the second copy.  Here and below we use the notation that the order of writing terms shows the order in which pieces are attached together (i.e, each term is connected to the term listed before and after it).

After performing surgery on the two sets of curves, we obtain:
\ba \label{Eq_SurgResults}
&& H^- \mycup H^+ \#_{h_0} \overline{H}^+ \mycup \overline{H}^- \#_{h_3} H^- \mycup \left(S^3 / \mbox{ Int} (H^-) \right) \n
&=& M^{(3)} \#_{h_0} \overline{M}^{(3)} \#_{h_3} S^3
\ea
where $M^{(3)} \equiv H^- \mycup H^+ $ is the $3$ manifold whose handle decomposition we have used to construct the \chainmail link.  (As argued by Ref.~\onlinecite{Martins2}, the multiple connect sums do not change the partition function of the resulting manifold, and can be dropped).  Note that crucially, the  $M^{(3)}$ can be traced back to $H^- \mycup (\Sigma \times I)_+$ in Eq. \ref{eq:divideS3} whereas the $\overline{M}^{(3)}$ is traced back to $(\Sigma \times I)_+ \mycup (\Sigma \times I)_-$.  We refer the reader to Ref.~\onlinecite{Martins2} for more details of this construction.

Hence surgery on the \chainmail link produces two copies of $M$, of opposite chirality, joined in a connected sum. (Or rather, by multiple connect sums, one for each vertex in the lattice).  Where do quasi-particle world-lines land after this surgery?  Right-handed quasi-particle strings are linked only through the edge loops, and hence can be continuously deformed such that they live in the $H^- \mycup (\Sigma \times I)_+$ component of the original decomposition.  After surgery, they therefore land in $M^{(3)}$.  Left-handed strings are linked through both edge and plaquette loops, and hence visit both $(\Sigma \times I)_+ $ and $(\Sigma \times I)_- $ portions of the decomposition.  Since the first  $H^{-}$ in Eq. (\ref{Eq_SurgResults}) by construction contains no quasi-particle strings (and since the $0$- and $3$-cells which are deleted to take the connect sum also cannot contain quasi-particle strings), these must reside in the $\overline{H}^+ \mycup \overline{H}^-$ portion of the final manifold.   After surgery, they are therefore found in $\overline{M}^{(3)}$.  Thus right-handed strings land in the right-handed copy of $M^{(3)}$, and left-handed strings in the left-handed copy, $\overline{M}^{(3)}$.  This gives them opposite statistics.

\section{Lattice models in non-braided categories} \label{NonModApp}

Here we will briefly explain how the construction outlined in the main text can be applied to lattice models based on tensor categories which are not braided -- that is, to tensor categories which have a consistent set of fusion rules, but no matrix $R$ (c.f. Eq. \ref{RFig}) specifying how to un-twist the strands.  Thus although these categories have a well-defined fusion
structure, there is no notion of a braiding structure, ie. of how to evaluate over- and under- crossings of strings.   Much less is known about these types of categories, and no analogue of surgery exists; however we note that a similar construction to that of Sect.~\ref{LWSect} can be used to obtain the ground-state partition function, and comment on how quasi-particles can be incorporated into the theory.

\subsection{Ground state partition function}

First, we will explain how the ground state partition function of a Levin-Wen Hamiltonian can be evaluated using our pictorial representation.  The idea is very similar to the case of doubled anyon theories, except that we must replace the diagram in Fig. \ref{CHFig} with one in which no strings are linked: in Sect.~\ref{LWSect}, we expressed both vertex and plaquette projectors in the pictorial model using $\Omega$ loops.  When the category is not braided this is no longer possible, as we cannot resolve diagrams in which these edge loops are linked around the plaquette loops.

To circumvent this difficulty, we may replace the edge loops in the \chainmail diagram with an appropriate formulation of the projector onto the $0$ string\cite{ThanksParsa}.  The action of the vertex projectors is then carried out by the appropriate projector in the center of each edge, and the action of the plaquette projector $B_P$ can be implemented with an $\Omega$ loop encircling the plaquette, as before.  Interestingly, this suggests that the
symmetry between lattice and dual lattice descriptions for models constructed from anyon theories is no longer possible for categories which have no braiding structure.

To evaluate such diagrams, we must further stipulate that we first act with all projectors on the edges, then slide the separated (but un-linked) diagrams apart if necessary, before evaluating them using the rules of the fusion category.  Further, though for the ground state all such diagrams {\em can} be drawn as planar graphs (i.e. with no strings crossing), in any particular projection of the knot the diagrams may not appear planar.  As the category contains no rules for un-doing crossings, one loses the general notion of projection independence, and must evaluate the diagrams in a projection in which they are planar.  We emphasize that both of these choices are external to the rules of the category, and must be made such that the diagrams to be evaluated contain no crossings.

With this caveat in mind, we can follow the steps for evaluating the ground state partition function exactly as in Sect.~\ref{LWSect}.  Again, we find that in the absence of plaquette strings, edge variables propagate unaltered upwards in time, provided that they satisfy the constraint imposed by $B_V$ at each vertex.  Strings associated to space-like plaquettes act by fusing with the edge variables (to satisfy the constraint of fusing to the $0$ string along {\em space-like} edges), producing the Levin-Wen action (Eq. (\ref{Eq_PlStr})) for $\mathcal{{B}}_P(s)$.  As before, in evaluating the partition function, summing over edge variables with appropriate weights leads us to label the time-like plaquette strings with $\Omega$ as well.

Since the diagram is no longer a link diagram, only some of the interesting features described in Sect.~\ref{CHSect} apply in this case.  We note, however, that it is still possible to establish a connection between the Levin-Wen partition function and  the Turaev-Viro invariant.
Specifically, if the lattice is chosen such that all vertices in the {\em $3$D} lattice have valence $4$ (in other words, if it is chosen such that its dual lattice gives a triangulation of the space-time $\cal M$), then acting with the edge projectors reduces the resulting partition function to the evaluation of a tetrahedral diagram at each vertex.  Hence in this case, $Z_{LW}$ is clearly the Turaev -Viro invariant of $\cal M$.  Though we expect that these models also should depend only on the topology of $\cal M$, and not the choice of lattice, we hence do not furnish a proof of this fact here-- the proofs of independence of handle decomposition (which translates, for our purposes, to independence of the $3$D lattice chosen) given by Roberts\cite{RobertsThesis}, do not trivially generalize to unbraided categories.

\subsection{Quasi-Particles} \label{QP2Sect}

A remaining question is whether the pictorial construction can
accommodate quasi-particles in the case that the category does not describe an anyon theory.  In this section we consider whether quasi-particles can be defined similarly in these more general cases.  We find that the minimal way in which quasi-particles can be introduced is by assuming that the category has a half-braided structure -- that is, that for each quasi-particle type, one type of crossing (either over or under) may be defined in a manner consistent with the rules of fusion.  Interestingly, there are categories of this type which indeed cannot be endowed with a full braiding structure\footnote{for example, see Ref. \onlinecite{HalfBraid}}, so that these models are nonetheless more general than those derived from anyon theories.

The principle difficulty here is that, as quasi-particle strings must cross between different $3$ cells in the lattice, in general it is not possible to find a projection in which quasi-particle strings do not cross the plaquette strings.   In at least some of the diagrams, after all edge projectors are applied, crossings will remain.  In principle this is less of a problem for R quasi-particles, as they are not in fact linked with the plaquette strings.  However, in a projection where R particles do not cross any plaquette strings, the L quasi-particles cross both under and over the plaquette loops with which they are linked. To evaluate such diagrams requires a rule for how to undo both over and under crossings.  By definition, such a structure cannot be consistently assigned unless the category has a braiding structure, and hence such diagrams can only be evaluated if the category is in fact an anyon theory.

It is instructive to first understand how this issue is resolved in the Levin-Wen construction.  If the theory is constructed from an anyon theory, then the quasi-particle excitations are right- and left- handed copies of the initial category.  For example, if strings represent world lines of particles in the of $SU(2)_k$, then the final model describes doubled Chern-Simons theory with a gauge group $SU(2)_{k,R} \times SU(2)_{k,L}$  (or equivalently $SU(2)_k \times SU(2)_{-k})$.  Strings representing particles in the right- and left- handed sectors respectively correspond exactly to the right- and left- handed quasi-particles identified in the previous section.

If the tensor category has no braiding structure, however, the process of `doubling' is more complex.  In practice, the construction of Ref.~\onlinecite{LW} requires assigning a phase to a given string type for each crossing -- though in the $2D$ picture, one need not specify whether these crossings are over-crossings or under-crossings.  Mathematically speaking, then, the tensor category is not braided  -- but quasi-particles cannot be introduced in these models without specifying some additional information about crossings.  We can think of this extra data as specifying how to resolve either over- or under- crossings, but not both.

We may use this additional data to specify rules for resolving under-crossings of left-handed quasi-particles, and over-crossings of right-handed particles.  In this way, the pictorial construction can reproduce the more general models without full braiding structure -- though the algorithm for evaluating the partition function apparently cannot be stated in a projection-independent fashion.
However, as noted above, in this more general case one does not expect the evaluation of the diagrams to be projection independent, and hence it is not surprising that a specific projection (in this case, the projection which looks down at a time slice from the positive $\tau$ direction) must be chosen.

This construction leaves open many interesting questions about the nature of these more general theories.  When the model is not built from an anyon theory, the final quasi-particles consist not of individual  right- and left- strings, but rather of specific linear combinations which yield quasi-particles for which braiding is well-defined.  Mathematically speaking, the result is the {\it Dreinfeld double} of the initial category\cite{Kassel}.   It would be interesting to understand, on grounds other than mathematical consistency of the category, how these preferred combinations arise.  Further, one is tempted to ask whether an analogue of the connection to a continuum topological gauge theory in the anyon theory case is possible here: is there a modified surgery procedure which allows for such a connection to be made?  We will address these questions in a future publication.

\end{appendix}

{\bf Acknowledgements}

The authors wish to thank M. Freedman, K. Walker, M. Levin, P. Bonderson, S. Trebst, S.L. Sondhi, and especially Z. Wang for numerous helpful discussions during the course of this work.  FJB is grateful for the support of KITP, All Souls College, and Microsoft Station Q.  SHS is grateful for the hospitality of KITP and Station Q where some of this work was completed.

\bibliography{CHBib}

\end{document}